\documentclass[iop]{emulateapj}
\usepackage{rotating} 
\usepackage{color}

\newcommand{\kms}       {\mbox{km s$^{-1}$}}%
\newcommand{\kmsMpc}    {\mbox{km s$^{-1}$ Mpc$^{-1}$}}%

\newcommand{\msun}      {\mbox{$M_\odot$}}

\newcommand{\tabphot}   {\mbox{1}} 
\newcommand{\tabhikin}   {\mbox{2}} 

\shortauthors{Kannappan et al.}
\shorttitle{Transitions}
\slugcomment{draft}
 
\begin{document}

\title{Connecting Transitions in Galaxy Properties to Refueling}

\author{Sheila J. Kannappan\altaffilmark{1},
	David V. Stark\altaffilmark{1},
	Kathleen D. Eckert\altaffilmark{1},
	Amanda J. Moffett\altaffilmark{1},
	Lisa H. Wei\altaffilmark{2,3},
	D. J. Pisano\altaffilmark{4,5},
	Andrew J. Baker\altaffilmark{6},
	Stuart N. Vogel\altaffilmark{7},
	Daniel G. Fabricant\altaffilmark{2},
	Seppo Laine\altaffilmark{8},
	Mark A. Norris\altaffilmark{1,9},
	Shardha Jogee\altaffilmark{10},
        Natasha Lepore\altaffilmark{11},
        Loren E. Hough\altaffilmark{12},
        Jennifer Weinberg-Wolf\altaffilmark{1}}
\altaffiltext{1}{Department of Physics and Astronomy, 
  University of North Carolina, 290 Phillips Hall CB 3255, 
  Chapel Hill, NC 27599, USA; sheila@physics.unc.edu}
\altaffiltext{2}{Harvard-Smithsonian Center for Astrophysics, 60
  Garden St. MS-20, Cambridge, MA 02138, USA}
\altaffiltext{3}{Atmospheric and Environmental Research, 131 Hartwell
  Avenue, Lexington, MA 02421, USA}
\altaffiltext{4}{Department of Physics, West Virginia University, 
  P.O. Box 6315, Morgantown, WV 26506, USA}
\altaffiltext{5}{Adjunct Assistant Astronomer, National Radio Astronomy 
  Observatory, P.O. Box 2, Green Bank, WV 24944, USA}
\altaffiltext{6}{Department of Physics and
  Astronomy, Rutgers, the State University of New Jersey, 136
  Frelinghuysen Road, Piscataway, NJ 08854-8019, USA}
\altaffiltext{7}{Department of Astronomy, University of Maryland,
  College Park, MD 20742-2421, USA}
\altaffiltext{8}{Spitzer Science Center, Caltech, MS 220-6, Pasadena,
  CA 91125, USA}
\altaffiltext{9}{Max Planck Institut fur Astronomie, Konigstuhl 17,
  D-69117, Heidelberg, Germany}
\altaffiltext{10}{Department of Astronomy, University of Texas at
  Austin, Austin, TX 78712, USA}
\altaffiltext{11}{Department of Radiology, University of Southern
  California and Children's Hospital Los Angeles, 4650 W Sunset Blvd,
  MS\#81, Los Angeles, CA 90027, USA}
\altaffiltext{12}{Department of Physics, University of Colorado at
  Boulder, Boulder, CO 80309-0390, USA}

\begin{abstract}
We relate transitions in galaxy structure and gas content to
refueling, here defined to include both the external gas accretion and
the internal gas processing needed to renew reservoirs for star
formation. We analyze two $z=0$ data sets: a high-quality
$\sim$200-galaxy sample (the Nearby Field Galaxy Survey, data release
herein) and a volume-limited $\sim$3000-galaxy sample with reprocessed
archival data. Both reach down to baryonic masses
$\sim$10$^9$\,\msun\ and span void-to-cluster environments. Two
mass-dependent transitions are evident: (i) below the ``gas-richness
threshold’’ scale ($V\sim125$\,\kms), gas-dominated quasi-bulgeless
Sd--Im galaxies become numerically dominant, while (ii) above the
``bimodality’’ scale ($V\sim200$\,\kms), gas-starved E/S0s become the
norm. Notwithstanding these transitions, galaxy mass (or $V$ as its
proxy) is a poor predictor of gas-to-stellar mass ratio $M_{\rm
  gas}/M_*$. Instead, $M_{\rm gas}/M_*$ correlates well with the ratio
of a galaxy’s stellar mass formed in the last Gyr to its preexisting
stellar mass, such that the two ratios have numerically similar
values. This striking correspondence between {\it past-}averaged star
formation and {\it current} gas richness implies routine refueling of
star-forming galaxies on Gyr timescales. We argue that this refueling
underlies the tight $M_{\rm gas}/M_*$ vs.\ color correlations often
used to measure ``photometric gas fractions.’’ Furthermore, the
threshold and bimodality scale transitions reflect mass-dependent
demographic shifts between three refueling regimes --- {\it accretion
  dominated, processing dominated,} and {\it quenched}. In this
picture, gas-dominated dwarfs are explained not by inefficient star
formation but by overwhelming gas accretion, which fuels stellar mass
doubling in $\la$1~Gyr. Moreover, moderately gas-rich bulged disks
such as the Milky Way are transitional, becoming abundant only in the
narrow range between the threshold and bimodality scales.
\end{abstract}


\section{Introduction}
\label{sec:intro}

Galaxies grow both by merging and by fresh gas accretion.
Hierarchical models that follow the merger histories of galaxies and
their host dark matter halos successfully explain the large-scale
structure of the universe, yet these models have difficulty
reproducing the relative abundance of disk-dominated
vs.\ bulge-dominated galaxies across a broad range of environments
\citep[e.g.,][]{navarro.white:simulations,abadi.navarro.ea:simulations,donghia.burkert:bulgeless,stewart.bullock.ea:merger,martig.bournaud.ea:diversity}.
Broadly speaking, this failure reflects the disk-destroying nature of
stellar-mass dominated mergers \citep[and can therefore be mitigated
  by ``quiet'' merger histories such as may be found in low-density
  environments;][]{weinzirl.jogee.ea:bulge,fontanot.de-lucia.ea:other}.\footnote{The
  failure may also be compounded by the extreme loss of gas angular
  momentum in some simulations, although implementing star formation
  feedback and/or higher mass and force resolution can counteract this
  problem
  \citep[e.g.,][]{weil.eke.ea:formation,governato.willman.ea:forming}.}
On the other hand, gas-rich mergers are much less destructive and may
even help to build disks. Such mergers are expected to predominate at
low galaxy masses and/or early epochs
\citep[e.g.,][]{robertson.bullock.ea:merger-driven,hopkins.somerville.ea:effects,stewart.bullock.ea:gas-rich}.
Fresh gas accretion may also rebuild disks in low-mass E/S0 merger
remnants, potentially restoring late-type morphologies
\citep[e.g.,][]{cox.sparke.ea:stars,morganti..ea:neutral,stark.kannappan.ea:fueling}.

Observations point to significant cold gas accretion onto galaxies
\citep[][and references
  therein]{sancisi.fraternali.ea:cold}. Moreover, the dynamics of halo
gas suggest that this gas can in principle provide the angular
momentum needed for the rapid growth of disks
\citep{stewart.kaufmann.ea:orbiting}.  Cosmological hydro simulations
show large-scale ``cold'' ($\sim$10$^5$--$10^6$K) gas flows that
travel along the filaments and walls of the cosmic web
\citep[e.g.,][]{kere-s.katz.ea:how}. Observational signs of such flows
have indeed been found
(\citealt{zitrin.brosch:ngc,stanonik.platen.ea:polar,narayanan.wakker.ea:cosmic};
see also
\citealt{giavalisco.vanzella.ea:discovery,churchill.kacprzak.ea:quenched}). It
is not yet clear to what extent these flows remain cold or shock-heat
upon halo entry (compare \citealt{nelson.vogelsberger.ea:moving}
vs.\ \citealt{birnboim.dekel:virial} and
\citealt{kere-s.katz.ea:how}).  These details may affect angular
momentum delivery as well as the onset of rapid accretion, which
typically occurs when the cooling radius exceeds the virial radius,
below a characteristic mass scale that depends on the model
\citep{lu.kere-s.ea:on}.  Regardless of these specifics, it is a
general feature of recent models that cosmic gas accretion accounts
for a larger percentage of galaxy growth in low-mass halos than can be
attributed to merging.  Thus the physics of accretion can dramatically
change the balance of bulges and disks within the hierarchical merging
paradigm.

Clearly mass-dependent gas physics affects patterns of growth by both
mergers and accretion, and transitions in gas physics may lie at the
heart of understanding the disky morphologies and overall growth
histories of galaxies.  Two galaxy mass scales have been previously
noted as important transition points in morphology, gas richness
(defined as gas-to-stellar mass ratio in this paper), and star
formation history: the ``bimodality scale'' and the ``gas-richness
threshold scale.''

The bimodality scale is typically identified with stellar mass
$M_*\sim10^{10.5}$\,\msun, which corresponds to rotation velocity
$V\sim200$\,\kms\ (see \S\ref{sec:disting} herein). This scale marks
the crossover point in relative abundance of young disk-dominated
vs.\ old spheroid-dominated stellar populations
\citep[][]{kauffmann.heckman.ea:dependence}.  As traced by late-type
vs.\ early-type morphology, this transition appears to shift downward
in mass over cosmic time \citep{bundy.ellis.ea:mass*1}.  Equivalently,
the bimodality scale marks a shift in the relative number density of
galaxies on the red and blue sequences in $u-r$ color vs.\ stellar
mass $M_*$ parameter space \citep{baldry.glazebrook.ea:quantifying},
which are associated with ``red and dead'' galaxies that have a strong
4000\AA\ break and blue star-forming systems, respectively. AGN
activity in early-type galaxies peaks up just below the bimodality
scale in a population that may be evolving toward the red sequence,
suggesting black hole growth in tandem with spheroid formation
\citep[][]{schawinski.urry.ea:galaxy}.  The slope of the gas-phase
metallicity vs.\ $M_*$ relation flattens above the bimodality scale,
indicating changes in the interplay of gas cooling/infall, gas
consumption, and gas loss in metal-enriched outflows
\citep{tremonti.heckman.ea:origin}.  Hot gas halos become common above
the bimodality scale \citep[][translating their $K$-band
  magnitudes to equivalent stellar masses]{mulchaey.jeltema:hot},
potentially enhancing the efficacy of AGN feedback
\citep{dekel.birnboim:galaxy}.

A second, lower-mass transition scale was previously highlighted by
\citet{dekel.silk:origin}, in a scenario explaining low-metallicity
diffuse dwarf galaxies as the result of global gas loss caused by
supernova winds acting in the shallow potential wells of
$V\la100$\,\kms\ dark matter halos. More recent work has shown that
such ``blowaway'' (as distinct from local ``blowout'') should occur
only in much smaller halos, near
$V\sim30$\,\kms\ \citep{mac-low.ferrara:starburst-driven}.  Blowaway
near $V\sim100$\,\kms\ would be in any case hard to reconcile with the
fact that ``high-mass dwarf'' galaxies are typically gas rich rather
than gas poor \citep[e.g.,][hereafter
  K04]{bettoni.galletta.ea:new,kannappan:linking}.  In fact gas-{\it
  dominated} galaxies become typical of the blue sequence below
$M_*\sim10^{9.5-10}$\,\msun\ (corresponding to
$V\sim125$\,\kms; \S\ref{sec:disting} herein), as gas fractions
rise on both sequences (K04\footnote{Identifying this shift in gas richness with the threshold scale requires shifting the stellar mass
zero point from \citealt{bell.mcintosh.ea:optical} as used by K04 to
coincide with that of \citealt{kauffmann.heckman.ea:dependence}.};
\citealt{kannappan.wei:galaxy,kannappan.guie.ea:es0}, hereafter
  KGB). We therefore refer to the $V\sim125$\,\kms\ scale as the ``gas-richness
  threshold scale,'' following KGB.  Calculations by
  \citet{dalcanton:metallicity} suggest that an increase in gas richness
  is essential to explain changes in metallicity at the
  threshold scale (specifically, the drop in effective yields below
  $V\sim125$\,\kms\ reported by
  \citealt{garnett:luminosity-metallicity}), and that a second
  essential ingredient is low star formation efficiency (in the sense
  of star formation rate divided by gas mass; we will revisit this
  concept in relation to cosmic accretion in  \S\ref{sec:accdom}).

In a separate study of edge-on, ``bulgeless'' disk galaxies,
\citet{dalcanton.yoachim.ea:formation} reported another, presumably
related change in interstellar medium (ISM) physics at the threshold
scale: thin, concentrated dust lanes emerge abruptly above
$V\sim120$\,\kms.  Moreover, despite the authors' best efforts to
select for bulgeless morphology, in practice their high-quality
follow-up imaging reveals a small ``three-dimensional'' bulge in every
sample galaxy above $V\sim120$\,\kms, suggesting a link between
changes in gas physics and galaxy structure.  The onset of inevitable
bulges above the threshold scale (also seen by \citealt{bell:galaxy})
occurs simultaneously with a sharp {\it decline} in the population of
``blue-sequence E/S0s,'' identified by KGB as gas-rich merger remnants
rebuilding disks (although higher-mass blue E/S0s are more often
quenching, especially above the bimodality scale;
\citealt{schawinski.lintott.ea:galaxy};
KGB). \citet{fisher.drory:demographics} report transitions in bulge
demographics at both the threshold and bimodality scales, with
bulgeless galaxies dominant below $M_*\sim10^{9.5}$\,\msun,
pseudobulges dominant between $10^{9.5}$--$10^{10.5}$\,\msun, and
classical bulges/elliptical galaxies dominant above
$10^{10.5}$\,\msun. Interestingly, the E/S0 mass--radius relation bifurcates
into two loci below the threshold scale (e.g., KGB;
\citealt{misgeld.hilker:families}).

Given the variety of physical processes and galaxy properties changing
at the threshold and bimodality scales, these closely spaced mass
scales have been conflated by multiple authors, including K04 and
\citet{dekel.birnboim:galaxy}, and the blurring of the two has been
further exacerbated by systematic differences in stellar mass zero
points between investigators
\citep[see][]{kannappan.gawiser:systematic,kannappan.wei:galaxy}.  Yet
appreciating the distinction between the threshold and bimodality
scales is of key interest, since we will demonstrate herein that only
in the narrow mass range between the threshold and bimodality scales
do galaxies like our Milky Way --- intermediate between gas-dominated
bulgeless disks and gas-starved spheroids --- become typical in the
galaxy population.  An exploration of the transitions occurring at the
threshold and bimodality scales may therefore shed light on our
Galaxy's past and future.

In what follows we employ multi-wavelength data from two complementary
samples described in \S\ref{sec:data} to explore structural and gas
richness ($M_{\rm gas}/M_*$) transitions across the threshold and
bimodality scales in \S\ref{sec:disting}. We show that notwithstanding
these notable transitions, there is far greater scatter in $M_{\rm
  gas}/M_*$ vs.\ galaxy mass than has been previously appreciated.  In
contrast, we demonstrate in \S\ref{sec:regimes} that $M_{\rm gas}/M_*$
correlates in a surprisingly one-to-one fashion with a quantity we
refer to as the ``long-term fractional stellar mass growth rate''
(FSMGR$_{\rm LT}$), which considers star formation integrated over the
last Gyr and is defined such that it can exceed one over the unit of
time, unlike a specific star formation rate.  We argue that the
FSMGR$_{\rm LT}$--$M_{\rm gas}/M_*$ correlation, and not the
Kennicutt-Schmidt Law, underlies the tight observed relation between
$M_{\rm gas}/M_*$ and ultraviolet/blue minus near-infrared colors (hereafter,
``U$-$NIR'' colors) previously reported by K04.  Moreover, we propose that
coordinated changes in morphology and gas richness as a function of
FSMGR$_{\rm LT}$ can be usefully understood in terms of changes in
cosmic accretion and internal gas processing between three ``refueling
regimes'': accretion-dominated, processing-dominated, and quenched.
Finally, we tie our results back to the threshold and bimodality
scales, showing that these scales represent transitions in the
relative numerical dominance of galaxies in the three regimes, likely
tied to the halo mass dependence of cosmic accretion.  Our results
suggest a reevaluation of dwarf galaxies as not ``inefficient'' gas
consumers but ``overwhelmed'' gas accretors, and of moderately
gas-rich bulged disks like our Milky Way as not ``normal'' but
``transitional'' in the galaxy population.

\section{Data and Methods}
\label{sec:data}

Our analysis relies on two samples. The highest quality HI data,
multi-band photometry, and kinematic data come from the Nearby Field
Galaxy Survey (NFGS, \citealt{jansen.franx.ea:surface,
  jansen.fabricant.ea:spectrophotometry, kannappan.fabricant:broad,
  kannappan.fabricant.ea:physical, wei.kannappan.ea:gas}), a broadly
representative sample of $\sim$200 galaxies spanning stellar masses
$M_*\sim10^8$--$10^{12}$\,\msun\ and all morphologies.  Better
statistics are offered by a volume-limited sample of $\sim$3000
galaxies, hereafter the ``V3000'' sample, with flux-limited HI data
and partial kinematic information from the blind 21cm Arecibo Legacy
Fast ALFA
\citep[ALFALFA,][]{giovanelli.haynes.ea:arecibo,haynes.giovanelli.ea:arecibo}
survey, which we combine with reprocessed photometry from the Sloan
Digital Sky Survey \citep[SDSS,][]{aihara.allende-prieto.ea:eighth}
Data Release 8 (DR8), the GALEX mission archive
\citep{morrissey.conrow.ea:calibration}, and the Two Micron All Sky
Survey \citep[2MASS,][]{jarrett.chester.ea:2mass}.  We detail new and
reprocessed archival data for the NFGS and V3000 samples below.

We assume $H_0=70$ \kmsMpc\ and $d=cz/H_0$ throughout this work.
Neglecting $\Lambda$ introduces negligible errors at the low redshifts
of our sample galaxies.

\subsection{The Nearby Field Galaxy Survey}
\label{sec:nfgs}

The NFGS was drawn from a $B$-selected parent survey, the CfA~1
Redshift Survey \citep{huchra.davis.ea:survey}, in approximate
proportion to the luminosity function, and it preserves the CfA~1
survey's relative frequency of morphological types at each luminosity
\citep{jansen.franx.ea:surface}. The 196 galaxies in the NFGS obey an
artificial luminosity-distance correlation imposed to ensure that
their apparent diameters do not vary too much, as an observational
convenience. Thus large-scale environments are not uniformly sampled
as a function of luminosity, although a wide variety of environments
are represented ranging from underdense regions to the Coma Cluster.
Our analysis makes use of 190/196 galaxies in the NFGS, after
rejection of six objects with nearly point source morphology due to
powerful AGN or in one case a superposed star.

\begin{figure*}[t]
\plottwo{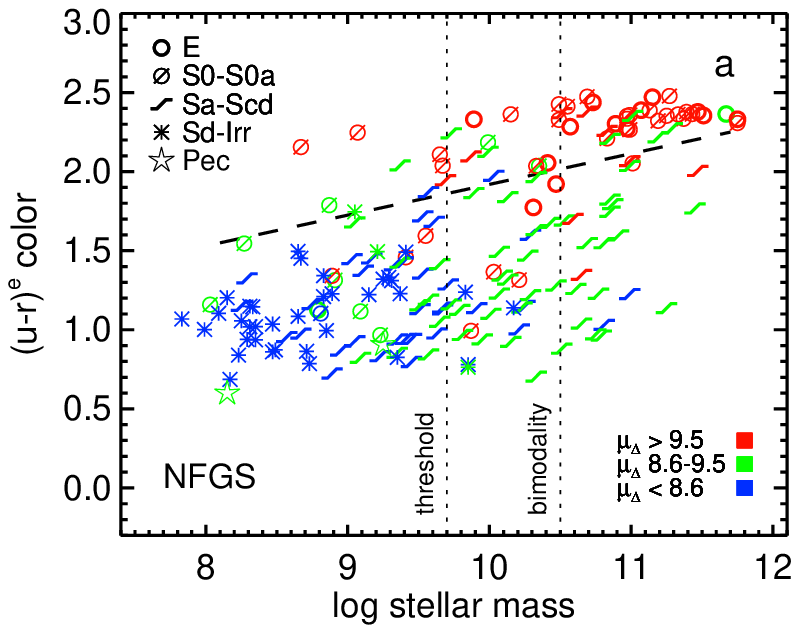}{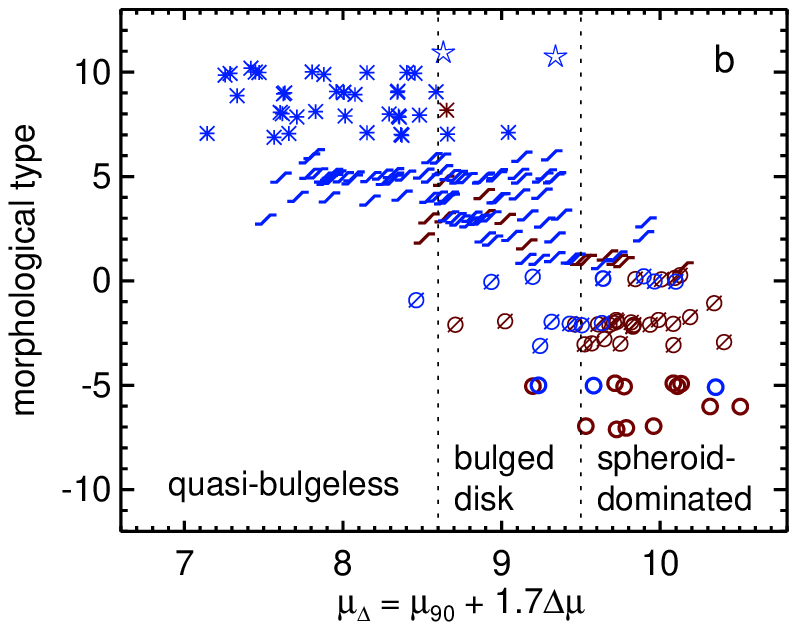}
\caption{Morphological type, color, and stellar mass distribution of
  the Nearby Field Galaxy Survey (NFGS). {\it (a)} Distribution of
  NFGS galaxy morphologies in $(u-r)^e$ color vs.\ stellar mass $M_*$
  parameter space, with symbol color corresponding to the $\mu_\Delta$
  classes defined by the dividing lines in panel {\it b}. Note that
  $(u-r)^e$ is a de-extincted color estimated by our stellar
  population modeling code and as such enhances the division between
  the red and blue sequences. However, it does not shift the basic
  locus of the red sequence significantly; rather, the bluer color of
  this locus compared to previous studies
  \citep[e.g.,][]{baldry.glazebrook.ea:quantifying} reflects
  improvements in our photometry compared to the SDSS pipeline, as
  described in \S\ref{sec:nfgsphot} and Fig.~\ref{fg:coltf}. {\it (b)}
  Calibration of the $\mu_\Delta$ parameter used to distinguish
  quasi-bulgeless, bulged-disk, and spheroid-dominated galaxies (see
  \S\ref{sec:nfgsphot}). Symbol color corresponds to the red/blue
  sequence division shown in panel {\it a}. Small random offsets have
  been applied to the morphological types to separate points.}
\label{fg:nfgscolmasstype}
\end{figure*}

Fig.~\ref{fg:nfgscolmasstype} illustrates the color, stellar mass, and
morphology distribution of these 190 galaxies.  We note that the
Hubble type classifications in Fig.~\ref{fg:nfgscolmasstype}a are
reliable but not perfect, due to the inclination-blind selection of
the NFGS as well as the fact that when the galaxies were
classified, \citet{jansen.franx.ea:surface} allowed fairly large
discrepancies between different classifiers to remain unresolved.
Following \citet{kannappan.guie.ea:es0}, we reclassify the polar ring
galaxy UGC~9562 as an S0. We also reclassify the very round dust-lane
galaxy NGC~3499 as an E.  Fig.~\ref{fg:nfgscolmasstype}b defines a new
quantitative morphology metric used as a complementary diagnostic,
discussed in \S\ref{sec:nfgsphot} below.

\subsubsection{Photometry and Stellar Masses}
\label{sec:nfgsphot}

Our analysis combines {\it UBR} photometry from
\citet{jansen.franx.ea:surface} and global (integrated slit-scanned)
spectrophotometry from \citet{jansen.fabricant.ea:spectrophotometry}
with our own custom reprocessed {\it GALEX} NUV, SDSS {\it ugriz},
2MASS {\it JHK}, and {\it Spitzer} IRAC 3.6$\mu$m photometry, provided
in Table~\tabphot.  For the new photometric measurements we
redetermine the position angle (PA) and axial ratio ($b/a$ = ratio of
semi-minor to semi-major axis) without reference to the Jansen et
al.\ values, but our kinematic observations and analysis predate the
new photometry and thus make use of measurements from Jansen et
al.\ as detailed in \S\ref{sec:optkin} and Table~\tabhikin. We adopt
foreground Milky Way extinction corrections from
\citet{schlegel.finkbeiner.ea:maps}, except at 3.6$\mu$m where such
corrections are negligible, and we adjust the Jansen et
al.\ photometry and spectrophotometry to match.

Initially, each galaxy is run through an optical photometric pipeline
that produces masks and freely determines elliptical apertures from a
deep {\it gri} coadded image (Eckert et al., in preparation).  The
outer disk PA and axial ratio are then fixed and used to determine a second set of elliptical apertures imposed on all bands
NUV+{\it ugrizJHK}+IRAC 3.6$\mu$m, enabling robust extrapolation of
total magnitudes, even at low S/N.  Final magnitudes and systematic
errors are calculated from a comparison of different methods (exponential profile
fitting, curve-of-growth, outer-disk color correction, and
large-aperture magnitude) for each band.  Our SDSS magnitudes are
measured using the newly optimized background sky estimates provided
with all DR8 images (though not actually incorporated in DR8 catalog
photometry), as described in \citet{blanton.kazin.ea:improved}. The
extended sky coverage of DR8 includes 177 of our 190 galaxies.  For
{\it GALEX} we adopt the background estimation provided by the mission
pipeline \citep{morrissey.conrow.ea:calibration}, and the available coverage yields NUV
magnitudes for 93 of the 190 galaxies.

In the near IR, custom background subtraction is necessary. IRAC
3.6$\mu$m images are available for 107 of our sample galaxies, mainly
courtesy of the {\it Spitzer} Survey of Stellar Structure in Galaxies
\citep[S4G,][]{sheth.regan.ea:spitzer} and our own program targeting
low-mass E/S0 galaxies (GO-30406, PI Kannappan).  We calculate 3.6$\mu
$m magnitudes using the level 2 (PBCD) images produced by the {\it
  Spitzer} pipeline, which have a residual low-level, non-uniform
background.  To remove this, we mask the primary galaxy and any other
bright objects in each frame, and the remaining image is convolved
with a median filter of size roughly four times the optical size of
the galaxy to create a smooth background map.  This map is subtracted
from the original image to yield the image on which we perform
photometric measurements.  The resulting magnitudes are in good
agreement with those from \citet{moffett.kannappan.ea:extended}, who
applied a similar background subtraction technique, with typical
differences of $\sim$0.04 magnitudes.  In turn, our 2MASS pipeline
incorporates background subtraction methods optimized with reference
to the IRAC imaging, which includes deep S4G imaging of dwarf
galaxies.  With this careful background subtraction and the imposition
of ellipses determined from the optical profile fits (using the PAs and
ellipticities from \citealt{jansen.franx.ea:surface} for the 13 galaxies
lacking SDSS data), we have found it possible to compute reliable {\it
  JHK} magnitudes for the entire sample, albeit sometimes with large
error bars.  As in the optical, we use multiple extrapolation
techniques to estimate systematic errors in the NIR, although we find
a curve of growth approach is generally most robust, especially for
shallow 2MASS data (see \citealt{stark.kannappan.ea:fueling},
hereafter S13, for further details on our NIR pipelines).  It is
noteworthy that our 2MASS magnitudes are well behaved in combined
stellar population fits of optical and near-IR data that include
high-quality spectrophotometry and deep IRAC photometry.

Table~\tabphot\ lists our NUV+{\it ugrizJHK}+3.6$\mu$m magnitudes
(including foreground Milky Way extinction corrections), with
uncertainties determined by combining Poisson errors with systematic
errors from profile extrapolation. The IRAC magnitude errors include
an extra 10\% uncertainty associated with the ``aperture corrections''
required for profiles extrapolated to infinity, which we have applied
as prescribed by the {\it Spitzer} IRAC instrument
handbook.\footnote{http://irsa.ipac.caltech.edu/data/SPITZER/docs/irac/
  iracinstrumenthandbook/}

\begin{figure*}[t]
\plottwo{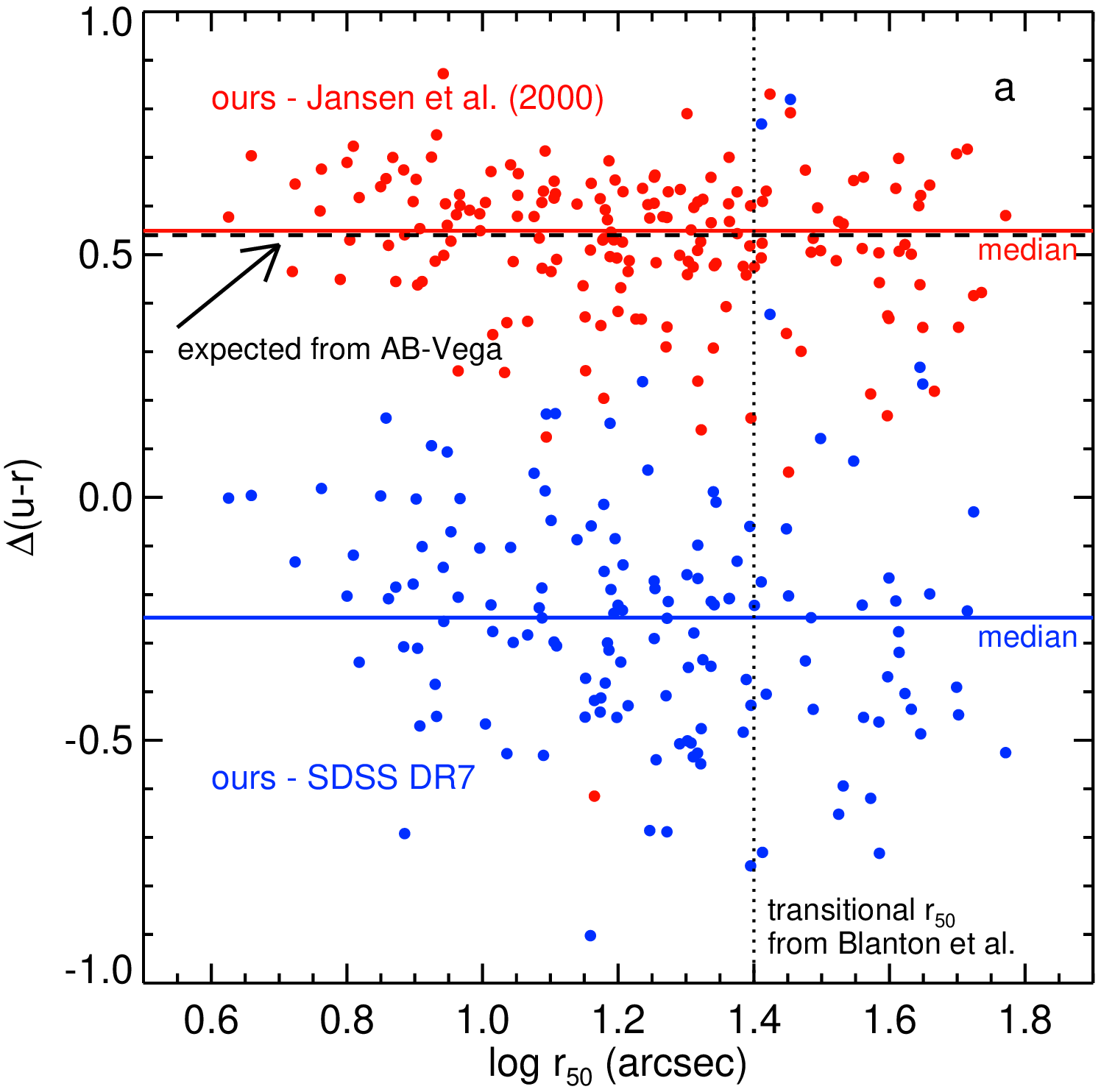}{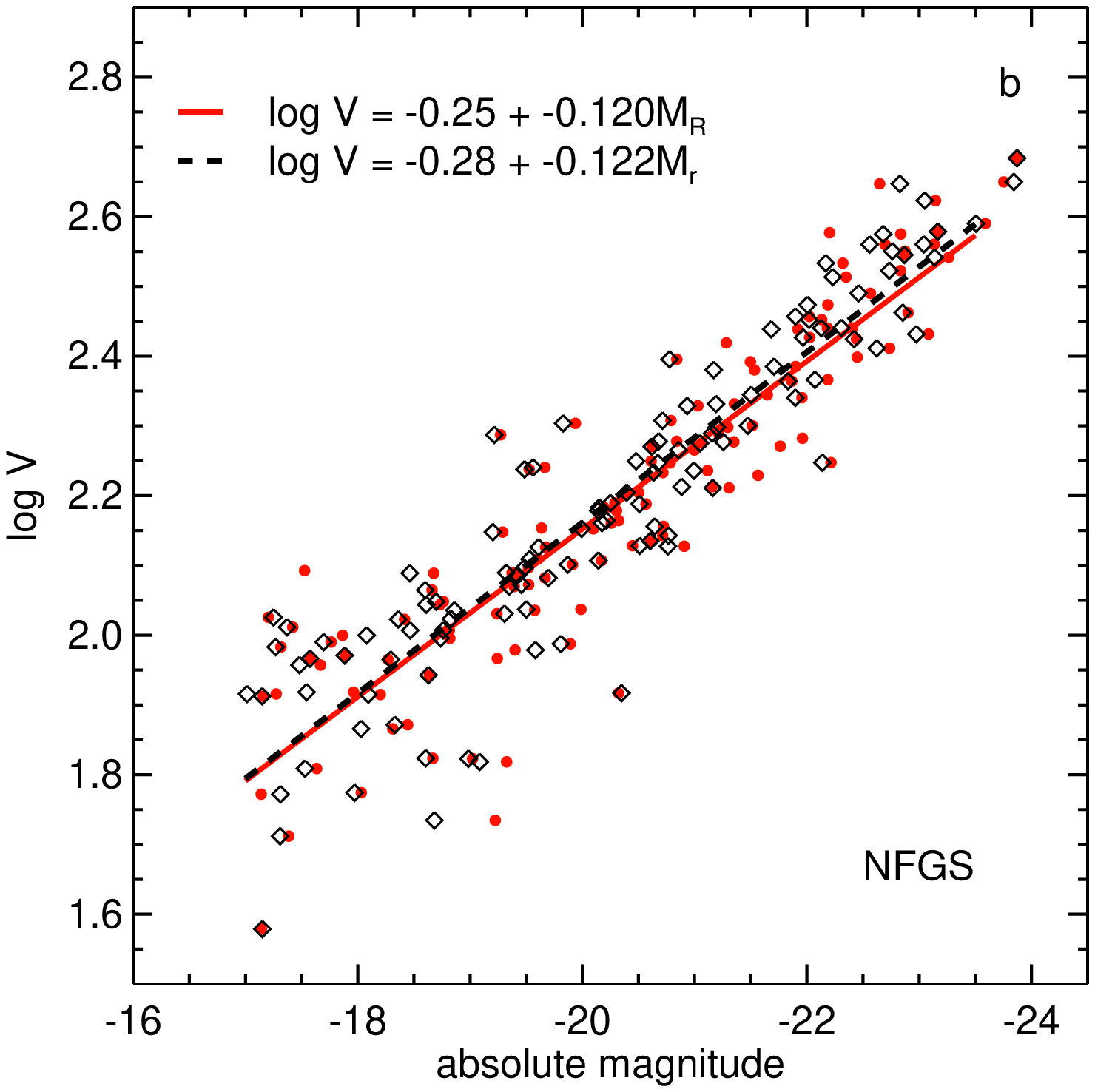}
\epsscale{0.5}
\caption{Comparisons of photometric and kinematic data for the NFGS.
  {\it (a)} Our newly measured $u-r$ colors compared to catalog $u-r$
  colors from SDSS DR7 \citep{abazajian.adelman-mccarthy.ea:seventh}
  and $U-R$ colors from \citet{jansen.franx.ea:surface}, as a function
  of $R$-band half-light radius (converted from the $\sqrt{ab}$
  convention of Jansen et al.\ to a major-axis convention).  Our
  colors are in excellent agreement with those of Jansen et al.\ after
  allowing for the known 0.04~mag AB offset of SDSS $u$ and the
  0.58~mag expected offset from Vega-to-AB conversions ($m_{\rm Vega,
    AB}$ of 0.79 and 0.21 for $U$ and $R$, respectively; see
  \citealt{blanton.roweis:k-corrections}).  The disagreement with SDSS
  catalog colors likely reflects (i) our use of the improved sky
  subtraction of \citet{blanton.kazin.ea:improved}, which those
  authors find to give bluer colors for $\log{r_{50}}\ga1.4$, (ii) the
  existence of color gradients, which are included in our photometry
  and that of Jansen et al.\ but forcibly set to zero by the SDSS
  model magnitude algorithm, and (iii) catastrophic errors in the SDSS
  pipeline, contributing to the large scatter including five extreme
  outliers outside the plot bounds, all with blue offsets
  $\Delta(u-r)<-1$. {\it (b)} Characteristic velocity $V$ (rotation
  speed or $\sqrt{2}\times$ stellar dispersion $\sigma$; see
  \S\ref{sec:optkin}) vs.\ $M_r$ and $M_R$.  Lines show forward fits
  minimizing residuals in $V$ with coefficients as indicated. The
  close coincidence of zero points is spurious, reflecting the
  canceling effects of the 0.21~mag Vega-to-AB conversion for $R$ and
  the $\sim$0.2~mag across-the-board zero point difference between the
  Jansen et al.\ {\it UBR} photometry and our own (see note
  \ref{fn:jansen}). Both $M_r$ and $M_R$ include foreground extinction
  and $k$-corrections but omit internal extinction corrections.}
\label{fg:coltf}
\end{figure*}

As illustrated in Fig.~\ref{fg:coltf}, our $u-r$ colors are
systematically $\sim$0.2~mag bluer than those determined from SDSS
catalog photometry, which is also evident in the fact that the red
sequence in Fig.~\ref{fg:nfgscolmasstype} is $\sim$0.2~mag bluer than
in previous studies using SDSS catalog data.  However,
the same figure shows that our colors are in close agreement with
those determined by \citet{jansen.franx.ea:surface}. We attribute
about half of the offset between our $u-r$ colors and those from the
SDSS catalog to the improved sky subtraction of
\citet{blanton.kazin.ea:improved}, who demonstrate that their new
protocol will yield $\sim$0.1~mag bluer colors for galaxies with
$\log{r_{50}}\ga1.4$ (marked in Fig.~\ref{fg:coltf}).  However, we
measure greater color differences than expected from the new sky
subtraction alone.  Some discrepancies certainly reflect catastrophic
failures of the SDSS pipeline (a few so extreme they lie outside the
plot boundaries), but some likely also reflect the fact that we (like
\citealt{jansen.franx.ea:surface}) measure magnitudes by methods that
permit color gradients, whereas the SDSS model magnitude algorithm
enforces a common profile in all bands, which is determined in the $r$
band.\footnote{http://www.sdss.org/dr7/algorithms/photometry.html\#mag\_model}
In the case of massive early type galaxies, for which the centers are
redder than the outskirts
\citep[e.g.,][]{la-barbera.de-carvalho.ea:spider}, the SDSS model
magnitude methodology would be expected to produce an ``overly red''
red sequence.

Stellar masses and other stellar population parameters are estimated
using a variant of the code described in
\citet{kannappan.gawiser:systematic} and improved by KGB, which
simultaneously fits both the spectral energy distribution (SED) and
the integrated spectrum (if available) of a galaxy with a suite of
composite stellar population models. The code combines old+young
simple stellar populations (SSPs) from \citet{bruzual.charlot:stellar}
in a grid of models with varying age, metallicity, and percentage of
young stars. The models allow for 11 different extinction/reddening
values ($\tau_{V}$ = 0, 0.12, 0.24... 1.2) applied to the young SSP
using the \citet{calzetti:dust} dust law.  Likelihoods and stellar
masses are computed for all models in the grid, and the median
of the likelihood-weighted stellar mass distribution provides
the most robust final stellar mass estimate, with uncertainties
determined from the 68\% confidence interval around the median.

In the present work we consider two different model grids.  The first,
also used by S13, assumes a ``diet'' Salpeter IMF
\citep{bell.mcintosh.ea:optical}.  It combines an old SSP (age 1.4,
2.5, 3.5, 4.5... 13.5~Gyr) with a young SSP (age 5, 25, 100, 290, or
1000 Myr), with the younger component contributing 0\%, 1\%, 2\%, 4\%,
8\%, 16\%, 32\%, or 64\% of the mass.  This grid differs from that of
KGB only in adding a young population age of 5 Myr.  We note also that
although not stated explicitly in KGB, the code {\it does} allow a
``middle-aged'' young population, i.e., one with any of the ``old''
ages younger than the designated old population age (for example,
2.5~Gyr combined with 13.5~Gyr).  We compensate for the
overrepresentation of $\la$1~Gyr SSPs in old+young pairings by
downweighting the likelihoods for these pairings such that all
$\la$1~Gyr SSPs together have equal weight to one ``middle-aged''
population option, approximating a uniform prior on the age of the
younger component.  Following KGB, this model set includes three
metallicities $Z$ = 0.008, 0.02, and 0.05 (solar/2.5, solar, and
solar$\times$2.5).

The second model grid is our primary one for the present work.  It is
designed to enable estimation of the ratio of stellar mass formed 
within the last Gyr to preexisting stellar mass (where this ratio equals the
long-term fractional stellar mass growth rate, FSMGR$_{{\rm
    LT}}$) and therefore includes both single-burst and continuous
star formation history (SFH) options for the young population. To
facilitate comparison with the work of \citet[][hereafter
  S07]{salim.rich.ea:uv} in \S\ref{sec:ukssfr}, our model grid
emulates theirs in including an additional low metallicity choice ($Z$
= 0.004) and adopting a Chabrier IMF. We further consider that even
continuous star formation can be bursty on $\sim$200~Myr timescales
(e.g., \citealt{weisz.johnson.ea:modeling}), which S07 choose to model
with superposition of random bursts.  To keep the grid size
manageable, since only the most recent bursts are likely to strongly
affect the model fits, we take the alternate approach of constructing
a set of young population models that have constant star formation
running from 1015~Myr ago to a turn-off point sometime between 0 to
195~Myr ago, sampled every 15~Myr.  The young population options also
include five models representing quenching bursts without subsequent
star formation: SSPs with ages 360, 509, 641, 806, and 1015~Myr. For
computational convenience, we restrict the old SSP choices in this
model grid to six ages (2, 4, 6, 8, 10, and 12~Gyr).  However, we now
consider 13 young stellar population mass fractions (0.001, 0.002,
0.005, 0.011, 0.025, 0.053, 0.112, 0.220, 0.387, 0.585, 0.760, 0.876,
0.941), equally spaced logarithmically in FSMGR$_{\rm LT}$.  Once
again, an old SSP can serve as the younger model in an old+young pair,
and FSMGR$_{\rm LT}$ $=$ 0 in such a case. The overrepresentation of
pairings with young age $\la$1~Gyr is approximately cancelled by
downweighting the likelihoods for these models, where we treat all of
the continuous SFH models together as having combined weight equal to
one of the five $\la$1~Gyr bursts, then further downweight both types
of young models relative to the ``middle-aged'' SSPs to approximate a
uniform prior on the age of the younger component.

As data are available, we simultaneously fit these model grids to
NUV+{\it ugrizJHK}+IRAC 3.6$\mu$m SEDs as well as optical
spectroscopy.  We apply the 0.04~mag AB offset for the $u$ band\footnote{http://www.sdss.org/dr7/algorithms/fluxcal.html\#sdss2ab} inside
the code.  To the individual magnitude errors we add extra photometric
uncertainties to account for variations between methods of foreground
extinction correction and sky level estimation; 0.1~mag in the NUV,
0.05~mag in $u$, 0.03~mag in $griz$, and 0.1~mag in $JHK$+IRAC
3.6$\mu$m, with an extra 0.1~mag in $JHK$ for faint blue galaxies
($M_r$ or $M_R$ $>$ $-19$ and $u-r<1.4$ or $U-R<0.7$). These choices
are motivated by the analyses of
\citet{abazajian.adelman-mccarthy.ea:second},
\citet{morrissey.conrow.ea:calibration}, and
\citet{blanton.kazin.ea:improved}, as well as our own analysis of the
agreement of our $JHK$ magnitudes with the best fits. When SDSS data
are unavailable, we substitute the {\it UBR} magnitudes from
\citet{jansen.franx.ea:surface}, applying a uniform 0.2~mag offset to
reconcile the $UBR$ zero points with our brighter SDSS magnitudes,
where this offset value is estimated by fitting both $ugriz$ and $UBR$
simultaneously when possible in an initial round of SED fits. Our
final fits do not include {\it UBR} when we have {\it ugriz} however,
to minimize systematics.  We note that our estimated across-the-board
$\sim$0.2~mag offset is opposite to the Vega-to-AB offset for the $R$
band and thus yields fortuitous zero point agreement between the $r$
and $R$ bands (as seen for example by comparing the merged NFGS
Tully-Fisher/Faber-Jackson relations shown in Fig.~\ref{fg:coltf}; see
\S\ref{sec:optkin}).\footnote{\label{fn:jansen} Applying the sky
  subtraction technique described in \citet{jansen.franx.ea:surface}
  to SDSS DR8 images (which are pre-sky-subtracted using the methods
  of \citealt{blanton.kazin.ea:improved}) reveals that $\sim$0.1~mag
  of the offset we measure is probably due to oversubtraction of sky
  by Jansen et al.\ compared to the new protocol (as expected from
  Fig.~12 of Blanton et al.). The remainder may be partly due to
  a mismatch between the {\it UBR} filter systems used in our
  stellar population fitting code and used by Jansen et al.,
  and/or partly due to differing profile extrapolation techniques.
  Regarding the latter, we note that the zero point discrepancy
  increases for low surface brightness galaxies, in the sense that our
  reprocessed SDSS magnitudes are brighter and the spurious agreement
  between the $V$--$M_r$/$M_R$ relations in Fig.~\ref{fg:coltf}
  becomes tighter.}

Stellar masses estimated with the second model grid are provided in
Table~\tabphot.  With so many independent data points, these mass
estimates are quite robust; estimates derived using the first model
grid are offset 0.1~dex higher than those derived using the second
model grid but otherwise agree within 0.1~dex rms (less than the
typical uncertainty of 0.15~dex from the stellar mass
distributions). The only obviously unreliable case is UGC~4879, an
extremely nearby system for which our photometry is notably inconsistent,
possibly due to the technical difficulty of sky subtraction or
to a complex post-starburst spectral energy distribution
(the same galaxy lacks any emission or absorption features and thus
also defies kinematic analysis).  For reference, Table~\tabphot\ also
provides the stellar masses previously derived by KGB using a model
set very similar to our first model grid, but with inferior photometry
(uncorrected {\it UBR} + catalog 2MASS data); these masses show
0.2~dex scatter and $+$0.06~dex median offset relative to our
preferred masses. We include the KGB masses because they have been
used in several recent papers
\citep{wei.kannappan.ea:gas,wei.vogel.ea:relationship,moffett.kannappan.ea:extended}
and compared to stellar masses from
\citet{kauffmann.heckman.ea:stellar}; this comparison demonstrates similar stellar
mass zero points \citep{kannappan.wei:galaxy}.

To ensure uniform color data, we use our stellar population model fits
to interpolate likelihood-weighted $(u-r)^m$ and $(u-J)^m$ colors for
all galaxies, regardless of the availability of these specific bands,
where the superscript $m$ is a reminder that these colors come from
the models and thus include the AB correction to the $u$ band as well
as $k$-corrections to $z=0$.  Self-consistent internal extinction
corrections can also be determined with our newer model grid, enabling
us to examine the behavior of the de-extincted colors, denoted
$(u-r)^e$ and $(u-J)^e$.

Half-light and 90\% light radii are also given in Table~\tabphot, as
these radii are used to compute $\mu_\Delta$, a new quantitative
morphology metric introduced in this work to facilitate comparison of
the NFGS and V3000 samples.  We define $\mu_\Delta$ as
\begin{equation}
\mu_\Delta=\mu_{90}+1.7\Delta\mu
\end{equation}
combining an overall surface mass density
\begin{equation}
\mu_{90}=\log{\frac{0.9M_*}{\pi r_{90,r}^2}} 
\end{equation}
with a surface mass density contrast 
\begin{equation}
\Delta\mu=\log{\frac{0.5M_*}{\pi r_{50,r}^2}} -
\log{\frac{0.4M_*}{\pi r_{90,r}^2 - \pi r_{50,r}^2}}
\end{equation}
representing the difference between
the surface mass densities within the 50\% light radius and between
the 50\%--90\% light radii, where all radii are converted to physical
kpc units. The 1.7 multiplier helps to separate quasi-bulgeless,
bulge+disk, and spheroid-dominated types as illustrated in
Fig.~\ref{fg:nfgscolmasstype}b, yielding approximate divisions at
$\mu_\Delta$= 8.6 and 9.5 as shown.  For galaxies without $r_{50,r}$
measurements, we use the $R$-band half-light radii from
\citet{jansen.franx.ea:surface}, converted from the authors' geometric
mean aperture radius convention to a major axis radius convention.  We
find a one-to-one correspondence between the $r$ and $R$ band
half-light radii, with rms scatter $\la$15\% and a small 3.3\% offset, in
the sense of larger $r$-band radii.  When using $r_{50,R}$ we assume
the median value of $r_{90,r}/r_{50,r}$=2.6 to infer $r_{90,R}$.


\subsubsection{Gas Masses and HI Linewidths}
\label{sec:hi}

The HI data set presented in Table~\tabhikin\ expands on that of
\citet[][hereafter W10a]{wei.kannappan.ea:gas} with 30 new Robert
C.\ Byrd Green Bank Telescope (GBT) 21cm observations taken for the
NFGS under program GBT10A-070 in 2010 January, February, and July (PI
Kannappan).  We add these to the 27 GBT observations obtained by W10a
under programs GBT07A-072 and GBT07C-148 in 2007 March and October.
We have reprocessed the W10a data along with our new data since
discovering that the default GBTIDL flux calibration, which was used
by W10a, is $\sim$15--20\% lower than that obtained from the observed
flux calibrators, and also since finding an error in how W10a
estimated rms noise during linewidth measurement, which led to
overestimation of linewidths by $\ga$10\%.  Other than these
adjustments, our flux and linewidth measurements follow the methods of
W10a closely for unconfused detections.  In particular, we do not
correct for self-absorption, which is expected to alter total HI flux
estimates by $<$30\%, even for the most inclined systems
\citep[][]{giovanelli.haynes.ea:extinction}.  Note that the
masses denoted $M_{\rm HI}$ in Table~\tabhikin\ and the rest of this paper are measured directly
from 21cm fluxes using the equation $M_{\rm
  HI}=1.4\times2.36\times10^5f_{\rm HI}(\frac{cz}{H_0})^2M_\odot$,
which combines the expression from \citet{haynes.giovanelli:neutral}
with a $1.4\times$ correction factor for He.

For non-detections and/or galaxies confused with companions in the
beam, we have adopted a slightly different approach than W10a in our
definition of integration bounds for estimating 3$\sigma$ upper limits
and for dividing flux between objects.  Upper limits are now
calculated using the equivalent $W_{20}$ linewidths determined either
from (a) the optically derived $V$, where $V$ is based on H$\alpha$ or
stellar rotation curves extending beyond 1.3$r_e$ (see
\S\ref{sec:optkin}--\ref{sec:inc}), or, in the absence of such data,
from (b) the implied $V$ from the $R$-band absolute magnitude--$V$
relation $\log{V}=-0.25 - 0.120M_R$, which has been calibrated using
all galaxies with reliable $V$ and $M_R<-17$ in the NFGS
(Fig.~\ref{fg:coltf}).  Here $V$ is estimated from either stellar or
ionized gas kinematics as described in
\S\ref{sec:optkin}--\ref{sec:inc} and is defined to scale as
$W_{50}/(2\sin{i})$ (so we adjust for an assumed offset of $W_{20}-W_{50}=20$\,\kms, see
\citealt{kannappan.fabricant.ea:physical}).  We also account for the
projection factor $\sin{i}$, adopting a minimum $i$ of $30^{\circ}$
for E galaxies to avoid overly strong upper limits, given the lack of
reliable inclination information for such round objects.  Likewise, we
avoid overly strong upper limits for face-on disks by employing a
minimum final integration linewidth of 40~\kms\ reflecting
non-rotational line broadening.  These definitions ensure that we
compute conservative upper limits.

We have confirmed 11 cases of profiles confused with companion
galaxies in our GBT data from a thorough search within twice the GBT
half-power beam diameter of 9$\arcmin$, assuming typical redshift
uncertainties and linewidths and inspecting each of the 15 potential
cases by eye.  To isolate the flux for the primary target, we
typically assign the primary all flux within the equivalent $W_{50}$
linewidth derived from the measured or inferred $V$ as discussed
above, omitting the offset to the $W_{20}$ scale.  Alternatively, if
one half of the profile is obviously more contaminated, we integrate
the uncontaminated half and double that flux. In the case of
UGC~12265N, which is strongly interacting with a similar size
companion and thus more severely confused than usual, W10a employ a
Very Large Array (VLA) 21cm map to determine that only $\sim$25\% of
the flux belongs to the target galaxy, and we retain this flux
division.  Our flux separations are validated by the absence of
significant outliers in our analysis (see especially
\S\ref{sec:colorgs}).

Table~\tabhikin\ summarizes the final derived GBT $M_{\rm HI}$
values/uncertainties or 3$\sigma$ upper limits, plus linewidths and
velocity integration ranges.  For galaxies that share the beam with
one or more companions, linewidths represent the full HI profiles not
deblended to account for confusion.  Confused linewidths are thus
enclosed in brackets to indicate that they are unreliable. Likewise,
linewidths derived from profiles with peak signal-to-noise S/N $<$ 6
are bracketed.  Linewidth uncertainties are estimated using
$\sigma_{W_{50}}=4.1\left(\frac{P}{{\rm S/N}}\right)^{0.85}$ where $P$
is the steepness parameter defined as $P=\left(W_{50}-W_{20}\right)/2$
(this formula is derived as in S13 but assuming 5\,\kms\ spectral
resolution; see S13 for further details).

We have further augmented the 21cm inventory for the NFGS using the
literature compilation of W10a (not duplicated in Table~\tabhikin) and
the ALFALFA survey \citep{haynes.giovanelli.ea:arecibo}.  In
particular, 1 detection and 4 upper limits inferred from ALFALFA are
given in Table~\tabhikin, with the upper limits estimated using the
median rms noise as a function of declination \citep[typically
  $\sim$2.3 mJy;][]{haynes.giovanelli.ea:arecibo}.  Unlike our GBT
upper limits, ALFALFA upper limits are determined at 5$\sigma$,
matching the survey detection threshold.  The literature compilation
of W10a includes the remaining 128 galaxies, so together, these data
sets yield $M_{\rm HI}$ values or upper limits for all 190 galaxies in
the sample, with only one of the 26 upper limits weaker than 10\% of
that galaxy's stellar mass (\S\ref{sec:nfgsphot}).  As it is
infeasible to uniformly assess and decompose confusion in the
literature HI measurements, we have simply flagged likely cases of
confusion in literature data based on the presence of a companion,
interaction, or merger identified by
\citet{kannappan.fabricant.ea:physical}; such cases are marked in
Table~\tabhikin.

Molecular gas data are not uniformly available for the NFGS, but we
have made use of 39 CO-derived H$_2$ masses tabulated in
\citet{wei.vogel.ea:relationship} and/or in S13 to determine that
including molecular gas in the total gas inventory has negligible
impact on our conclusions except to reinforce them.  The effect of
molecular gas is illustrated in several figures in
\S\ref{sec:disting}--\ref{sec:regimes}, always with a $1.4\times$
correction factor for He.

\subsubsection{Optical Kinematics}
\label{sec:optkin}
Ionized gas and stellar rotation velocities and stellar velocity
dispersions for the NFGS are also given in Table~\tabhikin.  These
measurements are based on emission- and absorption-line observations
previously reported in \citet[][hereafter
  KFF]{kannappan.fabricant.ea:physical} and
\citet{kannappan.fabricant:broad}, coming from either the FAST
Spectrograph on the Tillinghast telescope or the Blue Channel
Spectrograph on the MMT telescope.  For stellar dispersions $\sigma$,
the instrumental resolution is at least $\sigma_{\rm
  instr}\sim60$\,\kms\ at the Mg I triplet near 5175\AA, improving to
$\sim$40 \kms\ for the MMT data.  We tabulate the stellar dispersion
$\sigma_{r_e{_{/ 4}}}$ measured within $\frac{1}{4}$ the $B$-band
half-light radius $r_e$.  Comparison of dispersions derived with
various template star spectra (broadened and shifted to match the
galaxy spectrum using the Fourier-space fitting code of
\citealt{.franx:new}) indicates $\sim$5\% systematic errors, which we
convolve into the reported uncertainties.  For ionized gas rotation
velocities, we adopt a non-parametric statistical estimator of the
maximum rotation velocity, the ``probable min$-$max'' $V_{\rm pmm}$ of
\citet{raychaudhury..ea:tests} as implemented in KFF.  The raw error
for $V_{\rm pmm}$ is set to the 11\,\kms\ systematic uncertainty
estimated by KFF from comparison to radio linewidths (scaled from the
20\,\kms\ scatter in the conversion $W_{\rm 50}=33+0.92(2V_{\rm pmm})$
from KFF Appendix B3). Stellar rotation velocities $V_*$ are also
available for some galaxies, where we have applied the probable
min$-$max technique to a rotation curve extracted with the IRAF
cross-correlation task {\bf xcsao} \citep{kurtz.mink.ea:xcsao}.
Table~\tabhikin\ also provides the extent and asymmetry of the gas
rotation curves \citep[see][]{kannappan.barton:tools}, as well as the
slit PAs for both gas and stellar observations, with an asterisk in
case the slit alignment with the galaxy major axis is uncertain or
misaligned by more than $\sim$10$^{\circ}$.  Comparisons using (i)
different instrumental setups and (ii) different rotation curve
extraction techniques indicate that both asymmetry and $V_{\rm pmm}$
measurements are quite repeatable for rotation curves of reasonable
S/N \citep{barton.kannappan.ea:rotation, kannappan.barton:tools},
while extent measurements are somewhat more dependent on S/N.

Values of $V_{\rm pmm}$ may be bracketed as unreliable in
Table~\tabhikin\ if the ionized gas rotation curve is unlikely to
adequately probe the potential well due to limited radial extent
\citep[e.g.,][]{pisano.kobulnicky.ea:gas,kannappan.barton:tools}. For
morphological types earlier than Sc a reliable rotation curve is
defined to extend past 1.3$r_e$ (using the average extent of the two
sides), which corresponds to the canonical turnover radius for large
spiral galaxy rotation curves (KFF).  For types Sc or later we relax this
criterion slightly in requiring only extent $>$$r_e$, because
rotation curves that do not turn over are typical for very late type
galaxies and do not generally yield low-$V$ outliers in the
Tully-Fisher relation.  No values of stellar $\sigma$ are bracketed in
Table~\tabhikin\ because data with inadequate S/N or resolution to
obtain reliable $\sigma$'s are not reported; likewise, we do not
report $V_*$ when the stellar rotation curve extends to $<$1.3$r_e$.
Although all stellar kinematic measurements we report should therefore
be reliable, dispersions marked with an {\it l} should be used
cautiously as dynamical mass estimators, due to the likelihood of
substantial rotational support not accounted for in the stellar
$\sigma$ alone (see next section).

\subsubsection{Characteristic Velocity Assignments and Inclinations}
\label{sec:inc}

For each galaxy, we assign the largest reliable velocity derived from
either gas or stellar kinematics as the characteristic internal
velocity $V$ listed in Table~\tabhikin. In principle $V$ should be
equal to the inclination-corrected gas rotation velocity for an
idealized, purely dynamically cold system.  Because in practice gas
turbulence and other non-circular motions included in HI linewidths
contribute to the dynamical support of galaxies, we adopt the
inclination-corrected HI linewidth $W_{50}/(2\sin{i})$ as our fiducial
gas-derived $V$, converting the ionized-gas $V_{\rm pmm}$ to a
pseudo-$W_{50}$ via the empirical optical-radio calibration in KFF.
This approach restores non-rotational gas support to our
ionized-gas-derived $V$ estimates in an average way (and is validated
by the tighter Tully-Fisher relation obtained by
\citealt{kassin.weiner.ea:stellar} when they explicitly measure and
include ionized gas dispersion for a high-redshift galaxy sample).  To
put stellar $V_*$ measurements on the same scale, we first multiply
them by 1.1 (the typical $V_{\rm pmm}/V_*$ ratio measured in cases
where both rotation curves extend to $>$1.3$r_e$), then scale them
just like ionized gas $V_{\rm pmm}$ measurements.

We further attempt to put rotation- and dispersion-derived estimates
of $V$ on the same scale by applying the rule that for a pure
dynamically hot system, $V$ equals $\sqrt{2}\times$ the stellar
$\sigma$ (an approximation based on the scaling for an isothermal
sphere, \citealt{burstein.bender.ea:global}).  However, low-mass E/S0
galaxies may be significantly supported by stellar rotation
\citep{davies.efstathiou.ea:kinematic}, and we cannot assess this
support uniformly with the data in hand.  In several cases, gas or
stellar rotation curves confirm substantial rotational support,
generally yielding strong outliers from the Faber-Jackson relation
below $V\sim125$ \kms\ (\S\ref{sec:disting}). Thus we consider
stellar dispersion-derived $V$'s unreliable for E--S0a galaxies with
$V<125$ \kms, except for two that are clearly dynamically hot, with
dispersions consistent with the Faber-Jackson relation for high-mass
galaxies (NGC~3605 and NGC~4308).  For the same reason we
consider stellar dispersion-derived $V$'s unreliable for galaxies of
type Sa or later.  We mark low-mass/late-type $\sigma$ measurements
with an {\it l} in Table~\tabhikin\ to indicate likely unreliability.
However, if such a $\sigma$ exceeds a {\it reliable} gas-derived $V$
then we accept the stellar-derived $V$ as reliable.

Superscripts {\it ir, nr, sr, sd} indicate the origin of the final $V$
estimates in Table~\tabhikin\ from ionized-gas rotation $V_{\rm pmm}$
(79 galaxies), neutral gas rotation $W_{50}$ (53 galaxies), stellar
rotation $V_*$ (12 galaxies), or stellar dispersion $\sigma_{r_e{_{/
      4}}}$ (47 galaxies), where 24, 17, 2, and 8 of the final $V$'s
in each category are bracketed as unreliable, respectively, leaving
140 reliable final $V$s.  In combining the four types of kinematic
data, we adopt two additional unreliability criteria besides those
detailed in \S\ref{sec:hi}--\ref{sec:optkin}: (1) although we assume
HI $W_{50}$ measurements probe the full potential well for late-type
galaxies, for types S0a and earlier we consider the HI $W_{50}$
measurement unreliable if the ionized gas rotation curve is truncated,
based on experience that HI in S0s often fails to probe the full
potential well (KFF); (2) for rotation-derived $V$'s, we treat
the inclination correction as reliable only for $i>40^{\circ}$.

Our current best inclination angle estimates are provided in
Table~\tabhikin, mostly derived from photometric axial ratios as in
KFF.  A few values have been updated, with the old values used by KFF
given in parentheses.  For the polar ring galaxy UGC~9562 we adopt
$i=68^{\circ}$ for the gas ring based on interferometric CO data
\citep{wei.vogel.ea:relationship}, and similarly for NGC~3499 we adopt
$i=90^{\circ}$ for the gas (but not the stars) given the appearance of an
edge-on dust lane in an otherwise round system. For
NGC~7077 we list the $i=40^{\circ}$ inclination estimate from
\citet{wei.vogel.ea:relationship} for the record, although it just
misses our inclination cut so the change does not affect our
analysis. For UGC~6206 we adopt $i=43^{\circ}$ as a compromise between
the highly discrepant axial ratios quoted by
\citet{mazzarella.boroson:optical} and
\citet{jansen.franx.ea:surface}.  For UGC~12265N, which has a very small
angular size, the Jansen et al.\ axial ratio would be calculated with
too few significant digits to give a meaningful constraint on the
inclination, so we use the axial ratio from the NASA Extragalactic
Database (NED) to estimate $i=57^{\circ}$.  Finally, for the strongly
distorted galaxy NGC~5993 we reduce the inclination to $i=32^{\circ}$
based on our analysis of the SDSS photometry (implying $i<40^{\circ}$
or ``unreliable'' status, consistent with this galaxy's extreme
Tully-Fisher outlier behavior discussed in KFF).

\subsection{The V3000 Sample}
\label{sec:v3000}

\begin{figure}[t]
\plotone{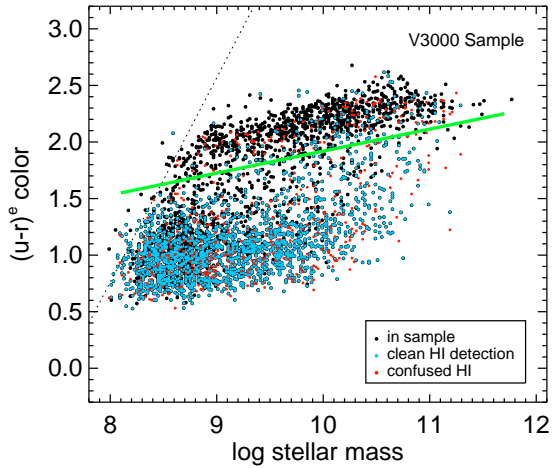}
\epsscale{0.5}
\caption{The V3000 sample shown in $(u-r)^e$ color
  vs.\ stellar mass parameter space. As in
  Fig.~\ref{fg:nfgscolmasstype}, we use de-extincted colors to enhance
  separation of the sequences. Galaxies with unconfused HI detections
  (blue dots) lie primarily on the blue sequence.  Red dots mark
  possibly confused HI detections, which are excluded from the
  remainder of our analysis.  The obviously slanting selection effect
  marked by the dashed line at the low-mass end of the plot reflects
  the fact that the $r$ band used to define the sample correlates most
  directly with {\it baryonic} rather than stellar mass, as shown in
  Fig.~\ref{fg:mlr} \citep[see also][]{kannappan.wei:galaxy}.}
\label{fg:v3000}
\end{figure}

\begin{figure*}[t]
\plotone{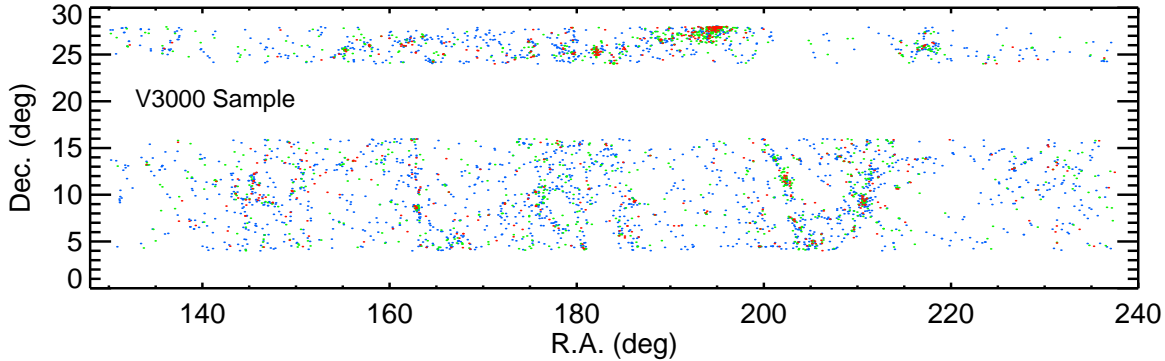}
\epsscale{0.5}
\caption{The V3000 sample on sky.  The sample includes all known
  galaxies with $M_r<-17$ in the two sky regions shown, from
  $cz$=2530--7000\,\kms. It spans a naturally diverse range of
  environments including part of the Coma Cluster (cut off at the
  top).  Blue, green, and red points indicate quasi-bulgeless, bulged
  disk, and spheroid-dominated galaxies based on $\mu_\Delta$ class
  (Fig.~\ref{fg:nfgscolmasstype}).  Points are overplotted in that
  order to highlight sites of spheroid-dominated galaxy
  concentration.}
\label{fg:v3000sky}
\end{figure*}

\begin{figure*}[t]
\plotone{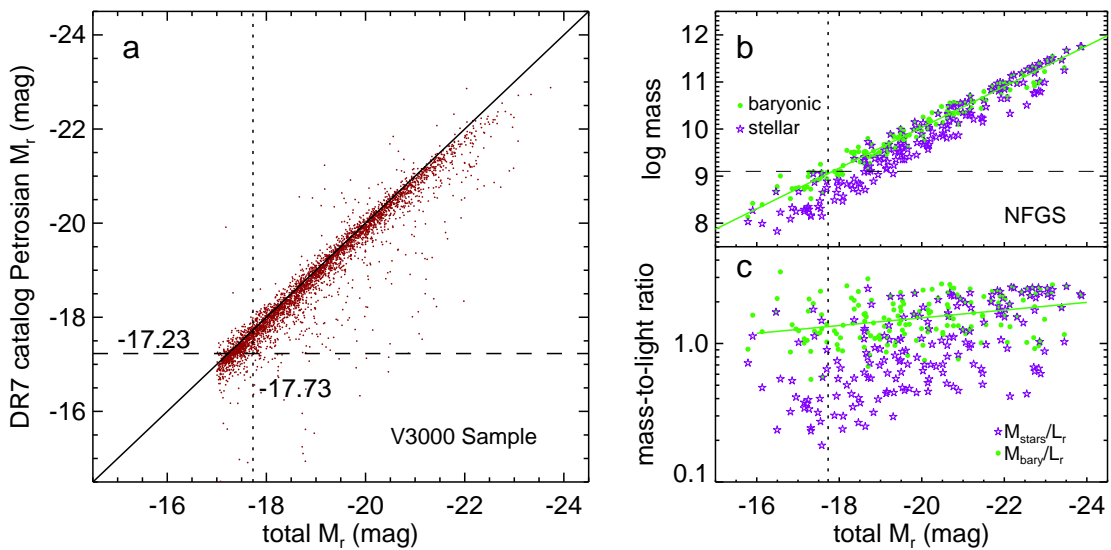}
\caption{Assessing completeness for the V3000 sample. {\it (a)}
  Comparison of SDSS DR7 catalog Petrosian magnitudes vs.\ total
  extrapolated magnitudes from our reprocessed DR8 photometry for the
  V3000 sample, omitting $k$-corrections for direct comparison. The
  one-to-one line is shown for reference. {\it (b)} Translation of the
  approximate completeness limit from panel {\it a} to a typical
  baryonic mass of $\sim$$10^{9.1}$\,\msun\ based on NFGS $r$-band
  magnitudes obtained with the same pipeline used for the V3000
  sample.  Note that the $r$ band correlates with baryonic mass better
  than with stellar mass, with 40\% smaller logarithmic scatter
  ($\sigma=0.14$ vs.\ 0.24~dex for galaxies with $M_r<-17$).  {\it
    (c)} Baryonic mass-to-light ratios in the $r$ band stay fairly
  flat (changing by a factor of $<$2 over 6 mag in $M_r$) and show
  modest 0.14~dex scatter, in contrast to the steeper slope and
  greater scatter seen for $r$-band stellar mass-to-light ratios.
  Thus $r$-band selection approximates selection on baryonic mass;
  nonetheless, scatter up to mass-to-light ratios of $\sim$3 implies a
  final baryonic mass completeness limit of $\sim$$10^{9.3}$\,\msun.}
\label{fg:mlr}
\end{figure*}

The V3000 sample spans a redshift range of 2530-7000~\kms\ within two
SDSS sky regions that offer uniform, public HI data from the blind
21cm ALFALFA survey in the $\alpha$.40 data release
(R.A. 130--237.5$^{\circ}$, Dec. 4--16$^{\circ}$ \& 24--28$^{\circ}$;
\citealt{haynes.giovanelli.ea:arecibo}).  Including all galaxies known
to be in the volume down to $M_r=-17$, the sample comprises 3834
galaxies, not including 57 galaxies with irretrievable photometry
(usually due to a missing/corrupted SDSS image or a very bright star).
Of these, 1911 have clean HI detections, 1527 have HI upper limits,
and 396 have possibly confused HI detections.  The distribution of HI
data in the color vs.\ stellar mass diagram for the V3000 sample is
shown in Fig.~\ref{fg:v3000}.  The high rate of upper limits
($\sim$40\%) is obviously unsatisfying, but increasing the survey
volume toward redshifts $<$2500\,\kms\ in order to add more detections
would present substantial challenges for robust photometry and stellar
mass estimation, while adding a modest number of galaxies and
introducing a strong bias toward the Virgo Cluster.  As defined, the
V3000 sample reflects the natural diversity of galaxy environments
ranging from voids to clusters (Fig.~\ref{fg:v3000sky}).

The V3000 sample has been designed as a superset of a sample that would
be complete to a limiting baryonic mass of $\sim10^{9.3}$\,\msun,
allowing for variable mass-to-light ratio. The $r$ band is optimal for
selection on baryonic mass due to the modest scatter and luminosity
dependence of $r$-band baryonic mass-to-light ratios
(Fig.~\ref{fg:mlr}; see also \citealt{kannappan.wei:galaxy}).
The nominal completeness limit of the SDSS redshift survey,
Petrosian $r=17.77$~mag (using SDSS DR7 catalog magnitudes corrected
for foreground extinction), corresponds to Petrosian $M_{r}=-17.23$ at
7000~\kms, the far side of the V3000 volume.  Our extrapolated
$r$-band magnitudes are systematically brighter than catalog Petrosian
magnitudes by $\sim$0.1 dex (with large outliers; see
Fig.~\ref{fg:mlr}), due to improved sky subtraction and robust
extrapolation (\S\ref{sec:nfgsphot}), so our equivalent limit is
$M_r\sim-17.33$.  We have furthermore reprocessed photometry for many
galaxies near the survey limit, extending the sample down to $M_r=-17$
where redshifts are available and recovering bright galaxies shredded
by the SDSS pipeline.  For this effort, we have made use of a merged
redshift catalog that draws on SDSS DR6/DR7/DR8, Updated Zwicky Catalog, HyperLEDA, ALFALFA,
6dF, 2dF, and GAMA data
\citep{adelman-mccarthy.agueros.ea:sixth,abazajian.adelman-mccarthy.ea:seventh,aihara.allende-prieto.ea:eighth,falco.kurtz.ea:updated,paturel.petit.ea:hyperleda,haynes.giovanelli.ea:arecibo,martin.papastergis.ea:arecibo,jones.read.ea:6df,colless.peterson.ea:2df,driver.hill.ea:galaxy}.
Roughly 8\% of the V3000 sample above $M_r=-17.33$ has been recovered
in this way.  It is worth noting that most of the extra redshifts come
from historical surveys with fairly bright limiting magnitudes, so our
inventory of low surface brightness dwarfs remains incomplete.

The scatter in Fig.~\ref{fg:mlr}a suggests high completeness to
$M_r=-17.73$, corresponding to a typical baryonic mass of
$\sim$$10^{9.1}$\,\msun, as shown in Fig.~\ref{fg:mlr}b.  However,
variations in baryonic mass-to-light ratio up to $\sim$3 would mandate
extension to $M_r\sim-17$ to retain unbiased completeness at M$_{\rm
  bary}\sim$$10^{9.1}$\,\msun.  As we have only partial completeness
from $M_r=-17$ to $-17.33$, we estimate that the V3000 sample is
complete to $M_{\rm bary}\sim10^{9.3}$\,\msun\ and
M$_*\sim$$10^{9.1}$\,\msun. We define an initial approximately
baryonic mass limited sample by $M_r< -17.73$, which yields 3020
galaxies, of which 1510 have unconfused HI detections and 1171 have
upper limits.  In \S\ref{sec:transitions} we attempt to define a more
precisely baryonic mass limited sample by estimating gas content via
``photometric gas fractions'' (K04) for galaxies with upper limits,
then selecting all galaxies with $M_{\rm
  gas}+M_*>10^{9.3}$\,\msun\ down to $M_r=-17$.

Photometric measurements for the V3000 sample have been performed as
part of the ongoing construction of two other volume-limited surveys
not restricted to the ALFALFA $\alpha$.40 footprint (RESOLVE: REsolved
Spectroscopy Of a Local VolumE, Kannappan et al., in prep., and a
larger survey encompassing both the RESOLVE and V3000 samples: Moffett
et al., in prep.).  These surveys will enable analysis within
environmental context, which we defer to future work. We reprocess
{\it GALEX}, SDSS DR8, and 2MASS imaging in the same way as for the NFGS
(\S\ref{sec:nfgsphot}); {\it GALEX} imaging is available for $\sim$30\% of
the V3000 sample.  Stellar mass estimates for the V3000 sample are
derived from the NUV+{\it ugrizJHK} magnitudes using the same updated
model grid used to determine new mass estimates for the NFGS
(\S\ref{sec:nfgsphot}).  From a comparison of colors and masses for
the V3000 sample and the SDSS-derived ``HyperLEDA+'' sample of KGB, we
conclude that our new photometry and updated model grid together yield
similar, slightly lower stellar masses and $\sim$$0.2$~mag bluer
$u-r$ colors than would be obtained with SDSS catalog photometry,
consistent with the discussion in \S\ref{sec:nfgsphot}.

For V3000 galaxies with 21cm detections we obtain HI fluxes and
$W_{50}$ linewidths from ALFALFA catalog measurements, implying likely
underestimation of the true characteristic velocity in the case of
gas-poor systems.  We therefore treat $W_{50}$ as unreliable for
detections below $M_{\rm HI}/M_*= 0.1$, which rejects a large fraction
of the outliers in the ALFALFA $W_{50}$-based Tully-Fisher relation.
Formal errors from the ALFALFA catalog typically imply
$\la$5--10\,\kms\ error on $W_{50}/(2\sin{i})$, so we treat cases with
$>$20\,\kms\ error as unreliable as well.  In either case, when our
analysis requires a velocity we use the equivalent $V$ inferred from
the $r$-band absolute magnitude--velocity relation $\log{V}=-0.29 -
0.123M_r$ (calibrated using all galaxies with reliable $V$ and
$M_r<-17$ in the NFGS; this calibration is very slightly different from that in Fig.~\ref{fg:coltf} due to evolution in the photometry after the HI measurements were performed).  An $M_r$-inferred $V$ is
also used to compute $M_{\rm HI}$ upper limits (defined at 5$\sigma$); we combine
the inferred $V$ with the measured inclination angle and the
declination-dependent ALFALFA rms noise to derive the limit value as
well as provide a $W_{50}/(2\sin{i})$ surrogate.  Finally, because we
estimate inclinations for the V3000 sample from axial ratios
determined by our automated pipeline, we cannot individually vet
inclination estimates in the borderline $i=40$--50$^{\circ}$ range as
we could for the NFGS, so $W_{50}/(2\sin{i})$ is designated unreliable
for $i<50^{\circ}$ and again replaced with an $M_r$-inferred $V$.

Possible confusion of HI sources is flagged within a 3$\arcmin$
radius, which combines the 2$\arcmin$ half-power radius of the
smoothed ALFALFA resolution element with another 1$\arcmin$ to allow
for the typical $\sim$0.5--1$\arcmin$ diameters of galaxies in this
redshift range. We assume catalog redshift uncertainties (minimum
50\,\kms) and $\sim$100\,\kms\ profile widths (the approximate average
of the HI velocity width function of
\citealt{papastergis.martin.ea:velocity}) to assess potential overlap,
employing the same merged redshift catalog described above.
Potentially confused detections are excluded from our HI analysis but
are shown as red dots in Fig.~\ref{fg:v3000}.

\section{Distinguishing the Gas-Richness Threshold and Bimodality Scales}
\label{sec:disting}

\begin{figure*}[t]
\plotone{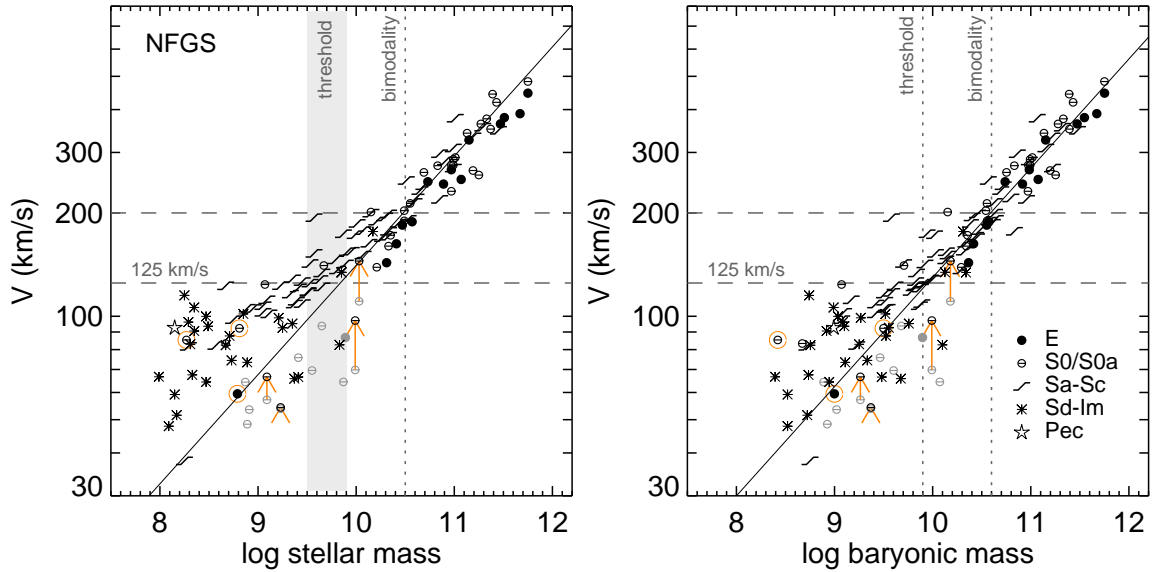}
\caption{Combined Faber-Jackson/Tully-Fisher relation for the NFGS,
  plotting characteristic velocity $V$ as in Fig.~\ref{fg:coltf}
  vs.\ {\it (a)} stellar mass and {\it (b)} baryonic mass (stars +
  neutral atomic gas). The gas-richness threshold and bimodality scales
  discussed in \S\ref{sec:disting} are marked.  The fit in panel {\it
    b} is performed for all galaxies above baryonic mass $M_{\rm
    bary}=10^{10}$\,\msun, minimizing residuals in $V$, and yields
  $\log{V}=-1.07+0.318\log{M_{\rm bary}}$ (implying $M_{\rm bary} \sim
  V^{3.1}$).  The same fit is repeated in panel {\it a}, shifted by
  the $\sim$0.1~dex mean mass offset for the same subsample, to
  highlight deviations from the $M_*$--$V$ relation due to increasing
  gas content, especially below
  $V\sim125$\,\kms\ (Fig.~\ref{fg:goversv}). Gray symbols indicate
  E/S0s whose $V$'s computed from stellar $\sigma$'s alone are treated
  as unreliable, due to likely mixed rotation+dispersion support.
  Orange arrows connect some of these to reliable $V$'s determined
  from gas rotation; in the remainder of this work, the rest are given
  $V$'s determined from the baryonic $M$--$V$ fit to all reliable data
  points in panel {\it b}: $\log{V}=-0.49+0.264\log{M_{\rm bary}}$
  (implying $M_{\rm bary} \sim V^{3.8}$). Orange circles indicate
  additional E/S0s with gas-derived $V$'s, in this case with no
  stellar $\sigma$'s for comparison.}
\label{fg:massv}
\end{figure*}

\begin{figure*}[t]
\plottwo{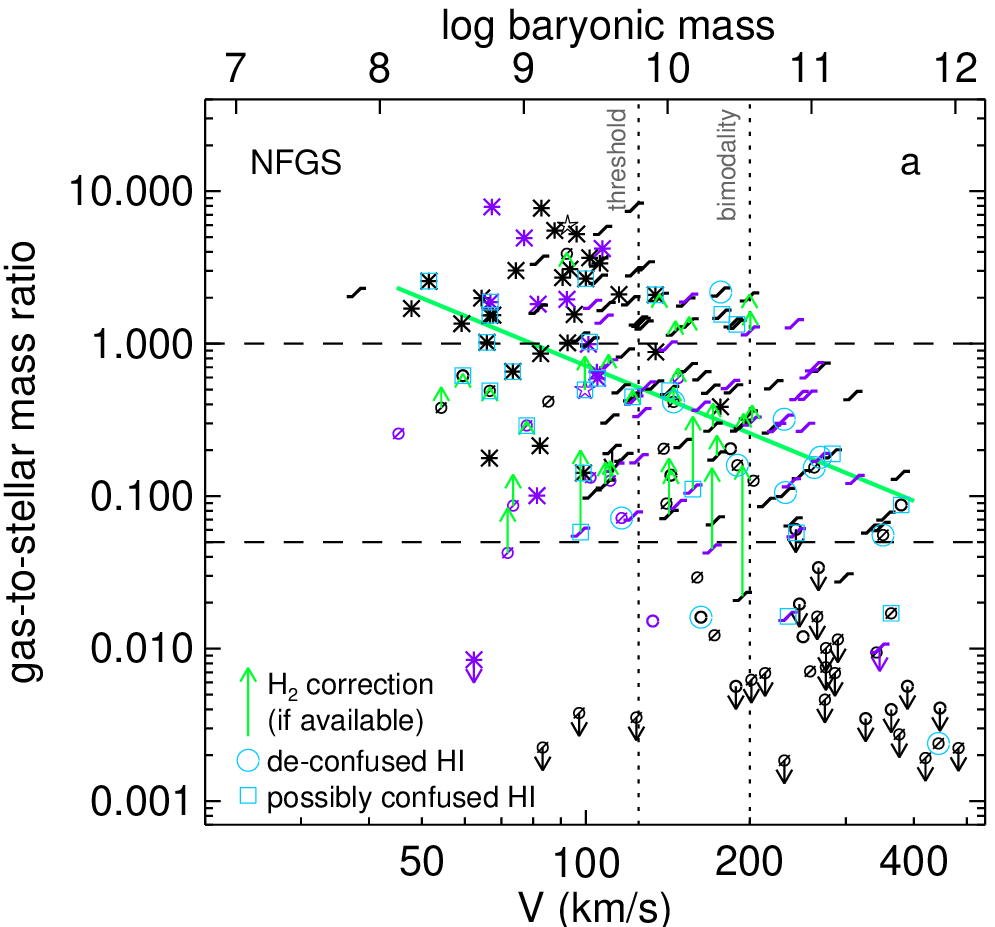}{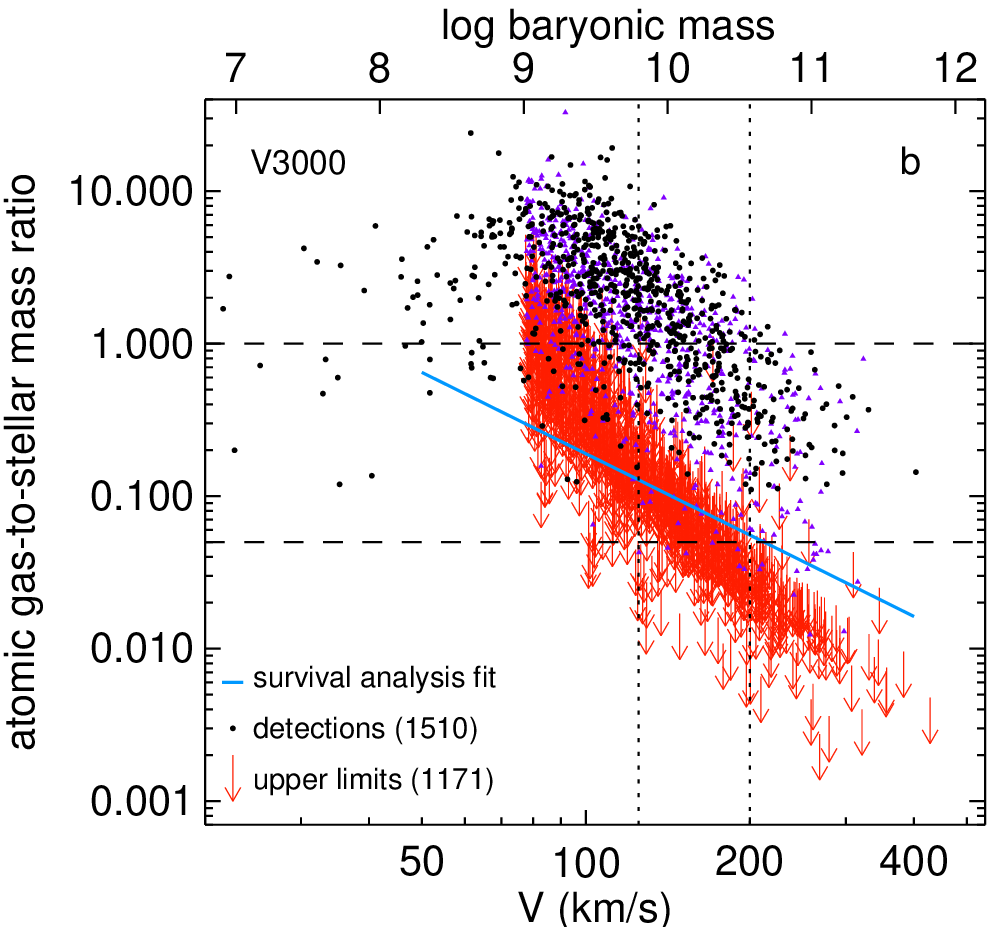}
\caption{$M_{\rm gas}/M_*$ vs.\ internal velocity $V$ for galaxies in
  {\it (a)} the NFGS (HI + H$_2$ where available) and {\it (b)} the
  V3000 sample limited at $M_r=-17.73$ (HI only).  Horizontal and
  vertical lines repeated in both panels mark key mass and gas
  richness scales for reference.  Downward arrows mark upper limits in
  both samples; the dramatic difference in scatter between the two
  samples illustrates the importance of strong limits. The blue line
  in panel {\it b} indicates a survival analysis fit to the V3000
  data, which attempts to take into account upper limits; this fit is
  unrealistic in lying below the fit to the NFGS data in panel {\it a}
  (green line) but nonetheless shows that the apparently tight trend
  seen for the V3000 detections is misleading (see \S\ref{sec:disting}
  for further discussion).  Galaxies without reliable direct $V$
  measurements are plotted using the $V$ inferred from the baryonic
  $M$--$V$ relation (NFGS galaxies) or $M_r$--$V$ relation (V3000
  galaxies) and colored purple.  The baryonic mass conversion
  determined over the full mass range of the NFGS is shown on both top
  axes (see Fig.~\ref{fg:massv} caption). {\it Further notes for panel
    a:} Morphology symbols are as in
  Fig.~\ref{fg:nfgscolmasstype}. Green upward arrows indicate
  molecular gas corrections where available. Blue circles mark
  galaxies with known companions in the GBT beam; for these galaxies
  we attempt separation of any confused flux in the profile using
  optical kinematic data. Blue boxes mark possibly confused literature
  data (\S\ref{sec:hi}). The green solid line is a fit minimizing
  residuals in $M_{\rm HI}/M_*$, excluding extremely gas-starved
  galaxies (those with $M_{\rm HI}/M_* < 0.01$ or an upper limit),
  which yields $\log{M_{\rm HI}/M_*}=2.80 -1.47\log{V}$ with 0.58~dex
  scatter in $\log{M_{\rm HI}/M_*}$. Excluding galaxies below $M_{\rm
    HI}/M_* \sim 0.05$ reduces the scatter to 0.50~dex but yields an
  even shallower slope: $\log{M_{\rm HI}/M_*}=2.40 -1.24\log{V}$.}
\label{fg:goversv}
\end{figure*}

Fig.~\ref{fg:massv} combines the Tully-Fisher and Faber-Jackson
relations for the NFGS into a unified mass--velocity ($M$--$V$)
relation for stellar mass in panel {\it a} and baryonic mass in panel
{\it b}, illustrating multiple equivalent definitions of the threshold
and bimodality scales in terms of $V$, $M_*$, and $M_{\rm bary}$
(stars + cold gas).  The characteristic velocity $V$ used is the
optimal choice of stellar or gas-derived internal velocity for each
galaxy as described in \S\ref{sec:inc}.  The bimodality scale
corresponds to the $M_*\sim10^{10.5}$\,\msun\ scale
highlighted by \citet{kauffmann.heckman.ea:dependence}, which we
identify with $V\sim200$\,\kms\ in Fig.~\ref{fg:massv}.  The threshold
scale corresponds to the $V\sim125$\,\kms\ scale highlighted by
\citet{garnett:luminosity-metallicity} and
\citet{dalcanton.yoachim.ea:formation}\footnote{Dalcanton et
  al.\ adopt a slightly different $V=120$\,\kms\ value as compared to
  Garnett's 125\,\kms\ value adopted here.}, which we identify with
$M_{\rm bary}\sim10^{9.9}$\,\msun\ in Fig.~\ref{fg:massv}. The wide
range of gas-to-stellar mass ratios at the threshold scale
(Fig.~\ref{fg:goversv}) permits no well defined stellar mass
equivalent, but a reasonable fiducial is $M_*\sim10^{9.7}$~\msun\ in
the middle of the 0.4~dex band shown.

The gas-richness threshold and bimodality scales are distinguished by
simultaneous changes in structure (morphology/dynamics) and gas
richness.  Fig.~\ref{fg:nfgscolmasstype}a shows that quasi-bulgeless
Sd--Im morphologies (``dwarf'' late types) become notably more common
below the threshold scale, while spheroid-dominated E--S0a types
become similarly abundant above the bimodality scale.  This pattern is
consistent with previous work showing the emergence of bulges above
the threshold scale \citep{dalcanton.yoachim.ea:formation,bell:galaxy}
and the transition from disk-dominated to spheroid-dominated systems
above the bimodality scale \citep{kauffmann.heckman.ea:dependence}.
In the NFGS, traditional Sa--Sc spirals are most prominent in the
narrow mass range between the threshold and bimodality scales, but
whether they numerically predominate is unclear.  We defer
consideration of this question to \S\ref{sec:procdom}, where we
will use the volume-limited V3000 sample to examine what defines the
transition range between the threshold and bimodality scales.

Symbol colors in Fig.~\ref{fg:nfgscolmasstype}a indicate
quasi-bulgeless, bulged disk, and spheroid-dominated galaxies
according to our $\mu_\Delta$ metric (\S\ref{sec:nfgsphot} and
Fig.~\ref{fg:nfgscolmasstype}b), which combines overall surface mass
density with central vs.\ outer-disk surface mass density contrast.
We see that below the threshold scale, most NFGS galaxies typed as
spirals are quasi-bulgeless, while earlier type galaxies, particularly
those on the blue sequence, often have $\mu_\Delta$ comparable to that of
Sa--Sc spirals. The latter result is consistent with the idea that
low-mass blue-sequence E/S0s are associated with disk rebuilding after
gas-rich mergers, likely leading to the regeneration of late-type
morphologies (KGB; W10a; \citealt{
  wei.vogel.ea:relationship,moffett.kannappan.ea:extended}; S13). Like
spirals, S0s display the full range of $\mu_\Delta$ classes \citep[as
  also emphasized by the classification system of][]{:new}.

E/S0s above the bimodality scale are generally dynamically hot (as
judged by the fact that stellar velocity dispersions alone can provide
most of the virial support required for the $M$--$V$
relation\footnote{It is interesting that massive S0 and early-type
  spiral galaxies may have photometrically prominent disks and yet
  still have sufficiently high stellar $\sigma$'s to account for most
  of their virial support.
  \citet{barway.kembhavi.ea:lenticular,barway.wadadekar.ea:near-infrared}
  find that lenticular galaxies above a $K$-band magnitude of -24.5
  (which roughly equates to $M_*\sim10^{10.8-10.9}$\,\msun; KGB) have
  bulges consistent with formation involving mergers rather than pure
  secular evolution.  The final analysis of \citet[][updating their
    earlier papers]{laurikainen.salo.ea:photometric} also confirms
  that lenticular galaxies above this luminosity behave more like
  ellipticals in the central surface brightness vs.\ effective radius
  relation than like pseudobulge-dominated S0s, which are more typical
  at lower luminosities.}; Fig.~\ref{fg:massv}), but the E/S0
population diversifies at lower masses, with only a minority remaining
dynamically hot. For E/S0s just below the bimodality scale, the
scatter in dispersion-derived $V$ estimates stays within $\sim$20\% of
the baryonic and stellar $M$--$V$ relations, but below the threshold
scale, severe $M$--$V$ outliers would be common if we accepted stellar
$\sigma$'s as reliable estimators of $V$ (light gray symbols in
Fig.~\ref{fg:massv}a).  Significant rotational support most naturally
explains this outlier behavior, as confirmed for a few cases that have
reliable gas-derived $V$'s (see $V$ estimates connected with orange
arrows in Fig.~\ref{fg:massv}b).  In keeping with this interpretation,
the low-mass E/S0s whose {\it only} reliable $V$ is gas-derived are
not outliers (circled in orange in Fig.~\ref{fg:massv}).  We infer
that most E/S0s are rotation-supported below the threshold scale (and
therefore treat their $V$ measurements as unreliable unless they are
unambiguously dynamically hot or have reliable gas-derived
velocities).  This population includes many blue-sequence E/S0s in
low-density environments, consistent with a formation scenario
involving increasingly gas-dominated mergers below the threshold
scale, as expected from Fig.~\ref{fg:goversv} (and logically extending
the wet/dry merger scenario used by
\citealt{emsellem.cappellari.ea:sauron} to explain the fast/slow rotator
dichotomy for E/S0s above the bimodality scale). 

Late-type galaxies also deviate from the stellar $M$--$V$ relation
below the threshold scale, in the sense of {\it higher} $V$ at a given
mass.  However this deviation is diminished by the inclusion of $M_{\rm HI}$ to recover a tight ``baryonic'' Tully-Fisher relation
\citep[][]{mcgaugh.schombert.ea:baryonic}.  The onset of significant
deviations is most easily seen by comparing the stellar $M$--$V$
relation to the fit from the baryonic $M$--$V$ relation, which is
derived for all galaxies with ${M_{\rm bary}}>10^{10}$\,\msun\ in
Fig.~\ref{fg:massv}b and shifted to the mean $M_*$ for the late
types among these galaxies in Fig.~\ref{fg:massv}a.  The gas mass correction becomes important below $V\sim125$\kms, exactly
where gas-dominated galaxies become abundant (Fig.~\ref{fg:goversv}).
It is intriguing that even after we add $M_{\rm HI}$, the $M$--$V$
relation remains slightly offset to higher $V$ below $V\sim125$\,\kms.
The remaining offset could reflect either a change in relative
baryonic-to-dark matter concentration below the threshold scale or an
additional undetected gas reservoir.  Previous studies of the baryonic
Tully-Fisher relation have argued that adding a multiple of the
HI-derived gas mass yields the tightest $M$--$V$ relation,
implying the possibility of undetected gas in ionized, CO-dark
molecular, or optically thick atomic form
\citep{pfenniger.revaz:baryonic,begum.chengalur.ea:baryonic}.

The threshold and bimodality scales are associated with two physically
significant transitions in galaxy gas content, seen most clearly in
the NFGS (Fig.~\ref{fg:goversv}a): below the threshold scale,
gas-dominated galaxies with gas-to-stellar mass ratios $M_{\rm
  HI}/M_*>1$ become noticeably more abundant, while above the
bimodality scale, gas-starved galaxies with $M_{\rm HI}/M_*<0.05$
start to predominate.\footnote{The choice of $M_{\rm HI}/M_*\la0.05$
  as a transitional value is motivated by the fact that nearly all
  spiral-type systems are detected with $M_{\rm HI}/M_*>0.05$;
  however, the preponderance of upper limits for E/S0s even down to
  $M_{\rm HI}/M_*\la0.002$ leaves open the possibility that typical
  $M_{\rm HI}/M_*$ values for gas-starved E/S0s may actually be much
  lower than 0.05.} Molecular gas corrections are provided where CO
data are available (green arrows) and do not significantly affect the
trends; H$_2$ contributions generally become significant only when
overall gas content is low.  To a significant extent, bulgeless and
bulge-dominated galaxies correspond to gas-dominated and gas-starved
galaxies, respectively.  However, S0--S0a galaxies are
not always bulge-dominated (see Fig.~\ref{fg:nfgscolmasstype}b), and they may have significant
gas \citep[as reported by many authors;
  e.g.,][]{hawarden.longmore.ea:neutral,sage.welch:cool}, which
complicates analysis by Hubble type.

Patterns of gas content are obscured by weak upper limits in the V3000
sample (Fig.~\ref{fg:goversv}b), creating a misleadingly tight trend.
To illustrate the extent of the problem, the blue line in
Fig.~\ref{fg:goversv}b shows a fit using survival analysis to include
upper limits. Specifically, we make use of the Buckley-James estimator
(discussed as most robust to non-Gaussian scatter by
\citealt{isobe.feigelson.ea:statistical}), as implemented in the ASURV
package \citep[Rev. 1.2,][]{lavalley.isobe.ea:asurv}.  The survival analysis fit
lies well below the detected data and in fact even below the
conventional fit for the NFGS data in Fig.~\ref{fg:goversv}a.
However, we stress that the survival analysis result should not be
over-interpreted, because survival analysis assumes that the incidence
of upper limits is uncorrelated with the underlying values, an
assumption often treated as roughly valid for flux-limited surveys but
clearly violated once distance limits are imposed as in the
volume-limited V3000 sample.  In \S\ref{sec:transitions} we will use
an updated calibration of the photometric gas fraction technique (K04)
to show that the full V3000 $M_{\rm HI}/M_*$ distribution likely looks
quite similar to that of the NFGS.

In fact, the {\it lack} of a tight relationship between $M_{\rm
  HI}/M_*$ and $V$ for the NFGS may be the most important result seen
in Fig.~\ref{fg:goversv}.  We stress that the extreme diversity of
$M_{\rm HI}/M_*$ in the NFGS is {\it real}, reflecting its selection
as a broadly representative sample of the galaxy population, whereas
the seemingly tighter $M_{\rm HI}/M_*$ vs.\ $V$ relation for the V3000
sample is created by weak upper limits as just discussed.  Moreover,
other data sets that may have {\it seemed} to show a tight dependence
of gas content on galaxy mass have also been inherently
detection-biased and/or selection-biased (for example, as in the work
of K04 and \citealt{mcgaugh:baryonic*1}, both widely used to estimate
gas mass from stellar mass, e.g., as in \citealt{stewart:galmass}). We
also note that past studies have generally plotted $M_{\rm HI}/M_*$
vs.\ $M_*$, in which case the usual $\ga$0.2~dex errors in $M_*$
combined with the covariance between plot axes will spuriously enhance
the impression of a correlation.  Fig.~\ref{fg:goversv} shows gas
richness transitions using $V$ as our preferred mass proxy, since it
correlates tightly with baryonic mass and avoids covariance with
$M_{\rm HI}/M_*$.  With this choice, we do see {\it some} measurable
trend in $M_{\rm HI}/M_*$ vs.\ $V$, but the large scatter around the
fit ($\ga$0.5~dex excluding quenched galaxies) implies little
predictive power not already contained in the statement that
gas-dominated galaxies start to appear in large numbers below the
threshold scale.  Furthermore, compared to gas-dominated galaxies,
gas-starved galaxies appear less closely tied to their characteristic
mass scale, with examples at all galaxy masses in spite of the
potential bias against gas-poor dwarfs in the $B$-selected NFGS.

Previous work has suggested abrupt changes in ISM physics at the
threshold scale \citep{dalcanton.yoachim.ea:formation}, so the
question of whether gas richness is changing sharply at
$V\sim125$\,\kms\ in Fig.~\ref{fg:goversv} is of interest.  The small
number statistics of the NFGS do not allow us to establish whether the
transition is sharp or continuous in that sample.  The V3000 sample is
larger, but also fails to confirm an abrupt transition, although a
sharp transition could perhaps be hidden by its larger measurement
uncertainties. Regardless, we will argue in \S\ref{sec:regimes} that
galaxy mass (or its proxy $V$) is {\it not} the most fundamental
variable underlying transitions in gas richness, and that the broad
scatter in $M_{\rm HI}/M_*$ at low $V$ implies additional physics.

\section{Gas Richness, Long-Term FSMGR, and Refueling Regimes}
\label{sec:regimes}
We have argued that distinct transitions occur at the threshold and
bimodality scales, coupling gas richness changes with structural
changes.  Yet we have also seen considerable diversity in galaxy
properties at any given mass.  Here we demonstrate that U$-$NIR colors
perform far better than galaxy masses in predicting gas richness and
morphology.  We show that these colors can be understood as proxies
for the long-term fractional stellar mass growth rate, FSMGR$_{\rm
  LT}$, averaged over the last Gyr.  We further argue that the most
natural interpretation of the tight $M_{\rm HI}/M_*$ vs.\ FSMGR$_{\rm
  LT}$ relation involves routine fresh gas accretion, and we identify
three distinct regimes in FSMGR$_{\rm LT}$, $M_{\rm HI}/M_*$, and
morphology that we link to qualitatively different states of external
gas accretion and internal gas processing (``refueling regimes'').
Finally, we examine the mass dependence of the refueling regimes to
show how they give rise to the threshold and bimodality scales.

\begin{figure*}[t]
\plotone{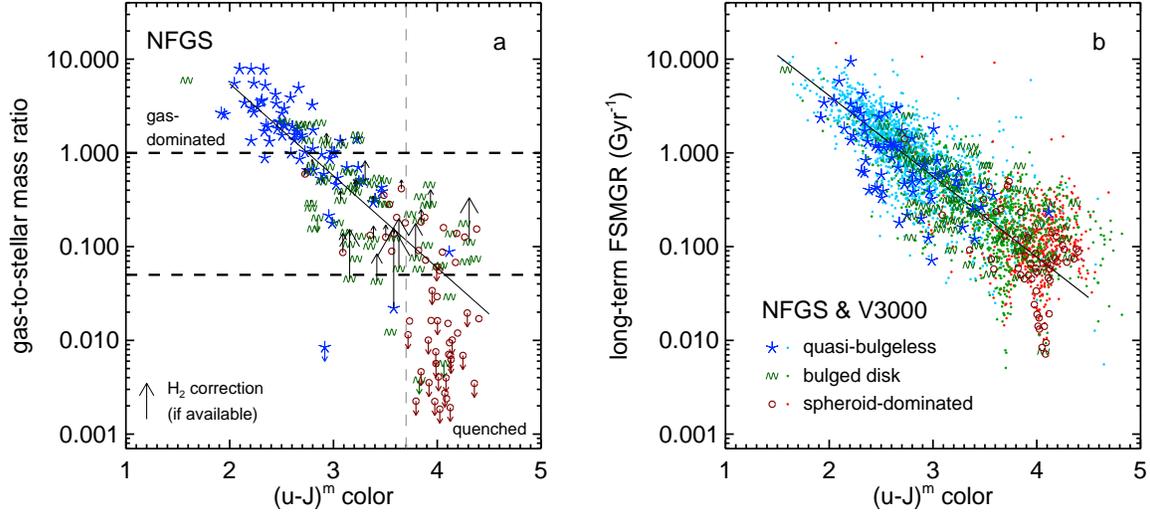}
\caption{Gas richness and long-term fractional stellar mass growth
  rate FSMGR$_{\rm LT}$ in relation to U$-$NIR color, with symbols
  indicating $\mu_\Delta$ morphology class as defined in
  Fig.~\ref{fg:nfgscolmasstype}b.  {\it (a)} $M_{\rm HI}/M_*$
  vs.\ $(u-J)^m$ color for the NFGS.  The solid line is a fit
  minimizing residuals in $M_{\rm HI}/M_*$ for all galaxies bluer than
  $(u-J)^m=3.7$ (excluding the outlier, see note~\ref{foot:ugc}),
  which yields $\log{M_{\rm HI}/M_*}=2.70-0.98(u-J)^m$ with 0.34~dex
  scatter.  If available, molecular gas corrections are shown with
  black arrows but are not used in the fit. {\it (b)} Mapping of
  $(u-J)^m$ color to FSMGR$_{\rm LT}$, determined by fitting to a
  suite of two-component old+young stellar population models as
  described in \S\ref{sec:ukssfr}. Small dots show V3000 sample
  galaxies with $M_r<-17.73$.  We measure 0.30~dex rms scatter around
  the V3000 sample fit (black line: $\log$ FSMGR$_{\rm LT}$ =
  $2.33-0.861(u-J)^m$ for galaxies bluer than $(u-J)^m=3.7$), which is
  comparable to the scatter in panel {\it a}.}
\label{fg:uk}
\end{figure*}

\subsection{The Tight Relation of $M_{\rm HI}/M_*$ vs.\ U$-$NIR Color}
\label{sec:colorgs}

U$-$NIR colors define a surprisingly tight correlation with $M_{\rm
  HI}/M_*$, providing the basis for the photometric gas fraction
technique of K04.  In contrast to the $\sigma\sim0.58$~dex scatter in
the $M_{\rm HI}/M_*$ vs.\ $V$ relation (Fig.~\ref{fg:goversv}a), we
measure 0.36~dex scatter in the $M_{\rm HI}/M_*$ vs.\ $(u-J)^m$ color
relation (Fig.~\ref{fg:uk}a), using the same subset of NFGS galaxies
(i.e., excluding extremely gas-starved systems with $M_{\rm HI}/M_*$
$<$ 0.01 or upper limits). We obtain 0.30~dex scatter in $M_{\rm
  HI}/M_*$ vs.\ $(u-J)^m$ when excluding gas-starved systems with
$M_{\rm HI}/M_*$ $<$ 0.05, as compared to the 0.50~dex scatter
measured for the same galaxies in $M_{\rm HI}/M_*$ vs.\ $V$.
Moreover, since exclusions based on gas content are impractical from
the point of view of predicting gas content, it is especially
interesting that gas-starved galaxies are {\it tightly confined} in
Fig.~\ref{fg:uk}a to colors redder than $(u-J)^m=3.7$.  U$-$NIR color
alone does not separate these gas-starved, spheroid-dominated galaxies
from the reddest bulged disks (which have non-negligible gas and
continue the $M_{\rm HI}/M_*$ vs.\ $(u-J)^m$ correlation).  Galaxies
bluer than $(u-J)^m=3.7$ display 0.34~dex scatter around the fit
shown, excluding the one outlier, whose nature is
uncertain.\footnote{\label{foot:ugc}The plot location of the outlier
  in Fig.~\ref{fg:uk}a, UGC~4879, is either spurious or intriguing, as
  it is both (i) a post-starburst galaxy, with strong Balmer
  absorption and no detectable H$\alpha$ emission, and (ii) very
  nearby, with individually resolved star clusters and clumpy,
  irregular structure, making our photometry highly unreliable
  (\S\ref{sec:nfgsphot}).}  As the typical stellar mass
uncertainty for the NFGS is 0.15~dex, and systematic errors in stellar
mass estimation are likely to be significantly larger
\citep[see][]{kannappan.gawiser:systematic}, measurement uncertainties
likely contribute much of the observed scatter in Fig.~\ref{fg:uk}a,
implying an impressively tight underlying correlation.

The $M_{\rm HI}/M_*$ vs.\ $(u-J)^m$ relation given in
Fig.~\ref{fg:uk}a supersedes that of K04, with higher quality data
(reprocessed SDSS and 2MASS imaging; carefully vetted, largely new HI
measurements), superior stellar mass estimation (from SED modeling
rather than the $g-r$ vs.\ $M_*/L_K$ calibration of
\citealt{bell.mcintosh.ea:optical}), and a more representative galaxy
sample spanning the natural diversity of galaxy gas richness (as
opposed to the inhomogeneous literature compilation used by K04).
Given the far greater {\it intrinsic} diversity of the NFGS compared
to prior samples used to calibrate the photometric gas fraction
technique
\citep[K04;][]{zhang.li.ea:estimating,catinella.schiminovich.ea:galex,
  li.kauffmann.ea:clustering}, it is remarkable that the scatter in
our U$-$NIR relation is the same as has been obtained by these authors
only by combining multiple parameters (e.g., optical color,
surface brightness, stellar mass, and/or color gradient).  We infer
that U$-$NIR colors capture the physics of galaxy gas richness in an
essential way, to be discussed in
\S\ref{sec:ukssfr}. However, we also note that
internal extinction is responsible for some fortuitous straightening
and tightening of the $(u-J)^m$ relation: if replotted using
the extinction-corrected $(u-J)^e$, Fig.~\ref{fg:uk} appears more similar to the bowed NUV$-r$
relation \citep[e.g., as reported
  by][]{catinella.schiminovich.ea:galex}.  Moreover, our reprocessed $u-r$ colors (unlike SDSS catalog $u-r$ colors) are comparable to U$-$NIR colors in predictive power (Eckert et al., in prep.).

\subsection{U$-$NIR Color as a Long-Term FSMGR Metric}
\label{sec:ukssfr}

\begin{figure*}[t]
\plotone{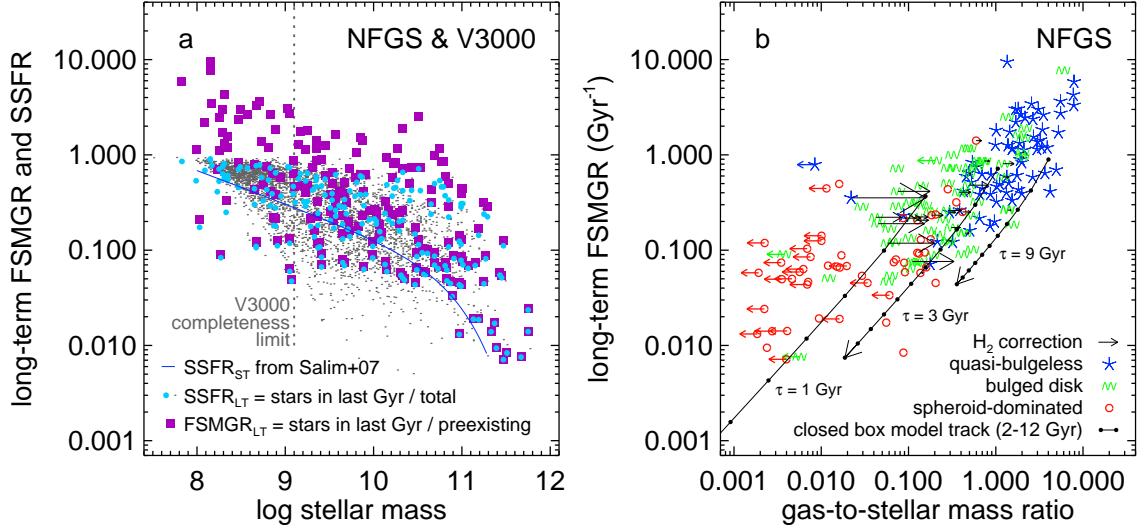}
\caption{FSMGR$_{\rm LT}$ as a function of stellar mass and $M_{\rm
    HI}/M_*$. {\it (a)} Comparison of FSMGR$_{\rm LT}$ and SSFR$_{\rm
    LT}$ derived from stellar population modeling of NFGS galaxies
  (large colored symbols) reveals the essential difference between
  these quantities: FSMGR divides by preexisting rather than by total
  mass, so it does not asymptote to one over the unit of time at high
  growth rates.  Small background dots represent SSFR$_{\rm LT}$'s for
  the V3000 sample (down to $M_r=-17$, hence incomplete below the
  $M_*$ limit shown, \S\ref{sec:v3000}). Comparison of the V3000 and
  NFGS distributions confirms that the NFGS is a broadly
  representative of the general galaxy population, albeit with
  overrepresentation of high SSFR galaxies consistent with its parent
  survey's $B$-band selection. Comparison of the V3000 distribution
  with the short-term SSFR trend (blue line) provided by S07 reveals
  an overall shift toward higher SSFRs in our long-term measurements.
  While some of this effect may reflect the timescale difference, we
  suspect that photometry differences are more important: as discussed
  in \S\ref{sec:nfgsphot}, our photometry yields significantly bluer
  $u-r$ colors than the SDSS DR7 catalog photometry used by S07 {\it
    (b)} FSMGR$_{\rm LT}$ and $M_{\rm HI}/M_*$ define a nearly
  one-to-one relation on Gyr timescales (note the identical axis
  ranges).  Closed box model tracks (whose arrows point in the
  direction of greater age) are inconsistent with the data except for
  the first few time steps, representing galaxies born $\sim$2--5~Gyr
  ago.  We interpret these results in terms of cosmic accretion in
  \S\ref{sec:interp}--\ref{sec:rr}.  We identify three refueling
  regimes: (i) a gas-dominated, stellar mass doubling (FSMGR $>$ 1)
  regime, (ii) a transitional regime with growth rates and gas
  fractions at tens of \% levels, and (iii) a gas-starved, minimal
  growth regime considered quenched.  Symbols indicate $\mu_\Delta$
  classes, which approximately map to these regimes, with interesting
  complexity in the transition range (see \S\ref{sec:rr}).  }
\label{fg:ssfrcheck}
\end{figure*}

As shown in Fig.~\ref{fg:uk}b, the primary physics underlying the
$(u-J)^m$--$M_{\rm HI}/M_*$ correlation is another tight correlation,
between $(u-J)^m$ and the long-term fractional stellar mass growth
rate FSMGR$_{\rm LT}$:
\begin{equation}
{\rm FSMGR}_{{\rm LT}} = \frac{mass_{\rm formed in last Gyr}}{1\,{\rm Gyr}\times(mass_{\rm preexisting})}.
\end{equation}
We measure FSMGR$_{\rm LT}$ as part of the same stellar population
modeling used to determine stellar masses (\S\ref{sec:nfgsphot}).
Our model grid is designed to sample FSMGR$_{\rm LT}$ uniformly in the
logarithm, with 13 values from $10^{-3}$, $10^{-2.65}$,
$10^{-2.3}$... up to $10^{1.2}$.  We estimate both the
likelihood-weighted mean and the likelihood-weighted median
FSMGR$_{\rm LT}$ for each galaxy, but since they are very similar, we
plot the means for visual clarity (the medians are discretized by
construction).

The definition of FSMGR$_{\rm LT}$ may seem superficially similar to a
specific star formation rate (SSFR, star formation rate normalized to
current stellar mass), e.g., as traced by EW(H$\alpha$) over short
timescales.  However, FSMGR$_{\rm LT}$ is not equivalent to a
long-term SSFR, because the newly formed stellar mass appears only in
the numerator.  In contrast, conventional definitions of SSFR (e.g.,
as in the work of S07) include new stellar mass in both the numerator
and the denominator, so that at high growth rates the SSFR cannot
exceed one over the unit of time in which the SFR is measured (see
Fig.~\ref{fg:ssfrcheck}a).  Thus SSFRs offer limited insight into star
formation in high fractional growth regimes.

Plotting FSMGR$_{\rm LT}$ directly against $M_{\rm HI}/M_*$, we find a
remarkably linear relationship for star-forming galaxies
(Fig.~\ref{fg:ssfrcheck}b). It is particularly striking that in per
Gyr units, the newly formed-to-preexisting stellar mass ratios for
quasi-bulgeless and bulged disk galaxies are not merely {\it
  proportional} to their gas-to-stellar mass ratios, but instead are
{\it almost the same.}  The numerical coincidence of scales on the
FSMGR$_{\rm LT}$ and $M_{\rm HI}/M_*$ axes has profound implications to
be discussed in \S\ref{sec:interp}, so it is worth noting that our
estimates of FSMGR$_{\rm LT}$ are higher than would be expected from
previous work.  Specifically, if we convert our FSMGRs to SSFRs (dots
in Fig.~\ref{fg:ssfrcheck}a), we find that both the V3000 sample and
the NFGS lie above the fit line for short-term SSFRs from S07.  A small
amount of this difference may be due to the short-term vs.\ long-term
measurement \citep[see][]{salim.dickinson.ea:mid-ir}, and the excess
of scatter for the NFGS may certainly reflect the blue selection of
its parent survey \citep{jansen.franx.ea:surface}, but the majority of
the effect is probably due to differences in photometry: as discussed
in \S\ref{sec:nfgsphot}, our reprocessed SDSS magnitudes yield
significantly bluer colors than the SDSS DR7 catalog photometry used
by S07.

\subsection{Interpreting the $M_{\rm HI}/M_*$ vs.\ FSMGR$_{\rm LT}$ Relation}
\label{sec:interp}

\begin{figure*}[t]
\plotone{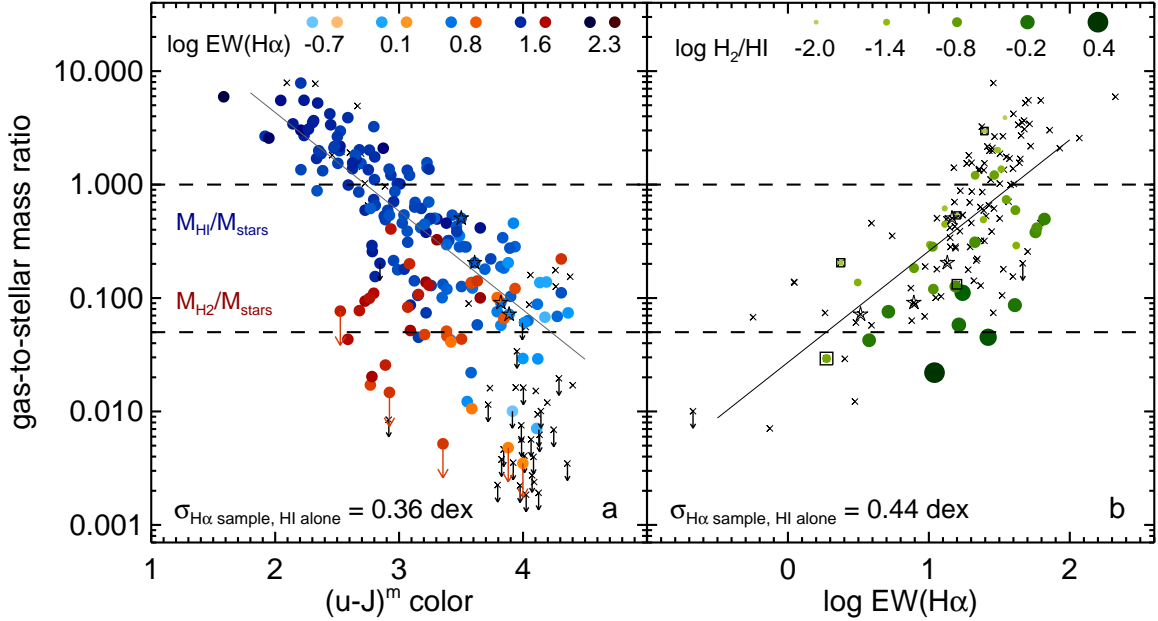}
\caption{Relationship of gas-to-stellar mass ratios to long-term and
  short-term $<$SSFR$>$, as parametrized by U$-$NIR color and
  EW(H$\alpha$) respectively, for the NFGS.  {\it (a)} $M_{\rm
    HI}/M_*$ (blue) and $M_{\rm H2}/M_*$ (red) vs.\ $(u-J)^m$ color as
  in Fig.~\ref{fg:uk}a, with points coded by EW(H$\alpha$) as shown in
  the legend ($\times$ symbols mark galaxies either lacking H$\alpha$
  data or undetected in H$\alpha$). Downward arrows indicate upper
  limits.  Stars mark AGN as classified by
  \citet{jansen.fabricant.ea:spectrophotometry}. {\it (b)} $M_{\rm
    HI}/M_*$ vs.\ EW(H$\alpha$), with points coded by H$_2$/HI mass
  ratio as shown in the legend. Additional symbols are as in panel
  {\it a}, except now $\times$ symbols mark galaxies lacking H$_2$
  data and boxed points denote H$_2$ upper limits. The scatter in
  panel {\it b} is notably higher than in panel {\it a}, relative to
  fits restricted to H$\alpha$-detected galaxies in both panels.}
\label{fg:halpha}
\end{figure*}

Since gas fuels star formation, it may seem self-evident that
FSMGR$_{\rm LT}$ should correlate with $M_{\rm HI}/M_*$. Indeed
\citet{zhang.li.ea:estimating} have argued that the Kennicutt-Schmidt
relation can be used to explain the correlation of another long-term
star formation tracer, $g-r$ color, with $M_{\rm HI}/M_*$.  Motivated
by the Kennicutt-Schmidt relation, they add $i$-band surface
brightness to $g-r$ color to obtain a correlation with $M_{\rm
  HI}/M_*$ with 0.31~dex scatter, albeit with a less
diverse/representative sample than we have presented (following K04,
they analyze galaxies with HI detections in HyperLEDA).  Despite the
appeal of their straightforward interpretation, there are two reasons
to rethink the underpinnings of the photometric gas fraction
technique.

First, most astronomers agree that stars form in molecular gas
\citep[e.g.,][]{krumholz.leroy.ea:which}.  Yet U$-$NIR colors
correlate only with atomic gas or the combination of atomic+molecular
gas --- a plot of U$-$NIR colors vs.\ $M_{\rm H_2}/M_*$ is a scatter
plot (red points in Fig.~\ref{fg:halpha}).  A typical plot of the
Kennicutt-Schmidt relation would show that the H$\alpha$-derived star
formation rate has a direct correlation with the mass in H$_2$, its
immediate precursor, where both are typically divided by surface areas
to yield surface densities
\citep[e.g.,][]{bigiel.leroy.ea:star,schruba.leroy.ea:molecular}.  The
timescales involved in the Kennicutt-Schmidt relation are measured in
tens of Myr, so long-lasting $U$ band flux is not an ideal
real-time tracer of star formation and molecular gas consumption. In
contrast, U$-$NIR color tracks FSMGR$_{\rm LT}$ over timescales long
enough to register whether molecular gas is resupplied from a larger
reservoir. Because HI can resupply H$_2$ and makes up most of the gas
in a typical galaxy (H$_2$-rich galaxies are relatively less common
and also typically gas poor; Figs.~\ref{fg:goversv}a, \ref{fg:uk}a,
and \ref{fg:halpha}), HI measurements generally trace a galaxy's
potential for long-term molecular gas consumption better than do H$_2$
measurements, which reflect only current H$_2$.  For the few galaxies
in Fig.~\ref{fg:uk}a with large H$_2$ corrections (green arrows), the
H$_2$ represents a large fraction of the total gas reservoir and thus
helps tighten the U$-$NIR vs.\ $M_{\rm gas}/M_*$ relation.  Of course,
short-term processes that affect H$_2$ and EW(H$\alpha$) also
affect U$-$NIR colors --- we see that EW(H$\alpha$) varies with
$(u-J)^m$ in Fig.~\ref{fg:halpha}a --- but the scatter in the plot of
$M_{\rm HI}/M_*$ vs.\ EW(H$\alpha$) is $>$20\% higher than in the
plot of $M_{\rm HI}/M_*$ vs.\ $(u-J)^m$ (Fig.~\ref{fg:halpha}b vs.~a).  The
long timescale of U$-$NIR colors would seem optimal for predicting
$M_{\rm HI}/M_*$, implying that the underlying physics is distinct
from the short timescale physics driving the Kennicutt-Schmidt
relation.

Moreover, the long U$-$NIR timescale suggests a second level of
reinterpretation.  We have shown that U$-$NIR colors reflect
FSMGR$_{\rm LT}$ past-averaged over a Gyr (\S\ref{sec:ukssfr}).  Why
such long-term past-averaged star formation should correlate tightly
with {\it present} gas richness is not obvious in a ``gas reservoir
causes star formation'' picture.  One might suppose the correlation
could result from a well-behaved time lag between $M_{\rm HI}/M_*$ and
FSMGR$_{\rm LT}$ in a closed box scenario, but again we note the
surprising fact that these quantities are not just proportional, but
{\it nearly the same} when star formation is integrated over the last
Gyr.  This fact, along with the modest scatter for non-gas-starved
galaxies in Fig.~\ref{fg:ssfrcheck}b, rules out any reasonable closed
box model.  To illustrate the difficulty, we show three model tracks
in Fig.~\ref{fg:ssfrcheck}b, with gas depletion times of 1, 3, and
9~Gyr.  The tracks obey pure closed box evolution:
\begin{equation}
M_{\rm gas}/M_* = \frac{e^{-t/\tau}}{1-e^{-t/\tau}}
\end{equation}
\begin{equation}
FSMGR_{\rm LT} = \frac{e^{-(t-1)/\tau} - e^{-t/\tau}}{1-e^{-(t-1)/\tau}}
\end{equation}
where $\tau$ is measured in Gyr.  Intersection with the data requires
an implausibly young box (just a few Gyr), which should ideally also
have short gas depletion time.

The most natural way out of this cause-effect conundrum is to suppose
that most galaxies are routinely refueled, both in the sense of
external gas accretion and in the sense of internal gas transport and
HI-to-H$_2$ conversion.  In this view, the coincidence of axes in
Fig.~\ref{fg:ssfrcheck}b implies that the entire cold gas reservoir of
an unquenched galaxy is routinely turned into stars and fully
replenished over the timespan of a Gyr, within a factor of a few.  It
follows that Gyr timescales must be longer than the typical time
intervals between any discrete processes necessary to maintain
refueling and star formation, such as gas accretion events,
development of internal instabilities, and/or interactions with
companions.  Thus the fundamental physics underlying the $M_{\rm
  HI}/M_*$ and FSMGR$_{\rm LT}$ relation involves gas refueling on
roughly Gyr timescales.

\subsection{Refueling Regimes}
\label{sec:rr}

\begin{figure*}[t]
\plotone{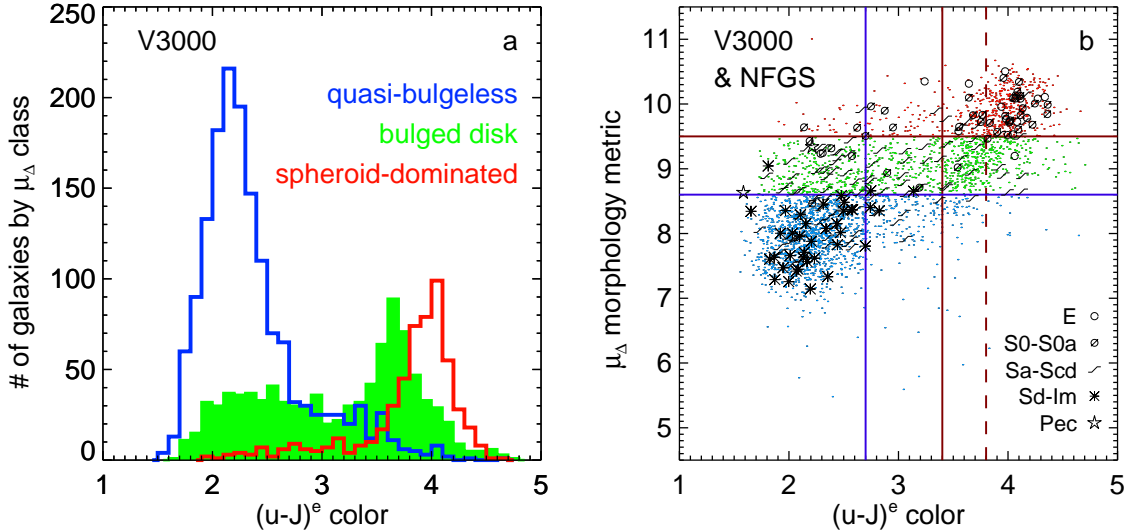}
\caption{Galaxy morphology in relation to de-extincted $(u-J)^e$ color,
  using $\mu_{\Delta}$ classes for the V3000 sample ($M_r<-17.73$) and
  traditional morphological types for the NFGS.  Horizontal lines in
  panel {\it b} mark the divisions used to define the three
  $\mu_\Delta$ classes (see Fig.~\ref{fg:nfgscolmasstype}) while
  vertical lines mark shifts in their relative dominance as seen in
  panel {\it a}.  Ultra-blue fast-growing quasi-bulgeless galaxies and
  red-and-dead spheroid-dominated galaxies occupy distinct loci.
  Bulged disk galaxies are transitional, displaying a broad range of
  colors, but also show a clear peak starting at $(u-J)^e \sim 3.4$,
  just shy of the quenched spheroid regime.  Visual inspection of
  galaxies in this peak reveals residually star-forming S0 or dying
  spiral galaxies.}
\label{fg:ujmorph}
\end{figure*}

\begin{figure*}[t]
\plottwo{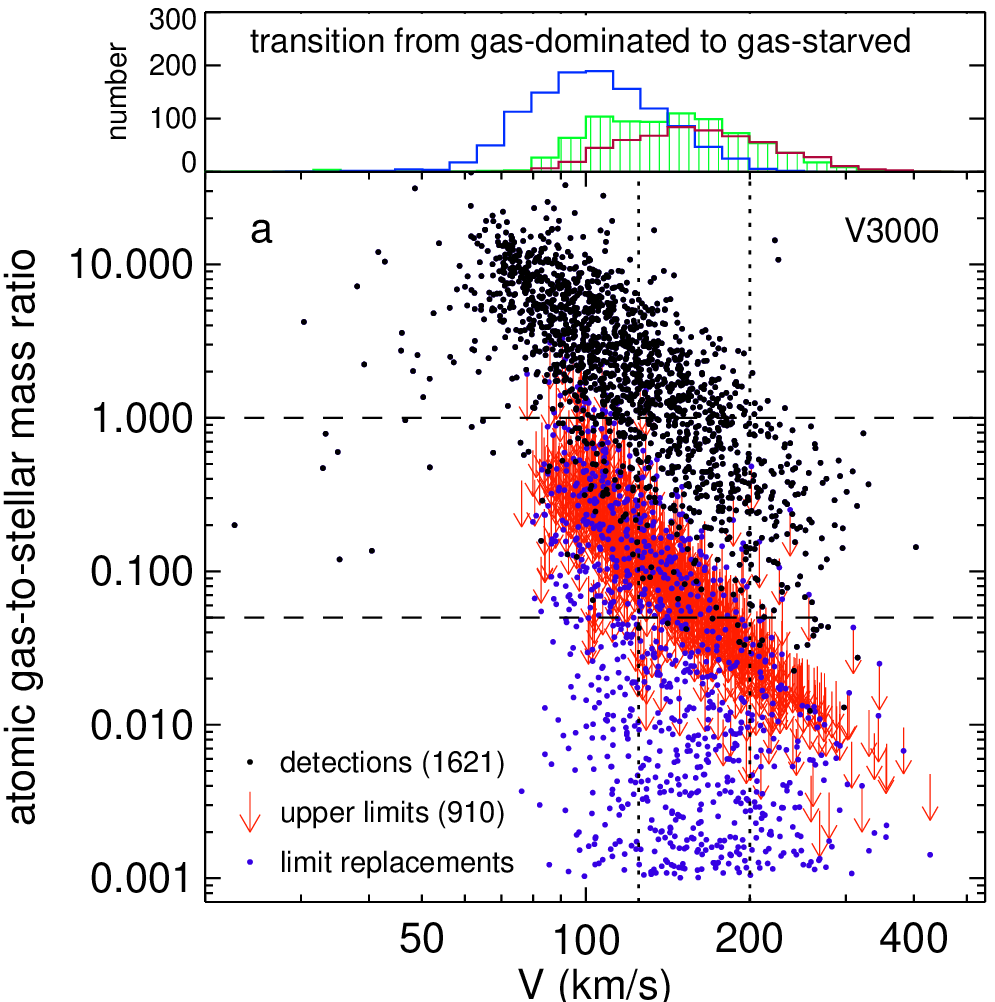}{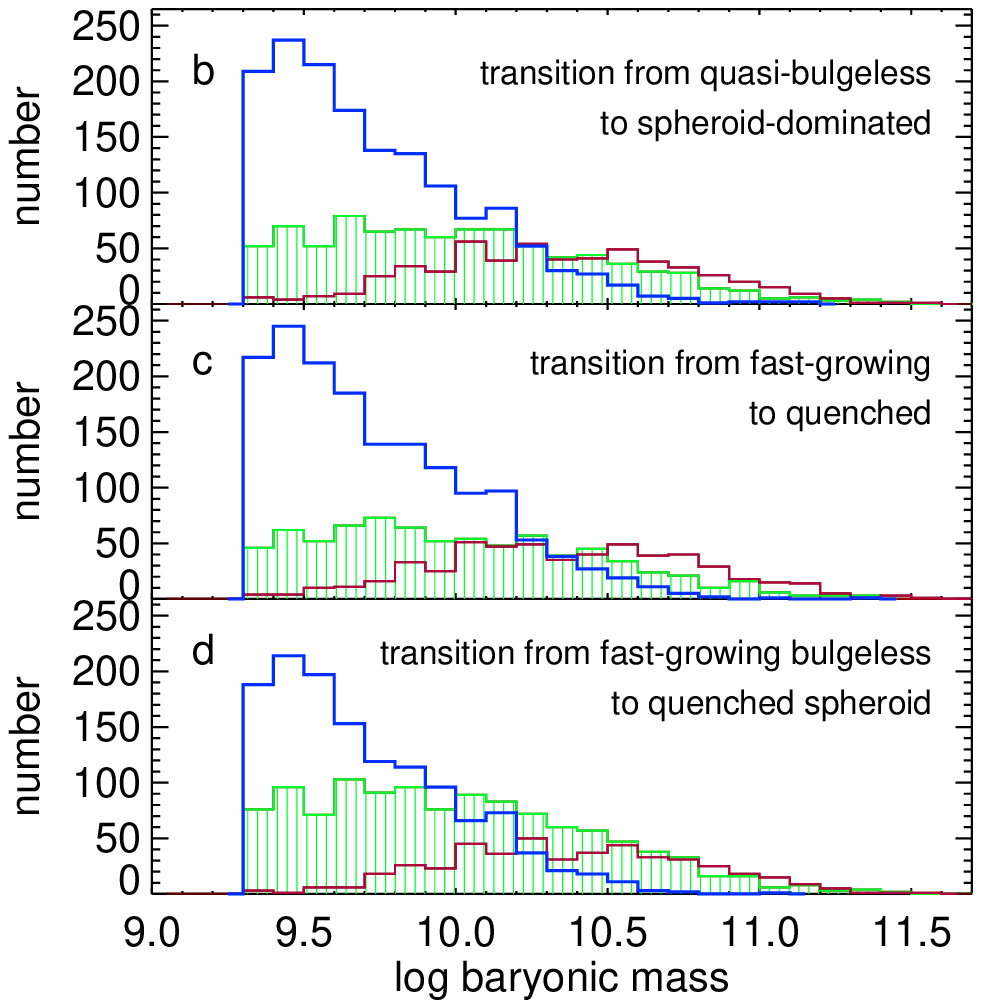}
\caption {{\it (a)} $M_{\rm HI}/M_*$ vs.\ $V$ for the V3000 sample,
  using the photometric gas fraction technique to replace upper limits
  with likely values based on $(u-J)^m$ color (see
  \S\ref{sec:transitions} for details).  We now draw from the full
  V3000 sample down to $M_r=-17$ to identify galaxies with measured or
  estimated $M_{\rm bary}>10^{9.3}$\,\msun, so the absence of quenched
  galaxies at low $M_{\rm bary}$ is plausibly correct (see
  \S\ref{sec:accdom} for further discussion).  The offset in $V$
  between detections and limits is not significant but reflects the
  use of the all-galaxy $M_r$--$V$ relation to provide $V$ for limits
  (\S\ref{sec:v3000}), which suppresses natural scatter. Vertical
  lines mark the threshold and bimodality scales and horizontal lines
  divide the gas-dominated, intermediate-gas, and gas-starved regimes.
  Histograms of these regimes (blue, green, and red, respectively) are
  shown at the top of the panel, illustrating their relationship to
  the threshold and bimodality scales.  {\it (b, c, d)} Mass
  distributions of galaxies in different color and morphology regimes
  for the baryonic mass limited sample shown in panel {\it a}. We
  consider three ways of identifying the transitional regime (green
  histogram) with reference to Fig.~\ref{fg:ujmorph}.  In panel {\it
    b}, transitional galaxies are those in the bulged disk
  $\mu_\Delta$ class.  In panel {\it c}, transitional galaxies are
  those whose colors fall in the range $(u-J)^e<2.7$--3.8, which is
  broadly inclusive of residually star-forming galaxies at the red
  end. In panel {\it d}, transitional galaxies are all those that do
  not occupy the regions of $(u-J)^e$ vs.\ $\mu_\Delta$ parameter
  space dominated by blue quasi-bulgeless systems ($(u-J)^e<2.7$,
  $\mu_\Delta<8.6$) or quenched spheroids ($(u-J)^e>3.4$,
  $\mu_\Delta>9.5$).  Only by this hybrid definition does the
  transitional population become ``typical'' (in the sense of ``more
  common than either other population considered separately'') within
  the range between the threshold and bimodality scales.}
\label{fg:ujregimes}
\end{figure*}

From a refueling perspective, three regimes emerge naturally from
Fig.~\ref{fg:ssfrcheck}b, reflecting coordinated shifts in gas
richness, growth rate, and morphology (as parametrized by $\mu_\Delta$
class in Figs.~\ref{fg:uk} and~\ref{fg:ssfrcheck}).  The {\it
  accretion-dominated} regime of gas-dominated quasi-bulgeless disks
is characterized by extremely rapid growth (FSMGR$_{\rm LT}$ $\sim$ 1
implies stellar mass doubling on Gyr timescales) and ultra-blue
colors.  The {\it processing-dominated} regime of ``normal'' bulged
disks is mostly characterized by moderate growth and gas content at
tens of \% levels, but these transitional systems show greater
diversity in color and gas content than quasi-bulgeless or
spheroid-dominated systems (Fig.~\ref{fg:ujmorph}).  We will argue in
\S\ref{sec:procdom} that this regime may also be considered to
include blue spheroid-dominated galaxies, to the extent that they are
likely to regrow disks rather than quench.  Finally, the {\it
  quenched} regime of red-and-dead spheroid-dominated galaxies is
characterized by gas poverty ($M_{\rm HI}/M_*\la0.05$) and slow growth
(FSMGR$_{\rm LT}\la0.1$), although the exact definition of this regime
is debatable, given the existence of a distinct population of {\it
  nearly} quenched bulged-disk systems with $(u-J)^e=3.4$--3.8
as seen in Fig.~\ref{fg:ujmorph}.

Below we examine the significance of these regimes in the context of a
broader picture of galaxy evolution. We first relate regime
transitions to the threshold and bimodality scales reviewed in
\S\ref{sec:disting}, with attention to the complicating factor of
environment, then go on to discuss implications of the
refueling regime picture for galaxies typical of each regime.

\subsubsection{Regime Transitions and Galaxy \& Halo Mass Scales}
\label{sec:transitions}

The existence of red disk and blue spheroid crossover systems in
Fig.~\ref{fg:ujmorph} points to the complexity of refueling and
quenching, likely reflecting the influence of large-scale environment as well
as local galaxy interactions and mergers. Environmental analysis of
the V3000 sample and its parent survey will be presented elsewhere
(Moffett et al., in prep.; Eckert et al., in prep.; Kannappan et al.,
in prep.), but for the present discussion we note that these studies
present a general picture in which galaxies that are central (most
massive) in their dark matter halos follow the simplest trajectories
from accretion-dominated to quenched, as judged by U-NIR colors, while
satellite galaxies display greater diversity including some of the
most extreme crossover states. This picture reinforces that of
\citet{peng.lilly.ea:mass}, who argue that central galaxies experience
essentially zero environmentally driven evolution, quenching solely as
a function of galaxy mass, while satellites are responsible for all
signatures of quenching that increase with group halo mass.  Thus
central/satellite differences offer a way to reconcile the scattered
dependence of gas richness on galaxy mass with the existence of clear
gas richness transitions at the threshold and bimodality scales, below
and above which, respectively, galaxies typical of the
accretion-dominated and quenched regimes become abundant (as seen in
\S\ref{sec:disting} and further confirmed for the V3000 sample below).

If we assume that observed transitions in gas richness, morphology,
and dynamics at the threshold and bimodality scales reflect the {\it
  en masse} transformation of central galaxies, then we can link these
scales to equivalent halo mass scales.  At the threshold scale, the
central-galaxy $M_*$ to $M_{\rm halo}$ calibration of
\citet{behroozi.wechsler.ea:average} indicates that
$M_*\sim10^{9.7}$\,\msun\ corresponds to $M_{\rm
  halo}\sim10^{11.5}$\,\msun.  Equivalently, the threshold velocity of
$\sim$125\,\kms\ matches a halo virial velocity of $V_{\rm
  vir}\sim100$\,\kms\ if $V_{\rm galaxy}/V_{\rm vir}\sim1.3$, as
reported by \citet{reyes.mandelbaum.ea:optical-to-virial}.  At the
bimodality scale, the same logic implies halo parameters of $M_{\rm
  halo}\sim10^{12.1}$\,\msun\ and $V_{\rm vir}\sim160$\,\kms.

These two halo mass scales are potentially important in relation to
cosmic accretion.  First, observational analysis suggests that {\it
  above} a halo mass of $\sim$10$^{12.1}$\,\msun, central galaxy $M_*$
rises much more slowly with growing halo mass
\citep{leauthaud.tinker.ea:new}, implying that the primary mode of
central galaxy growth that operates at lower halo mass is being valved
off above $M_{\rm halo}\sim$10$^{12.1}$\,\msun\ (corresponding to the
bimodality scale).  Second, several theoretical prescriptions suggest
that somewhere between $M_{\rm halo}$ =
10$^{11}$--10$^{12}$\,\msun\ (likely corresponding to the threshold
scale), halo gas cooling becomes much more efficient, such that
$\ga$50\% of baryons accreted onto the halo over the age of the
universe cool within the same time \citep{lu.kere-s.ea:on}.  In the
specific case of the ``cold-mode'' accretion picture
\citep{birnboim.dekel:virial,kere-s.katz.ea:how}, wherein slow
hot-mode accretion replaces rapid cold-mode accretion as shock-heating
becomes more effective with increasing halo mass, calculations by
\citet[][see their Figs.~2--3]{dekel.birnboim:galaxy} show that at
$z=0$, the shock-heating transition occurs in the {\it inner} halo at
0.1$\times$ the virial radius for halos with $V_{\rm
  vir}\ga$100\,\kms\ (the threshold scale for centrals), while the
transition expands to the outer halo past the virial radius for halos
with $V_{vir}\ga170$\,\kms\ (the bimodality scale for centrals).
Recent work using the moving-mesh AREPO code questions the details of
this picture, as $\sim$half of simulated cold-mode accretion may reflect
numerical artifacts \citep{nelson.vogelsberger.ea:moving}, yet overall
accretion rates onto halos are actually higher using AREPO, due to
faster cooling of the hot mode, and the general halo mass dependence
of cooling remains.  Furthermore, we note that the onset of a
``rapid'' accretion mode when the cooling radius exceeds the halo
virial radius is a common feature of many semi-analytic cooling
prescriptions, as reviewed by \citet{lu.kere-s.ea:on}, most of which do not
involve a cold-mode/hot-mode distinction {\it per se}.

In this context it is interesting to revisit the possibility of a {\it
  sharp} transition at the threshold scale as proposed by
\citet{dalcanton.yoachim.ea:formation}, which might constrain
theoretical models of accretion.  To construct a statistically robust
sample to test this possibility, we first estimate baryonic masses for
the full V3000 sample down to $M_r=-17$. 
We adopt the photometric gas fraction estimator 
presented in Fig.~\ref{fg:uk}a for galaxies with $(u-J)^m<3.7$:
\begin{equation}
\log{M_{\rm HI}/M_*}=2.70-0.98(u-J)^m
\end{equation}
with 0.34~dex scatter.  We assign gas masses using Monte Carlo methods
to produce realistic variations matching this 0.34~dex scatter in
$M_{\rm HI}/M_*$, while not allowing gas masses higher than the
measured upper limits.  For galaxies redder than $(u-J)^m=3.7$, we
assign gas fractions randomly in the logarithmic interval from $M_{\rm
  HI}/M_*=0.001$ to the limit value or 0.5, whichever is smaller.
Finally, we sub-select a baryonic mass limited sample from the V3000
data set with $M_{\rm bary}>10^{9.3}$\,\msun (the expected
completeness limit; see \S\ref{sec:v3000}). The $M_{\rm HI}/M_*$
vs.\ $V$ distribution for this sample is plotted in
Fig.~\ref{fg:ujregimes}a.  Note that the offset in $V$ between
detections and limit replacements is due to the fact that most
detections can scatter to low or high $V$, whereas limits (and some
detections, e.g., those with $i<50\deg$) must be plotted with an
inferred $V$ from the all-galaxy $M_r$--$V$ relation, which tends to
overestimate $V$ for gas-poor galaxies (Fig.~\ref{fg:massv}).

Based on this analysis, Fig.~\ref{fg:ujregimes}a shows no obviously
sudden onset of gas richness for the V3000 sample. We do see that
gas-dominated galaxies become the majority population roughly below
the threshold scale (histograms at the top of
Fig.~\ref{fg:ujregimes}a), and that the V3000 sample shows the same
large, strongly mass-dependent scatter in $M_{\rm HI}/M_*$ that we saw
for the NFGS in Fig.~\ref{fg:goversv}, lending credence to the NFGS
results.  Gas-starved dwarfs are now represented (based on $(u-J)^m$)
but drop sharply in relative frequency as gas-dominated galaxies
become more common.  Furthermore, the gas-starved dwarf population
resides predominantly in the highest-density environments seen in
Fig.~\ref{fg:v3000sky}, such as the Coma Cluster, implying
environmental quenching.  By contrast, the majority of dwarf galaxies
are isolated and gas rich.

\bigskip

\subsubsection{The Accretion-Dominated Regime} 
\label{sec:accdom}

For the quasi-bulgeless, gas-dominated disks typical of the accretion-
regime, fresh gas is accreted as fast or faster than it can be
consumed, based on Fig.~\ref{fg:ssfrcheck}b.  Most quasi-bulgeless
disks scatter around FSMGR$_{\rm LT}\sim1$\,Gyr$^{-1}$, suggesting
that they are capable of doubling their stellar masses each Gyr.  Thus
{\it typical gas-dominated Sd--Im galaxies are growing exponentially},
at least in the ``high-mass dwarf'' regime we explore
(\citealt{bouche.dekel.ea:impact} note the possibility of an accretion
``floor'' that could prevent effective gas cooling for lower mass
galaxies).  This conclusion is consistent with that of
\citet{moster.naab.ea:galactic}, who infer from a multi-epoch halo
abundance matching analysis that today's low-mass galaxies are still
increasing in star formation rate (where these authors' definition of
low mass roughly equates to below the threshold scale).

The standard counterargument supposes that dwarf galaxies have
intrinsically bursty star formation and therefore experience off
states that ensure low time-averaged SSFR despite high short-term SSFR
\citep[e.g.,][]{feulner.goranova.ea:connection}.  However, we have
measured high growth rates averaged over the last Gyr.  Moreover,
systematic searches for non-star-forming dwarf galaxies have shown
them to be exceedingly rare in field environments
(\citealt{lee.gil-de-paz.ea:galex}; see also
\citealt{lee.kennicutt.ea:dwarf}) and, when found, to be associated
with proximity to massive hosts \citep{geha.blanton.ea:stellar}.  In
the volume-limited V3000 sample, the sharply dropping relative
frequency of gas-starved galaxies with decreasing galaxy mass in
Fig.~\ref{fg:ujregimes} is likewise consistent with few (or brief)
truly off states, although we remind the reader that this sample
suffers from residual bias against low surface brightness dwarfs,
which might in principle include a high mass-to-light ratio population
(\S\ref{sec:v3000}).  Nonetheless from the evidence in hand it is
reasonable to conclude that most low-mass galaxies achieve high
FSMGR$_{\rm LT}$ through sustained growth.

A related concern is the effect of outflows on dwarfs. However, the
early idea that supernova feedback in dwarfs should expel their gas
and turn them off \citep[][]{dekel.silk:origin} appears most relevant
to galaxies much less massive than we consider, with
$V\sim30$\,\kms\ \citep{mac-low.ferrara:starburst-driven}.  Powerful
outflows are indeed expected --- the data of
\citet{martin.shapley.ea:demographics} reveal a significant increase
in the frequency of strong outflows for $z=0.4$--1.4 galaxies with
short-term SSFRs $\ga0.8$\,Gyr$^{-1}$ --- but they are
likely to regulate rather than terminate star formation (see the
comparison of outflow scenarios in \citealt{dave.katz.ea:neutral}).
The cosmological simulations of \citet{brooks.governato.ea:origin}
also provide another perspective, suggesting that even significant
outflows are less important than low star formation efficiency in
setting the gas and metal content of low-mass galaxies.  Of course,
``low'' star formation efficiency in systems experiencing cosmic
accretion may not imply low SSFRs in absolute terms, but simply star
formation that cannot keep up with the inflow rate.

As noted by \citet{khochfar.silk:specific} in the course of
interpreting high-redshift observations, galaxies with sustained high
SSFRs may pose a challenge for current models of galaxy evolution. For
example, the simulation-motivated analytic model of
\citet[][]{dave.finlator.ea:analytic}, which assumes an equilibrium
between mass inflow, mass outflow, and star formation, would predict
specific star formation rates reaching only $\sim$0.2--0.3\,Gyr$^{-1}$
for isolated $z=0$ galaxies near the threshold scale \citep[in rough
  agreement with the simulations of][in which feedback prevents
  accretion from reaching the ISM for such
  galaxies]{.schaye.ea:rates}.  In this context, it is interesting to
consider the possibility of excess accretion, leading to
non-equilibrium ``puddling'' of ionized gas, contrary to the
assumptions of \citet{dave.finlator.ea:analytic}.  While we have
argued that {\it cold} HI+H$_2$ gas cycles through galaxies on Gyr
timescales, nothing in Fig.~\ref{fg:ssfrcheck} requires that ionized
gas arrive at the same rate.  In fact, analyses of the baryonic
Tully-Fisher relation suggest that dwarf galaxies may contain
3--4$\times$ larger gas reservoirs than their HI-derived gas masses imply
\citep{pfenniger.revaz:baryonic,begum.chengalur.ea:baryonic}. Our own
baryonic $M$--$V$ relation in Fig.~\ref{fg:massv} may be ``missing''
such a multiplier, which would straighten it out below
$V\sim125$\,\kms.\footnote{Another interesting set of results in the
  literature centers on the possibility of substantial CO-dark
  molecular gas in dwarf galaxies
  \citep[e.g.,][]{madden.poglitsch.ea:c}.  If such undetected H$_2$
  gas were comparable in mass to the HI, it might explain some of the
  scatter in $M_{\rm HI}/M_*$ vs.\ FSMGR$_{\rm LT}$.}  

Any such excess gas would presumably end up being processed
intermittently in merger-driven bursts (distinct from internally
driven bursts as considered above), likely producing phenomena such as
blue compact dwarfs (BCDs). BCDs have previously been linked to
gas-rich dwarf-dwarf mergers
\citep[e.g.,][]{pustilnik.kniazev.ea:environment,bekki:formation}.  If
accretion rates remain high after such events, the remnants may enjoy rapid
disk regrowth.  For example, the BCD NGC~7077 (shown in
Fig.~\ref{fg:goversv} at $M_{\rm HI}/M_*\sim0.4$ and
$V\sim60$\,\kms\ and classed as an S0a in the NFGS) reveals a
combination of color gradients, molecular and atomic gas content, and
{\it GALEX}-detected extended UV disk structure that together suggest
this galaxy is a post-merger system just starting to rebuild its disk
(\citealt{moffett.kannappan.ea:extended}; S13). If such a galaxy 
enters a reduced accretion regime post-merger, it may give rise
to an Sa--Sc spiral, while if it stays in a high accretion regime, it
may soon revert to Sd--Im morphology.  Consistent with this picture,
kinematic signatures of past mergers are observed in the stellar
components of even very late-type dwarf disks \citep[e.g.,
  counterrotating stars and/or thick
  disks:][]{kannappan.fabricant:broad,yoachim.dalcanton:kinematics,leaman.cole.ea:stellar,leaman.venn.ea:resolved}.

The reader might reasonably object to the notion of exponentially
growing Sd--Im galaxies based on the implausibility of efficient gas
processing in such systems (although the rate of {\it molecular} gas
conversion to stars in such galaxies may actually be higher than in
spirals, e.g.,
\citealt{gardan.braine.ea:particularly,pelupessy.papadopoulos:molecular}).
However, we stress that regardless of any deficiency in gas transport
mechanisms, the fact that FSMGR$_{\rm LT}$ values remain high in the
accretion-dominated regime implies that accretion-dominated systems do
not have to form stars {\it efficiently} in order to form them {\it
  rapidly}: the rate of gas influx is apparently sufficient to
overwhelm any inefficiency in consumption.  Moreover, it is not
obvious that the conventional definition of star formation efficiency,
i.e., the inverse of the timescale for the current rate of star
formation to consume the current reservoir of gas, makes sense for
accretion-dominated galaxies.  \citet{huang.haynes.ea:arecibo} report
as a paradox the fact that the most gas-rich galaxies in ALFALFA have
both the highest SSFRs and the lowest star formation efficiencies, and
they attribute this paradox to a bottleneck in processing the gas.  We
would propose that the paradox arises from the fact that the English
words ``low efficiency'' connote galaxies lazily consuming gas: if
instead the gas is pouring in faster than even the highest-SSFR
galaxies can possibly consume it, such galaxies might be better termed
``overwhelmed'' than ``inefficient.''

\subsubsection{The Processing-Dominated Regime} 
\label{sec:procdom}

For galaxies in the processing-dominated regime, gas is still consumed
roughly as fast as it is accreted (Fig.~\ref{fg:ssfrcheck}b, in
agreement with \citealt[][]{fraternali.tomassetti:estimating}).  For
centrals whose halos are experiencing reduced cosmic accretion, the
transition to a processing-dominated state may be self-reinforcing.
Slowed dilution of the ISM with fresh gas and/or reduced metal loss in
outflows may lead to more efficient H$_2$ formation on dust grains
\citep{krumholz.dekel:metallicity-dependent}.  This process may in
turn increase consumption efficiency for the diminished infalling gas
and accelerate the galaxy's evolution to a processing-dominated state
(thus decelerating growth).  In addition, mild quenching may encourage
more prominent stellar bulges and more concentrated disks that further
promote higher gas densities and more efficient star formation
\citep[e.g.,][]{blitz.rosolowsky:role}.

Notably, galaxies in the processing-dominated regime show deviations
in $M_{\rm HI}/M_*$ that seem to correlate with high H$_2$/HI, based
on the incomplete H$_2$ data available (Fig.~\ref{fg:ssfrcheck}).  Our
previous analysis of NFGS galaxies has revealed bursty, correlated
variations in central and outer disk colors and H$_2$/HI ratios,
particularly for bulged disks (S13). These variations seem to reflect
externally triggered gas inflow events and central star formation
enhancements, with corresponding lulls in outer-disk growth
(\citealt{kannappan.jansen.ea:forming}; S13).  The most extreme events
lead to formation of blue-sequence E/S0s, which account for many of
the spheroid-dominated galaxies embedded in the main FSMGR$_{\rm
  LT}$--$M_{\rm HI}/M_*$ locus in Fig.~\ref{fg:ssfrcheck}.  Based on a
variety of evidence, including identification of temporal sequences in
a ``fueling diagram'' linking stellar populations and gas content, we
have argued elsewhere that blue-sequence E/S0s reflect a process of
gas-rich merger-induced bulge building followed by disk regrowth,
which may play a crucial role in the morphological evolution of disk
galaxies (KGB; W10a; \citealt{
  wei.vogel.ea:relationship,moffett.kannappan.ea:extended};
S13). Other authors have also documented a ``living'' population of
E/S0s
\citep[e.g.,][]{driver.allen.ea:millennium*1,cortese.hughes:evolutionary,thilker.bianchi.ea:ngc,huertas-company.aguerri.ea:evolution,marino.bianchi.ea:tracing}.
In the mass regime below the threshold scale, blue-sequence E/S0s
overlap with BCDs, suggesting a continuity of phenomena; such galaxies
are most common below the threshold scale, with a tapering population
up to the bimodality scale.

While bursty star formation has traditionally been ascribed to dwarf
galaxies, \citet{kauffmann.heckman.ea:gas} demonstrate that stochastic
star formation is in fact most prevalent for galaxies with the
intermediate masses and surface mass densities of bulged spirals
(which happen to coincide with those of blue-sequence E/S0s; KGB).
From a comparison to semi-analytic models, these authors infer that the
observed stochasticity reflects efficient processing of fresh gas
infall. \citet{ferreras.silk.ea:star} advance similar ideas in their
analysis of intermediate-redshift galaxies, concluding that efficient,
short-lived bursts are typical for galaxy rotation velocities
$V\ga140$\,\kms.  These authors' bursty accretion picture is roughly
consistent with ours, with the caveat that we emphasize the roles of
not only cosmic infall but also gas processing mechanisms in refueling.
The fundamental importance of both is illustrated by an ``exception
that proves the rule'': the gas-dominated S0 at $V\sim90$\,\kms\ and
$M_{\rm HI}/M_*\sim4$ in Fig.~\ref{fg:goversv}a is UGC~9562, a polar
ring galaxy for which the misalignment of fresh accretion relative to
the inner bulge/disk orientation likely inhibits normal gas transport,
creating a system hovering between the accretion-dominated and
processing-dominated regimes.\footnote{\citet{cox.sparke.ea:stars}
  make the case that the gas in this polar ring was accreted long
  prior to the galaxy's interaction with a neighbor. Independent
  evidence linking other polar rings to cosmic gas accretion includes
  low gas-phase metallicity \citep{spavone.iodice.ea:chemical} and
  alignment within a large-scale cosmic wall
  \citep{stanonik.platen.ea:polar}.}

Both externally driven and secular gas inflow mechanisms should become
more important above the threshold scale.
\citet{hopkins.kere-s.ea:when} perform an analytic timescale analysis
comparing growth by cosmic accretion, merging, and secular disk
evolution as a function of galaxy mass at $z=0$.  They find that
although merger rates are insignificant compared to gas accretion
rates for $M_{\rm bary}\la$10$^{10}$\,\msun, the two become comparable
at intermediate masses, as the merger rate (per Gyr) increases with
galaxy mass \citep[e.g.,][]{maller.katz.ea:galaxy}.  Furthermore,
\citet{sinha.holley-bockelmann:first} find that the flyby interaction
rate, normally ignored in theoretical merger histories, may exceed the
minor merger rate at intermediate mass scales.  At these mass scales
Hopkins et al.\ also find that reduced accretion enables secular
processes such as bar formation to become relatively more important.
Observationally, bars appear more common in lower-mass, later-type
galaxies than in Sa--Sc galaxies
(\citealt{barazza.jogee.ea:bars,nair.abraham:on}), but
\citet{ellison.nair.ea:impact} argue that strong bars in galaxies with
$M_*>10^{10}$\,\msun\ may be substantially more {\it effective} in
enhancing star formation than bars in lower mass galaxies (although
the possibility of differential time lags in enhanced star formation
complicates interpretation).  On the other hand, central star
formation enhancements strong enough to create blue-centered color
gradients do not obviously correlate with the presence of bars, but
instead with signs of minor mergers and interactions
\citep[][S13]{kannappan.jansen.ea:forming,gonzalez-perez.castander.ea:colour}.

While morphology correlates better with FSMGR or U-NIR
color\footnote{Interestingly, the bimodality in $\mu_\Delta$ as a
  function of $(u-J)^e$ is much less apparent as a function of
  FSMGR$_{\rm LT}$.  We suspect that the coarse binning of possible
  FSMGR$_{\rm LT}$ values in our stellar population models blurs the
  structure seen in $(u-J)^e$.} than with mass (compare Figs.~\ref{fg:ujmorph}a
and~\ref{fg:ujregimes}b; see also
\citealt{franx..ea:structure}), the definition of the
processing-dominated regime by morphology and/or U-NIR color is complicated by
the variety of transitional states.  The histograms in
Fig.~\ref{fg:ujregimes} panels {\it b, c, } and {\it d} show three
possible definitions, with only the most inclusive ({\it d:} all
bulged disks + bluish spheroid-dominated galaxies + reddish
quasi-bulgeless galaxies) yielding a numerically dominant population
between the threshold and bimodality scales, where intermediate gas
richness systems are the norm (Fig.~\ref{fg:ujregimes}a).  At the red
end, this population includes residually star-forming S0s and dying
spirals (Fig.~\ref{fg:ujmorph}a), not unreasonably since such galaxies
appear to extend the gas--star formation relation of ``normal'' Sa--Sc
spirals in Fig.~\ref{fg:uk}. On the other hand, this analysis suggests
that such ``normal'' spirals, although broad in their mass
distribution, are not actually typical at any mass.

\subsubsection{The Quenched Regime} 

Most galaxies in the quenched regime have negligible $M_{gas}/M_*$,
although values of FSMGR$_{\rm LT}$ as high as $\sim$$0.1$\,Gyr$^{-1}$
are not uncommon. The low gas-to-stellar mass ratios in this regime
may reflect not only slowed cosmic accretion but also ram-pressure
stripping and/or efficient AGN feedback in hot-gas dominated halos
\citep[e.g.,][]{somerville.hopkins.ea:semi-analytic,woo.dekel.ea:dependence}.
However, significant growth largely uncorrelated with cosmic gas
accretion may be possible via intermittent satellite mergers, which
might rejuvenate the galaxy either by briefly fueling star formation
or by injecting stellar mass formed relatively recently in the
satellite \citep{morganti..ea:neutral,martini.dicken.ea:origin}.  For
spheroid-dominated galaxies, \citet{martig.bournaud.ea:morphological}
argue that ``morphological quenching'' inhibits the gas processing
necessary to form stars, and indeed, the most massive
($V\ga325$\,\kms) early-type galaxies seem to lack recent star
formation regardless of gas content \citep{serra.trager.ea:stellar}.
Consistent with this idea, we see two degrees of quenching in
Fig.~\ref{fg:ujmorph}a, with a distinct population of {\it nearly}
quenched bulged disks forming a peak just blueward of the red-and-dead
peak for spheroid-dominated galaxies. Arguably, these systems
represent the extreme tail of the processing-dominated regime,
appearing as dying spirals or slightly blue S0s.  Previous studies
have shown that residual star formation is common for S0s over a wide
range in mass
\citep[e.g.,][]{salim.rich:star,moffett.kannappan.ea:extended}, even
in environments where Es are quenched
\citep{salim.fang.ea:galaxy-scale}. Stellar mass loss might account
for much of the gas supply
\citep{sage.welch:cool,leitner.kravtsov:fuel}.  The phenomenon of
partial quenching seen in Fig.~\ref{fg:ujmorph} reaffirms that both
accretion and processing are important to refueling: although the hot
gas in galaxy clusters greatly slows accretion (and also removes gas
by ram-pressure stripping), still gas buildup can occur, and disk
galaxies remain capable of processing it.

\section{Summary and Conclusions}

Our analysis has made use of two samples, both broadly representative
of the general galaxy population down to the ``high-mass dwarf''
regime (baryonic masses $\sim$10$^{9}$\,\msun).  These samples span a
wide range of void to cluster environments, with non-cluster galaxies
naturally predominant.  The NFGS comprises $\sim$200 galaxies with
newly complete HI data, partial CO data, and internal kinematic data
homogenized from a mix of stellar, ionized gas, and neutral gas
measurements. The NFGS HI database is uniquely powerful in providing
detections or strong upper limits ($M_{\rm HI}\la0.1M_*$) for all
galaxies.  The V3000 sample comprises $\sim$3000 galaxies in a
volume-limited union of the SDSS survey and the blind 21cm ALFALFA
survey, with HI upper limits computed for all ALFALFA non-detections.
We have combined SDSS and ALFALFA redshifts with redshifts from
several other surveys to achieve high completeness in the V3000 sample
down to baryonic masses $\sim$10$^{9.3}$\,\msun, although residual
incompleteness for dwarf galaxies remains.

Using these data sets, we have confirmed and clarified
observed shifts in gas richness and morphology at two key galaxy mass
scales:

\begin{itemize}

\item{Below the
{\it gas-richness threshold scale} ($V\sim125$\,\kms; baryonic mass
$\sim$10$^{9.9}$\,\msun), gas-dominated quasi-bulgeless disks become most numerous in the galaxy population.}

\item{Above the {\it bimodality scale} ($V\sim200$\,\kms; baryonic
  mass $\sim$10$^{10.6}$\,\msun), gas-starved spheroid-dominated
  galaxies rise to prominence, with numbers comparable to nearly
  quenched bulged disks.}

\end{itemize}

Notwithstanding these transitions, we have shown that both of our
samples display far greater scatter in $M_{\rm gas}/M_*$ as a function
of $V$ (a non-covariant mass proxy) than has been commonly assumed.
In particular, the NFGS reveals the natural diversity of gas richness
in a sample unbiased with respect to 21cm flux: its $M_{gas}/M_*$
vs.\ $V$ plot is striking in its complexity, with quenched galaxies at
all masses and scatter $\ga$0.5~dex even for unquenched galaxies. We
suggest that the diversity of $M_{gas}/M_*$ has been suppressed in
previous studies by selection and detection biases as well as the
practice of plotting $M_{gas}/M_*$ vs.\ $M_*$, which allows large
covariant errors in $M_*$ to enhance the apparent correlation.

In contrast to its complicated correlation with mass, $M_{\rm
  gas}/M_*$ shows a simple and clean correlation with another
fundamental quantity: the {\it long-term} fractional stellar mass
growth rate (FSMGR$_{\rm LT}$), which we define as the ratio of a
galaxy's stellar mass formed within the last Gyr to its preexisting
stellar mass.  The $M_{\rm gas}/M_*$--FSMGR$_{\rm LT}$ correlation
represents the physical underpinnings of the remarkably tight
($\sigma\sim0.3$~dex) correlation between $M_{\rm gas}/M_*$ and
U$-$NIR color, used in the ``photometric gas fraction'' technique of
K04, because U$-$NIR color directly predicts FSMGR$_{\rm LT}$.
Importantly, a fractional growth rate is related to, but distinct
from, a specific star formation rate, as it can exceed a value of one
over the unit of time.  Choosing Gyr as our unit of time reveals the
remarkable fact that for an unquenched galaxy, the mass of new stars
formed in the last Gyr is roughly {\it equal to} the mass of fresh gas
it has available for future star formation, within factors of a few.
Thus for galaxies, past performance is a predictor of future success.
We have demonstrated that this result is incompatible with any
reasonable closed box model and does not derive from the inherently
short-term ``gas fuels star formation'' physics of the
Kennicutt-Schmidt Law, contrary to previous interpretations of the
photometric gas fraction technique.  Instead, we have argued that it
reflects the physics of refueling: both cosmic accretion and the
mechanisms that drive internal gas processing must be routine on
$\la$1~Gyr timescales.

We have distinguished three refueling regimes:

\begin{itemize}

\item{Blue quasi-bulgeless galaxies are {\it accretion dominated}, enjoying
  largely continuous refueling and stellar mass growth of order 100\%
  per Gyr.  Contrary to their reputation for ``inefficient'' star
  formation, such gas-dominated dwarf galaxies appear to be growing
  exponentially.  We suggest that their high $M_{\rm gas}/M_*$ ratios
  may be more fairly interpreted as evidence that they are
  ``overwhelmed'' rather than ``inefficient.'' }

\item{Bulged disk galaxies are {\it processing dominated}, consuming a
  reduced rate of fresh accretion in efficient, often externally
  driven bursts with net growth rates at tens of \% levels.  Mergers
  and interactions in this regime create fluctuations in morphology
  and gas content, including transient blue E/S0 states that may
  regenerate spiral morphology.
}

\item{Red-and-dead spheroid-dominated galaxies are {\it quenched} (per
  the usual terminology), with negligible gas and $\sim$1--10\%
  stellar mass growth per Gyr.  They may grow slowly by satellite
  accretion.  A population of ``nearly quenched'' bulged disk galaxies
  lies at the interface of the processing-dominated and quenched
  regimes, extending the $M_{\rm gas}/M_*$--FSMGR$_{\rm LT}$ relation
  to low levels.}

\end{itemize}

High SSFRs in dwarf galaxies are sometimes discounted as evidence of
bursty star formation, which should average out to lower levels over
time.  We cannot rule out the possibility of an extremely high
mass-to-light ratio, low SSFR population missed in the V3000 sample
due to incompleteness in the SDSS redshift survey.  However, we note
that FSMGR$_{\rm LT}$ is averaged over the past Gyr.  Moreover, we
have reviewed evidence from recent studies suggesting that dwarf
galaxies are {\it not} especially bursty, except for the major bursts
obviously inherent in dwarf-dwarf mergers, such as might produce blue
compact dwarfs.  In fact, we have argued that bulged disk galaxies are
far more subject to bursty star formation than typical quasi-bulgeless
dwarfs.  As discussed, the burstiness of such processing-dominated
galaxies likely reflects increasingly efficient mechanisms of gas
processing (both gas transport and HI-to-H$_2$ conversion) that
develop in tandem with, and partly because of, reduced cosmic
accretion.

Connecting back to the threshold and bimodality scales, we have seen
that the mass distributions of accretion-dominated,
processing-dominated, and quenched galaxies cross at these two scales,
suggesting an indirect relationship between refueling and galaxy
mass. Focusing on galaxies that are central in their halos, we find a
plausible correspondence between these galaxy mass scales and two halo
mass scales previously linked to transitions in cosmic accretion ($\sim$10$^{11.5}$ and $\sim$10$^{12.1}$\,\msun).  In
future work considering environment data we will examine the behavior of centrals and satellites
separately, demonstrating that centrals evolve through the
processing-dominated regime in precisely the narrow mass range between
the threshold and bimodality scales. In fact the V3000 sample
demonstrates that ``normal'' galaxies like our Milky Way --- bulged
disks with intermediate gas richness --- are not in general typical,
but most nearly approach typical in this mass range.

Although the analysis in this paper has not used halo mass data, our
results clearly motivate the need for a full analysis of $M_{\rm
  gas}/M_*$ vs.\ $V$ as a function of halo mass and central
vs.\ satellite status, for a sample that is highly complete down to
galaxy masses below the threshold scale.  To date no such sample
exists, in that all surveys that probe the full diversity of galaxy
environments lack sufficiently sensitive 21cm data to probe the full
range of $M_{\rm gas}/M_*$ in the dwarf regime.  The NFGS offers a
glimpse of the power of truly unbiased HI observations for a broad
sample but lacks well characterized and fairly sampled environments,
leading us back to the need for a volume-limited sample with
high-quality 21cm and kinematic data for a cosmologically diverse
range of environments.  We aim to construct such a data set as part of
the RESOLVE Survey ({\it http://resolve.astro.unc.edu}), currently in
progress on the SOAR, SALT, Gemini, GBT, and Arecibo telescopes.

\acknowledgements We are grateful to Douglas Mar, Jacqueline van
Gorkom, John Hibbard, Adam Leroy, Ari Maller, Martha Haynes, Sadegh
Khochfar, and Eric Gawiser for illuminating conversations.  The
anonymous referee provided helpful feedback that improved this
manuscript.  We thank Perry Berlind, Barbara Carter, Marijn Franx, and
the Mt. Hopkins observatory staff for their help with the NFGS
kinematics observing program.  We thank the GBT operators and Green
Bank staff for their support of program 10A-070.  The National Radio
Astronomy Observatory is a facility of the National Science Foundation
operated under cooperative agreement by Associated Universities, Inc.
SJK, DVS, KDE, and MAN were supported in this research by NSF CAREER
grant AST-0955368.  DVS and KDE also acknowledge support from GAANN
Fellowships and North Carolina Space Grant Fellowships.  SJK
acknowledges the hospitality and financial assistance of the NRAO
Charlottesville visitor program during spring 2010.  AJM acknowledges
support from a NASA Harriet Jenkins Fellowship, the University of
North Carolina Royster Society of Fellows, and the North Carolina
Space Grant Program.  LHW was supported in part by the NSF under the
CARMA cooperative agreement and in part by an SMA Postdoctoral
Fellowship.  SJ acknowledges support from the Norman Hackerman
Advanced Research Program (NHARP) ARP-03658-0234-2009, NSF grant
AST-0607748, and $HST$ grant GO-11082 from STScI, which is operated by
AURA, Inc., for NASA, under NAS5-26555. This research has made use of
the HyperLEDA database (http://leda.univ-lyon1.fr).  This work has
used the NASA/IPAC Extragalactic Database (NED) which is operated by
the Jet Propulsion Laboratory, California Institute of Technology,
under contract with the National Aeronautics and Space Administration.
This work has made use of data products from the Two Micron All Sky
Survey (2MASS), which is a joint project of the University of
Massachusetts and the Infrared Processing and Analysis
Center/California Institute of Technology, funded by the National
Aeronautics and Space Administration and the National Science
Foundation.  This work is based in part observations and on archival
data obtained with the Spitzer Space Telescope, which is operated by
the Jet Propulsion Laboratory, California Institute of Technology
under a contract with NASA. Support for this work was provided by an
award issued by JPL/Caltech and by NASA.  We acknowledge use of the
Sloan Digital Sky Survey (SDSS). Funding for SDSS-III has been
provided by the Alfred P. Sloan Foundation, the Participating
Institutions, the National Science Foundation, and the U.S. Department
of Energy Office of Science. The SDSS-III web site is {\it
  http://www.sdss3.org/}. SDSS-III is managed by the Astrophysical
Research Consortium for the Participating Institutions of the SDSS-III
Collaboration including the University of Arizona, the Brazilian
Participation Group, Brookhaven National Laboratory, University of
Cambridge, Carnegie Mellon University, University of Florida, the
French Participation Group, the German Participation Group, Harvard
University, the Instituto de Astrofisica de Canarias, the Michigan
State/Notre Dame/JINA Participation Group, Johns Hopkins University,
Lawrence Berkeley National Laboratory, Max Planck Institute for
Astrophysics, New Mexico State University, New York University, Ohio
State University, Pennsylvania State University, University of
Portsmouth, Princeton University, the Spanish Participation Group,
University of Tokyo, University of Utah, Vanderbilt University,
University of Virginia, University of Washington, and Yale University.


\scriptsize
\tablecaption{Photometry and Stellar Mass Estimates Table Description}
\begin{deluxetable}{ll}
\tablewidth{0pt}
\tablehead{\colhead{Column} & \colhead{Description}}
\startdata
1 & NFGS ID Number\\
2 & Object Name\\
3 & {\it GALEX} NUV magnitude\tablenotemark{1,2}\\
4 & Error on {\it GALEX} NUV magnitude    \\ 
5 & SDSS $u$ magnitude\tablenotemark{1}    \\
6 & Error on SDSS $u$ magnitude    \\ 
7 & SDSS $g$ magnitude\tablenotemark{1}   \\
8 & Error on SDSS $g$ magnitude \\
9 & SDSS $r$ magnitude\tablenotemark{1}   \\
10 & Error on SDSS $r$ magnitude   \\
11 & SDSS $i$ magnitude\tablenotemark{1}  \\
12 & Error on SDSS $i$ magnitude\\
13 & SDSS $z$ magnitude\tablenotemark{1}  \\
14 & Error on SDSS $z$ magnitude   \\
15 & 2MASS $J$ magnitude\tablenotemark{1}  \\
16 & Error on 2MASS $J$ magnitude   \\
17 & 2MASS $H$ magnitude\tablenotemark{1}  \\
18 & Error on 2MASS $H$ magnitude   \\
19 & 2MASS $K$ magnitude\tablenotemark{1}\\
20 & Error on 2MASS $K$ magnitude   \\
21 & $Spitzer$ IRAC 3.6 micron magnitude   \\
22 & Error on $Spitzer$ IRAC 3.6 micron magnitude   \\
23 & Log Stellar Mass       \\
24 & Log Stellar Mass from KGB\tablenotemark{3}  \\
25 & SDSS $r$ band 50\% light radius    \\
26 & SDSS $r$ band 90\% light radius   \\   
\enddata
\tablecomments{$^{1}$NUVugrizJHK magnitudes are reported with foreground extinction corrections determined from the \citet{schlegel.finkbeiner.ea:maps} dust maps using the extinction curves of \citet{odonnell:rnu-dependent} and \citet{cardelli.clayton.ea:relationship} in the optical and UV, respectively.  $^{2}$ We assume an effective wavelength of 2271$\AA$ for the {\it GALEX} NUV filter. $^{\rm{3}}$ Stellar masses derived by \citet{kannappan.guie.ea:es0} are not used in this paper. }
\end{deluxetable}

\clearpage
\LongTables
\setlength{\tabcolsep}{0.05in}
\begin{deluxetable*}{lcccccccccccc}
\tabletypesize{\scriptsize}
\tablewidth{0pt}
\tablecaption{Kinematic and HI data for the NFGS}
\tablehead{
\colhead{ID} 
& \colhead{PA\tablenotemark{  a}} 
& \colhead{$i$\tablenotemark{  b}} 
& \colhead{$V_{\rm pmm}$\tablenotemark{  c}} 
& \colhead{$r_e$\tablenotemark{  d}}
& \colhead{extent} 
& \colhead{asym.} 
& \colhead{W$_{50}$\tablenotemark{  e}} 
& \colhead{$\sigma_{r_e{_{/ 4}}}$}
& \colhead{$V_*$\tablenotemark{  f}} 
& \colhead{$V$\tablenotemark{  g}} 
& \colhead{$\log{M_{\rm HI}}$\tablenotemark{  h}} 
& \colhead{HI Range\tablenotemark{  j}}
\\
\colhead{} 
& \colhead{$^{\circ}$} 
& \colhead{$^{\circ}$} 
& \colhead{km\,s$^{-1}$} 
& \colhead{$\arcsec$} 
& \colhead{$\arcsec$} 
& \colhead{\%} 
& \colhead{km\,s$^{-1}$} 
& \colhead{km\,s$^{-1}$} 
& \colhead{km\,s$^{-1}$} 
& \colhead{km\,s$^{-1}$} 
& \colhead{$\log{M_{\odot}}$} 
& \colhead{km\,s$^{-1}$}
}
\startdata
1 & 160/160 & 39.9 & \nodata &  7.1 & \nodata & \nodata & \nodata & 133$\pm$10 & 120 & 188$\pm$14$^{sd}$ & $<$8.32 & 4661--4904\\
2 & 10/10 & 56.9 & 52 & 12.1 & 13.6 &  4.8 & [W10a] & 22$\pm$8\tablenotemark{{\it l}} & 66 & 99$\pm$19$^{sr}$ & W10a$^{\ddagger}$ & \nodata\\
3 & 55/48 & 20.2 & \nodata & 28.8 & \nodata & \nodata & W10a & 188$\pm$11 & \nodata & 266$\pm$16$^{sd}$ & W10a & \nodata\\
4 & 168/160 &  0.0 & 58 & 11.7 & 22.7 &  2.9 & W10a & 61$\pm$8\tablenotemark{{\it l}} & \nodata & [87$\pm$12]$^{sd}$ & W10a & \nodata\\
5 & 24/25 & 78.0 & 217 & 32.1 & 47.6 &  5.3 & W10a & 106$\pm$10\tablenotemark{{\it l}} & \nodata & 221$\pm$11$^{ir}$ & W10a & \nodata\\
7 & -/48 & 34.3 & \nodata & 32.3 & \nodata & \nodata & W10a & 316$\pm$18 & \nodata & 447$\pm$25$^{sd}$ & W10a & \nodata\\
8 & 24/25 & 62.1 & 173 & 10.6 & 22.7 &  1.3 & W10a & 132$\pm$10\tablenotemark{{\it l}} & 171 & 215$\pm$28$^{sr}$ & W10a & \nodata\\
10 & 110*/110* &  0.0 & \nodata &  6.6 & \nodata & \nodata & \nodata & 174$\pm$11 & 17 & 247$\pm$16$^{sd}$ & $<$9.02\tablenotemark{k} & 5094--5339\\
11 & -/90 & 27.9 & \nodata &  4.5 & \nodata & \nodata & [54$\pm$40] & 115$\pm$9 & 28 & 163$\pm$13$^{sd}$ & 8.62$\pm$0.20$^{\dagger}$ & 5322--5471\\
12 & 90/90 & 51.4 & 154 &  9.5 & 12.5 &  1.1 & [318$\pm$42] & 121$\pm$12\tablenotemark{{\it l}} & 163 & 232$\pm$31$^{sr}$ & 9.67$\pm$0.10$^{\dagger}$ & 9650--10716\\
13 & 122.5/- &  0.0 & [37] & 23.2 & 23.8 &  9.0 & [W10a] & \nodata & \nodata & \nodata & W10a$^{\ddagger}$ & \nodata\\
14 & -/44 & 74.2 & \nodata & 17.3 & \nodata & \nodata & \nodata & 49$\pm$8\tablenotemark{{\it l}} & 76 & 97$\pm$17$^{sr}$ & $<$7.57 & 2355--2548\\
15 & 2.5/5 & 49.5 & 9 & 13.1 & 15.9 &  9.4 & W10a & 28$\pm$11\tablenotemark{{\it l}} & \nodata & 68$\pm$10$^{nr}$ & W10a & \nodata\\
16 & 90*/90* & 27.9 & 93 &  8.2 & 19.3 &  2.3 & W10a & 79$\pm$12\tablenotemark{{\it l}} & \nodata & [289$\pm$146]$^{nr}$ & W10a & \nodata\\
17 & 168/- & 39.9 & 87 &  6.0 & 20.4 &  5.8 & W10a & \nodata & \nodata & [150$\pm$44]$^{ir}$ & W10a & \nodata\\
18 & 30/- & 37.8 & [11] & 61.7 & 29.5 & 37.0 & W10a & \nodata & \nodata & [44$\pm$10]$^{ir}$ & W10a & \nodata\\
19 & 40/44 & 26.4 & 75 &  7.6 & 14.7 &  2.9 & W10a & \nodata & \nodata & [358$\pm$172]$^{nr}$ & W10a & \nodata\\
21 & 132.5/- & 62.1 & 61 & 45.2 & 69.2 &  4.9 & W10a & \nodata & \nodata & 83$\pm$10$^{ir}$ & W10a & \nodata\\
22 & -/135 & 59.7 & \nodata & 10.7 & \nodata & \nodata & [374$\pm$10] & 134$\pm$10 & \nodata & 190$\pm$14$^{sd}$ & 9.69$\pm$0.02$^{\dagger}$ & 3526--3829\\
23 & 54/55 & 62.1 & 164 & 12.9 & 28.3 &  3.8 & [W10a] & 87$\pm$8\tablenotemark{{\it l}} & \nodata & 189$\pm$22$^{ir}$ & W10a$^{\ddagger}$ & \nodata\\
24 & 132.5*/- & 17.3 & 39 & 22.2 & 36.3 & 12.1 & W10a & \nodata & \nodata & [187$\pm$61]$^{nr}$ & W10a & \nodata\\
25 & 43/43 & 62.1 & 128 & 16.6 & 27.2 &  0.9 & W10a & 60$\pm$8\tablenotemark{{\it l}} & \nodata & 163$\pm$19$^{nr}$ & W10a & \nodata\\
26 & 54/- & 66.4 & 223 & 15.5 & 62.4 &  2.7 & W10a & \nodata & \nodata & 249$\pm$15$^{nr}$ & W10a & \nodata\\
27 & 70*/- &  0.0 & 49 & 16.3 & 31.8 &  8.3 & W10a & \nodata & \nodata & \nodata & W10a & \nodata\\
28 & 12/- & 71.1 & 50 & 29.4 & 30.6 &  4.9 & W10a & \nodata & \nodata & 67$\pm$10$^{ir}$ & W10a & \nodata\\
29 & 120/- & 49.5 & [197] & 22.6 & 17.0 &  1.9 & W10a & \nodata & \nodata & 377$\pm$31$^{nr}$ & W10a & \nodata\\
30 & 70/70 & 75.8 & \nodata & 13.7 & \nodata & \nodata & [261$\pm$8] & 186$\pm$12 & 205 & 262$\pm$17$^{sd}$ & 9.88$\pm$0.01$^{\dagger}$ & 3428--3850\\
32 & 45*/- & 39.9 & 41 & 46.1 & 61.2 &  3.5 & W10a & \nodata & \nodata & [84$\pm$11]$^{ir}$ & W10a & \nodata\\
33 & -/70 & 24.1 & \nodata & 11.4 & \nodata & \nodata & [84$\pm$3] & 177$\pm$14 & \nodata & 250$\pm$19$^{sd}$ & 9.15$\pm$0.10 & 6367--6529\\
34 & 90*/90* &  0.0 & 113 &  8.9 & 19.3 &  2.3 & [203$\pm$6] & 122$\pm$11\tablenotemark{{\it l}} & 79 & [172$\pm$15]$^{sd}$ & 10.24$\pm$0.03$^{\dagger}$ & 8100--8418\\
35 & 110/110 & 42.5 & \nodata & 22.6 & \nodata & \nodata & [277$\pm$15] & 241$\pm$15 & \nodata & 341$\pm$21$^{sd}$ & 9.10$\pm$0.06 & 4598--4957\\
36 & 165/170 & 78.0 & \nodata & 19.8 & \nodata & \nodata & W10a & 170$\pm$11\tablenotemark{{\it l}} & \nodata & 240$\pm$15$^{sd}$ & W10a & \nodata\\
37 & 1/170 & 48.1 & 259 & 21.5 & 35.2 &  1.9 & W10a & 234$\pm$14\tablenotemark{{\it l}} & \nodata & 348$\pm$35$^{nr}$ & W10a & \nodata\\
38 & 1/- & 85.0 & 54 & 33.7 & 52.2 &  3.0 & [W10a] & \nodata & \nodata & 67$\pm$10$^{ir}$ & W10a$^{\ddagger}$ & \nodata\\
39 & 150/151 & 46.8 & 123 & 15.2 & 22.7 &  4.0 & W10a & 47$\pm$19\tablenotemark{{\it l}} & 99 & 194$\pm$39$^{nr}$ & W10a & \nodata\\
40 & 80/90 & 54.7 & \nodata & 40.9 & \nodata & \nodata & \nodata & \nodata & \nodata & \nodata & $<$6.58 & 522--675\\
41 & 130/- & 82.4 & 53 & 25.2 & 31.8 &  6.9 & [W10a] & \nodata & \nodata & 66$\pm$10$^{ir}$ & W10a$^{\ddagger}$ & \nodata\\
42 & -/170 & 54.7 & \nodata & 45.6 & \nodata & \nodata & [146$\pm$86] & 122$\pm$9 & \nodata & 172$\pm$13$^{sd}$ & 8.44$\pm$0.19 & 2570--2930\\
43 & 13/14 & 64.0 & 148 & 18.4 & 43.1 &  2.6 & W10a & 110$\pm$9\tablenotemark{{\it l}} & \nodata & 174$\pm$11$^{nr}$ & W10a & \nodata\\
44 & 52/50 & 27.9 & 59 &  6.7 & 18.1 &  5.4 & 159$\pm$4 & 54$\pm$7\tablenotemark{{\it l}} & 24 & [170$\pm$86]$^{nr}$ & 8.53$\pm$0.03 & 1420--1665\\
45 & 115*/- & 29.6 & 3 & 12.4 & 18.1 & 14.4 & W10a & \nodata & \nodata & [415$\pm$221]$^{nr}$ & W10a & \nodata\\
46 & 1*/- & 31.7 & 77 & 10.9 & 19.3 &  5.6 & [W10a] & \nodata & \nodata & [167$\pm$94]$^{ir}$ & W10a$^{\ddagger}$ & \nodata\\
47 & 76/- & 57.3 & [68] & 37.2 & 28.3 &  6.1 & W10a & \nodata & \nodata & 102$\pm$7$^{nr}$ & W10a & \nodata\\
48 & 130/- & 49.5 & 136 & 21.4 & 37.4 &  3.8 & W10a & \nodata & \nodata & 186$\pm$28$^{ir}$ & W10a & \nodata\\
49 & 103/100 & 66.1 & 127 & 13.8 & 29.5 &  0.6 & W10a & 66$\pm$15\tablenotemark{{\it l}} & \nodata & 146$\pm$12$^{ir}$ & W10a & \nodata\\
50 & 40/- & 51.7 & 59 & 46.8 & 60.1 & 10.2 & W10a & \nodata & \nodata & 91$\pm$11$^{ir}$ & W10a & \nodata\\
51 & 103/- & 62.1 & 61 & 25.1 & 31.8 &  1.6 & W10a & \nodata & \nodata & 82$\pm$11$^{ir}$ & W10a & \nodata\\
52 & 1/- & 62.1 & 53 & 20.8 & 28.3 &  6.1 & [W10a] & \nodata & \nodata & 73$\pm$11$^{ir}$ & W10a$^{\ddagger}$ & \nodata\\
53 & 144/- & 54.7 & 78 & 36.8 & 64.6 &  5.8 & W10a & \nodata & \nodata & 123$\pm$6$^{nr}$ & W10a & \nodata\\
54 & -/50 & 81.2 & \nodata & 13.6 & \nodata & \nodata & \nodata & 195$\pm$11 & 218 & 276$\pm$16$^{sd}$ & $<$8.97 & 6913--7433\\
55 & 101/- & 37.8 & 27 & 15.2 & 22.7 &  6.1 & [W10a] & \nodata & \nodata & [68$\pm$21]$^{ir}$ & W10a$^{\ddagger}$ & \nodata\\
56 & 130/140 & 44.5 & 64 & 18.7 & 29.5 &  3.1 & W10a & 36$\pm$10\tablenotemark{{\it l}} & \nodata & 109$\pm$21$^{nr}$ & W10a & \nodata\\
57 & 1/- & 67.2 & 55 & 41.7 & 76.0 &  9.4 & W10a & \nodata & \nodata & 74$\pm$3$^{nr}$ & W10a & \nodata\\
58 & 154/- & 90.0 & 170 & 43.3 & 45.4 &  1.5 & W10a & \nodata & \nodata & 173$\pm$10$^{ir}$ & W10a & \nodata\\
59 & 110/110 & 27.9 & 41 &  6.5 &  7.9 &  2.7 & 104$\pm$27 & 38$\pm$10\tablenotemark{{\it l}} & 73 & [193$\pm$99]$^{sr}$ & 8.60$\pm$0.05 & 3100--3600\\
60 & 1/- & 85.5 & 145 & 24.4 & 31.8 &  1.6 & W10a & \nodata & \nodata & 151$\pm$7$^{nr}$ & W10a & \nodata\\
61 & 114/- & 59.7 & 169 & 15.3 & 30.6 &  2.4 & W10a & \nodata & \nodata & 200$\pm$18$^{ir}$ & W10a & \nodata\\
62 & 54/- & 75.8 & 68 & 19.4 & 31.8 &  6.5 & W10a & \nodata & \nodata & 82$\pm$10$^{ir}$ & W10a & \nodata\\
63 & 40/- & 84.6 & 71 & 18.1 & 31.8 &  5.5 & W10a & \nodata & \nodata & 82$\pm$10$^{ir}$ & W10a & \nodata\\
64 & 40/- & 31.7 & 40 &  7.4 & 10.2 &  6.4 & W10a & \nodata & \nodata & [210$\pm$118]$^{nr}$ & W10a & \nodata\\
65 & 121/120 & 84.8 & 333 & 23.6 & 36.3 &  1.4 & 708$\pm$7 & 231$\pm$14\tablenotemark{{\it l}} & 306 & 355$\pm$4$^{nr}$ & 10.16$\pm$0.03 & 10367--11208\\
66 & 130/140 &  0.0 & 95 &  9.1 & 23.8 &  2.6 & W10a & 78$\pm$8\tablenotemark{{\it l}} & \nodata & [110$\pm$12]$^{sd}$ & W10a & \nodata\\
67 & 144/140 & 26.4 & 136 & 15.4 & 26.1 &  7.1 & 279$\pm$3 & 124$\pm$10\tablenotemark{{\it l}} & \nodata & [319$\pm$154]$^{ir}$ & 9.90$\pm$0.04 & 7920--8242\\
68 & 172/170 & 66.1 & 67 & 14.5 & 19.3 &  8.1 & 140$\pm$7 & \nodata & \nodata & 85$\pm$13$^{ir}$ & 7.89$\pm$0.02 & 535--869\\
69 & 100*/100* &  0.0 & 8 &  4.0 &  5.7 & 19.3 & 136$\pm$10 & 46$\pm$11\tablenotemark{{\it l}} & 21 & [64$\pm$16]$^{sd}$ & 9.64$\pm$0.02 & 5644--6023\\
70 & 116/- & 86.6 & 102 & 31.3 & 45.4 &  2.0 & W10a & \nodata & \nodata & 111$\pm$10$^{ir}$ & W10a & \nodata\\
71 & 54/- & 79.6 & 111 & 23.0 & 39.7 &  4.5 & W10a & \nodata & \nodata & 123$\pm$4$^{nr}$ & W10a & \nodata\\
72 & -/50* & 90(29.6) & \nodata &  7.6 & \nodata & \nodata & [257$\pm$11] & 61$\pm$7\tablenotemark{{\it l}} & 49 & [133$\pm$74]$^{sr}$ & 8.07$\pm$0.06 & 1358--1725\\
73 & 165/- & 85.5 & 86 & 42.9 & 88.5 &  6.4 & W10a & \nodata & \nodata & 96$\pm$10$^{ir}$ & W10a & \nodata\\
75 & -/122 & 56.0 & \nodata & 14.8 & \nodata & \nodata & [245$\pm$60] & 99$\pm$8 & \nodata & 141$\pm$12$^{sd}$ & 8.62$\pm$0.05 & 934--1489\\
76 & 90/- & 63.6 & 107 & 43.8 & 62.4 &  4.6 & W10a & \nodata & \nodata & 128$\pm$11$^{ir}$ & W10a & \nodata\\
77 & 165/170 & 69.3 & 163 & 27.0 & 65.8 &  1.0 & [W10a] & 74$\pm$12\tablenotemark{{\it l}} & \nodata & 178$\pm$12$^{ir}$ & W10a$^{\ddagger}$ & \nodata\\
78 & -/100 & 45.6 & \nodata &  8.1 & \nodata & \nodata & \nodata & 195$\pm$13 & \nodata & 276$\pm$18$^{sd}$ & $<$8.87 & 7368--7750\\
79 & 20/- & 43.0(58.1) & 113 &  8.4 & 15.9 &  5.2 & [263$\pm$16] & \nodata & \nodata & 177$\pm$31$^{ir}$ & 10.67$\pm$0.01$^{\dagger}$ & 12500--13076\\
80 & -/18 & 62.1 & \nodata & 20.6 & \nodata & \nodata & \nodata & 88$\pm$7 & \nodata & 124$\pm$10$^{sd}$ & $<$6.62 & 549--772\\
81 & 165/170 & 76.2 & 90 & 24.3 & 29.5 &  2.5 & W10a & 69$\pm$21\tablenotemark{{\it l}} & \nodata & 103$\pm$11$^{ir}$ & W10a & \nodata\\
82 & 76/- & 74.2 & [151] & 23.1 & 20.4 &  3.9 & [W10a] & \nodata & \nodata & [161$\pm$12]$^{ir}$ & W10a$^{\ddagger}$ & \nodata\\
83 & 125/125 & 42.5 & 11 &  6.0 &  7.9 & 21.6 & [W10a] & \nodata & 23 & 59$\pm$25$^{sr}$ & W10a$^{\ddagger}$ & \nodata\\
84 & 11/- & 52.1 & 51 & 64.6 & 72.6 &  4.1 & W10a & \nodata & \nodata & 83$\pm$4$^{nr}$ & W10a & \nodata\\
85 & 133/140 & 75.4 & 131 & 19.3 & 26.1 &  3.2 & W10a & 56$\pm$11\tablenotemark{{\it l}} & \nodata & 142$\pm$11$^{ir}$ & W10a & \nodata\\
86 & -/170 & 51.4 & \nodata & 19.9 & \nodata & \nodata & \nodata & 257$\pm$15 & \nodata & 363$\pm$21$^{sd}$ & $<$9.07 & 9245--9786\\
87 & 122/122 & 62.1 & [23] &  9.5 &  7.9 &  2.4 & 100$\pm$6 & 49$\pm$9\tablenotemark{{\it l}} & \nodata & [70$\pm$13]$^{sd}$ & 8.65$\pm$0.02 & 1489--1794\\
88 & 172/- & 90.0 & 94 & 32.0 & 56.7 &  2.8 & W10a & \nodata & \nodata & 105$\pm$4$^{nr}$ & W10a & \nodata\\
89 & 80/- & 37.8 & 46 & 37.9 & 54.4 & 11.2 & W10a & \nodata & \nodata & [95$\pm$13]$^{ir}$ & W10a & \nodata\\
90 & 54/- & 29.6 & 115 &  9.2 & 22.7 &  4.0 & 171$\pm$5 & \nodata & \nodata & [248$\pm$132]$^{ir}$ & 10.64$\pm$0.01 & 10660--11110\\
91 & 52/50 & 80.2 & 106 & 25.6 & 53.3 &  1.9 & W10a & \nodata & \nodata & 116$\pm$10$^{ir}$ & W10a & \nodata\\
92 & 73/73 & 66.1 & 48 &  7.6 & 11.3 &  0.9 & [W10a] & 40$\pm$14\tablenotemark{{\it l}} & 34 & 67$\pm$12$^{ir}$ & W10a$^{\ddagger}$ & \nodata\\
93 & 18/18 & 54.7 & [4] &  7.6 &  9.1 & 19.5 & 51$\pm$3 & \nodata & \nodata & [31$\pm$8]$^{nr}$ & 7.44$\pm$0.02 & 682--824\\
94 & 40/- & 31.7 & 60 & 23.1 & 44.2 &  4.8 & W10a & \nodata & \nodata & [143$\pm$24]$^{nr}$ & W10a & \nodata\\
95 & 131/- & 65.4 & 82 & 36.3 & 45.4 &  7.7 & [W10a] & \nodata & \nodata & 102$\pm$11$^{ir}$ & W10a$^{\ddagger}$ & \nodata\\
96 & 100/100 & 45.6 & [24] &  5.6 &  5.7 & 20.7 & 102$\pm$10 & 34$\pm$12\tablenotemark{{\it l}} & 14 & [71$\pm$21]$^{nr}$ & 7.83$\pm$0.05 & 992--1252\\
97 & 165/- & 21.5 & [23] & 40.7 & 39.7 &  7.7 & W10a & \nodata & \nodata & [102$\pm$22]$^{ir}$ & W10a & \nodata\\
98 & 170/170 & 79.4 & [93] & 15.5 & 11.3 &  3.0 & W10a & \nodata & \nodata & 136$\pm$6$^{nr}$ & W10a & \nodata\\
99 & 146/- & 29.0 & 82 & 39.5 & 61.2 &  3.1 & W10a & \nodata & \nodata & [189$\pm$31]$^{ir}$ & W10a & \nodata\\
100 & 90.5/100 & 20.2 & 96 & 16.4 & 36.3 &  3.1 & W10a & 174$\pm$15\tablenotemark{{\it l}} & \nodata & [304$\pm$114]$^{ir}$ & W10a & \nodata\\
101 & 40*/- & 17.5 & 53 & 48.5 & 79.4 &  9.8 & W10a & \nodata & \nodata & [217$\pm$36]$^{ir}$ & W10a & \nodata\\
102 & 80/- & 51.1 & 67 & 21.3 & 36.3 &  2.6 & [W10a] & \nodata & \nodata & 100$\pm$14$^{ir}$ & W10a$^{\ddagger}$ & \nodata\\
103 & 1/6 & 44.9 & 91 & 25.2 & 37.4 &  3.5 & [W10a] & \nodata & \nodata & 142$\pm$19$^{ir}$ & W10a$^{\ddagger}$ & \nodata\\
104 & 172/- & 77.3 & 81 & 25.2 & 44.2 &  3.9 & W10a & \nodata & \nodata & 94$\pm$10$^{ir}$ & W10a & \nodata\\
105 & 100/100 & 62.1 & 11 &  8.4 & 10.2 &  6.7 & 73$\pm$4 & 38$\pm$14\tablenotemark{{\it l}} & 31 & 54$\pm$18$^{sr}$ & 8.81$\pm$0.01 & 1389--1663\\
106 & 18/18 & 72.2 & [46] & 16.8 &  7.9 &  3.2 & [253$\pm$39] & 66$\pm$8\tablenotemark{{\it l}} & \nodata & [94$\pm$11]$^{sd}$ & 8.51$\pm$0.04$^{\dagger}$ & 873--1150\\
107 & 165/- & 84.3 & 100 & 30.4 & 45.4 &  3.4 & W10a & \nodata & \nodata & 109$\pm$10$^{ir}$ & W10a & \nodata\\
108 & 80/- & 53.6 & [49] & 16.9 & 13.6 &  3.6 & [W10a] & \nodata & \nodata & [77$\pm$13]$^{ir}$ & W10a$^{\ddagger}$ & \nodata\\
109 & 80/- & 51.4 & 72 & 18.2 & 35.2 &  3.2 & W10a & \nodata & \nodata & 106$\pm$16$^{ir}$ & W10a & \nodata\\
110 & 40/- & 69.3 & 75 & 18.4 & 34.0 &  7.6 & W10a & \nodata & \nodata & 93$\pm$8$^{nr}$ & W10a & \nodata\\
111 & 146/- &  0.0 & 19 & 61.5 & 70.3 &  9.3 & W10a & \nodata & \nodata & \nodata & W10a & \nodata\\
112 & 40/- & 81.2 & 118 & 26.9 & 44.2 &  2.8 & W10a & \nodata & \nodata & 126$\pm$10$^{ir}$ & W10a & \nodata\\
113 & 108/110 & 68.8 & [25] & 40.7 & 37.4 &  6.1 & W10a & \nodata & \nodata & [43$\pm$10]$^{ir}$ & W10a & \nodata\\
114 & 40/- & 77.3 & 146 & 26.3 & 46.5 &  1.9 & W10a & \nodata & \nodata & 154$\pm$11$^{ir}$ & W10a & \nodata\\
115 & -/110* & 26.4 & \nodata & 15.6 & \nodata & \nodata & [227$\pm$14] & 248$\pm$15 & \nodata & 351$\pm$21$^{sd}$ & 10.11$\pm$0.02$^{\dagger}$ & 8312--8634\\
116 & 131/- & 54.7 & 65 & 32.3 & 64.6 &  8.0 & W10a & \nodata & \nodata & 106$\pm$6$^{nr}$ & W10a & \nodata\\
117 & -/46* & 29.6 & \nodata &  8.4 & \nodata & \nodata & \nodata & 59$\pm$7 & 24 & 83$\pm$10$^{sd}$ & $<$6.02 & 607--703\\
118 & 1*/- & 23.1 & 10 & 25.0 & 30.6 & 15.4 & W10a & \nodata & \nodata & [86$\pm$37]$^{nr}$ & W10a & \nodata\\
120 & 63/- & 29.6 & [20] & 59.1 & 47.6 &  3.9 & W10a & \nodata & \nodata & [70$\pm$12]$^{ir}$ & W10a & \nodata\\
121 & 89/- & 52.8 & 24 & 15.9 & 29.5 & 21.2 & W10a & \nodata & \nodata & 48$\pm$11$^{ir}$ & W10a & \nodata\\
122 & 18/- & 39.4 & 31 & 24.3 & 39.7 &  3.1 & [W10a] & \nodata & \nodata & [71$\pm$13]$^{ir}$ & W10a$^{\ddagger}$ & \nodata\\
123 & 40/- & 78.9 & 105 & 51.6 & 93.0 &  5.1 & W10a & \nodata & \nodata & 115$\pm$10$^{ir}$ & W10a & \nodata\\
124 & 33/- & 54.7 & 28 & 10.5 & 21.5 & 30.2 & [W10a] & \nodata & \nodata & 52$\pm$12$^{ir}$ & W10a$^{\ddagger}$ & \nodata\\
125 & 100/100 & 45.6 & 99 & 11.7 & 21.5 &  6.0 & 242$\pm$4 & 61$\pm$13\tablenotemark{{\it l}} & \nodata & 169$\pm$49$^{nr}$ & 9.97$\pm$0.02 & 6680--7124\\
126 & 80/- & 44.9 & 27 & 30.1 & 35.2 & 25.6 & W10a & \nodata & \nodata & 59$\pm$12$^{ir}$ & W10a & \nodata\\
127 & 159/- & 79.6 & [86] & 47.6 & 57.8 &  6.9 & W10a & \nodata & \nodata & [97$\pm$10]$^{ir}$ & W10a & \nodata\\
128 & 115/116 & 31.7 & 34 & 22.9 & 10.2 & 28.7 & [W10a] & 160$\pm$10\tablenotemark{{\it l}} & 178 & [374$\pm$64]$^{sr}$ & W10a$^{\ddagger}$ & \nodata\\
129 & -/30* & 42.5 & \nodata &  8.2 & \nodata & \nodata & \nodata & 203$\pm$12 & \nodata & 286$\pm$17$^{sd}$ & $<$8.83 & 6801--7176\\
130 & -/135* &  0.0 & \nodata &  8.0 & \nodata & \nodata & \nodata & 172$\pm$11 & \nodata & 243$\pm$15$^{sd}$ & $<$9.67\tablenotemark{k} & 6170--6408\\
131 & -/71* & 25.2 & \nodata & 13.1 & \nodata & \nodata & \nodata & 266$\pm$15 & \nodata & 376$\pm$22$^{sd}$ & $<$8.77 & 7710--8075\\
132 & 100/100 & 47.8 & 105 & 17.1 & 36.3 &  2.7 & W10a & 49$\pm$12\tablenotemark{{\it l}} & \nodata & 152$\pm$20$^{ir}$ & W10a & \nodata\\
133 & 80/- & 75.8 & 85 & 23.3 & 35.2 &  5.7 & 208$\pm$2 & \nodata & \nodata & 107$\pm$4$^{nr}$ & 9.79$\pm$0.01 & 2340--2760\\
134 & 172/- & 90.0 & 88 & 35.7 & 57.8 &  4.8 & W10a & \nodata & \nodata & 98$\pm$10$^{ir}$ & W10a & \nodata\\
135 & 40/- & 29.6 & 65 & 11.1 & 19.3 &  7.4 & 156$\pm$2 & \nodata & \nodata & [158$\pm$84]$^{nr}$ & 9.39$\pm$0.02 & 3179--3596\\
136 & 153/- & 65.4 & 101 & 27.4 & 53.3 &  4.9 & W10a & \nodata & \nodata & 121$\pm$11$^{ir}$ & W10a & \nodata\\
137 & 100*/100* & 39.7 & [82] &  9.4 &  6.8 & \nodata & W10a & 101$\pm$8 & \nodata & 143$\pm$12$^{sd}$ & W10a & \nodata\\
138 & 1/170* & 75.1 & 285 & 26.3 & 55.6 &  4.9 & W10a & 196$\pm$12\tablenotemark{{\it l}} & \nodata & 309$\pm$24$^{nr}$ & W10a & \nodata\\
139 & -/170 & 71.1 & \nodata & 18.0 & \nodata & \nodata & [402$\pm$10] & 257$\pm$15 & \nodata & 363$\pm$22$^{sd}$ & 9.50$\pm$0.05$^{\ddagger}$\tablenotemark{k} & \nodata\\
140 & 1*/- & 31.7 & 23 & 28.1 & 46.5 &  9.9 & W10a & \nodata & \nodata & [141$\pm$23]$^{nr}$ & W10a & \nodata\\
141 & 100/- & 77.3 & 69 & 27.4 & 31.8 &  4.0 & W10a & \nodata & \nodata & 95$\pm$5$^{nr}$ & W10a & \nodata\\
142 & 60/56 & 72.4 & 290 & 21.7 & 32.9 &  1.4 & W10a & 175$\pm$12\tablenotemark{{\it l}} & 249 & 298$\pm$17$^{ir}$ & W10a & \nodata\\
143 & 168/- & 76.2 & 114 & 20.3 & 30.6 &  3.2 & W10a & \nodata & \nodata & 125$\pm$11$^{ir}$ & W10a & \nodata\\
144 & 100/100 & 57.3 & [21] & 22.9 &  4.5 & 11.5 & 46$\pm$16$^{\dagger \dagger}$ & 45$\pm$11\tablenotemark{{\it l}} & \nodata & [64$\pm$16]$^{sd}$ & 7.50$\pm$0.06 & 668--925\\
145 & 18/16 & 79.6 & [139] & 52.9 & 62.4 &  3.5 & W10a & 34$\pm$9\tablenotemark{{\it l}} & \nodata & [146$\pm$10]$^{ir}$ & W10a & \nodata\\
146 & -/30 & 78.7 & \nodata & 10.9 & \nodata & \nodata & \nodata & 142$\pm$10 & 153 & 201$\pm$14$^{sd}$ & $<$7.94 & 1900--2282\\
147 & -/100 & 45.6 & \nodata & 19.1 & \nodata & \nodata & \nodata & 194$\pm$12 & \nodata & 275$\pm$16$^{sd}$ & $<$8.50 & 5230--5610\\
148 & 127/- & 82.6 & 109 & 28.1 & 48.8 &  1.4 & W10a & \nodata & \nodata & 118$\pm$10$^{ir}$ & W10a & \nodata\\
149 & 40/36 & 83.3 & \nodata & 14.7 & \nodata & \nodata & \nodata & 151$\pm$10 & 164 & 213$\pm$14$^{sd}$ & $<$8.39 & 3632--4040\\
150 & 65/56 & 90.0 & [120] & 46.2 & 59.0 &  2.1 & W10a & 30$\pm$15\tablenotemark{{\it l}} & \nodata & [127$\pm$10]$^{ir}$ & W10a & \nodata\\
151 & 80/- & 52.8 & 227 & 17.1 & 44.2 &  1.9 & [W10a] & \nodata & \nodata & 283$\pm$28$^{ir}$ & W10a$^{\ddagger}$ & \nodata\\
152 & -/140 &  0.0 & \nodata & 22.3 & \nodata & \nodata & \nodata & 341$\pm$19 & \nodata & 483$\pm$27$^{sd}$ & $<$9.10 & 7214--7676\\
153 & 1/- & 39.9 & 156 & 10.3 & 18.1 &  2.0 & [339$\pm$13] & \nodata & \nodata & [250$\pm$71]$^{ir}$ & 10.19$\pm$0.02$^{\dagger}$ & 7540--7860\\
154 & -/100 & 42.5 & \nodata &  9.4 & \nodata & \nodata & [71$\pm$143] & 113$\pm$10 & \nodata & 160$\pm$14$^{sd}$ & 8.80$\pm$0.11 & 2934--3588\\
155 & 100/- & 57.2 & 41 & 33.2 & 42.0 &  8.1 & W10a & \nodata & \nodata & 64$\pm$10$^{ir}$ & W10a & \nodata\\
156 & 100/100 & 62.1 & 98 & 16.3 & 31.8 &  2.9 & W10a & \nodata & \nodata & 129$\pm$9$^{nr}$ & W10a & \nodata\\
157 & -/100 & 33.4 & \nodata & 19.8 & \nodata & \nodata & \nodata & 164$\pm$11 & \nodata & 231$\pm$16$^{sd}$ & $<$8.24 & 3981--4234\\
158 & 140/140 & 42.5 & 80 & 19.2 & 47.6 &  2.8 & W10a & 75$\pm$7\tablenotemark{{\it l}} & \nodata & 134$\pm$18$^{ir}$ & W10a & \nodata\\
159 & 68/- & 75.8 & 28 &  7.4 & 22.7 & 47.4 & W10a & \nodata & \nodata & 93$\pm$6$^{nr}$ & W10a & \nodata\\
160 & 29/- & 68.0(0.0) & 61 & 10.7 & 22.7 & 16.4 & W10a & \nodata & \nodata & 92$\pm$7$^{nr}$ & W10a & \nodata\\
161 & 68/75 & 56.9 & 93 & 17.5 & 34.0 &  3.4 & [W10a] & 44$\pm$7\tablenotemark{{\it l}} & \nodata & 122$\pm$15$^{ir}$ & W10a$^{\ddagger}$ & \nodata\\
162 & 80/- & 66.6 & 19 & 11.6 & 17.0 & 10.6 & W10a & \nodata & \nodata & 38$\pm$3$^{nr}$ & W10a & \nodata\\
164 & 60/- & 46.8 & 110 & 12.7 & 22.7 &  3.7 & 279$\pm$7 & \nodata & \nodata & 192$\pm$37$^{nr}$ & 9.98$\pm$0.02 & 6240--7040\\
165 & 55/54 & 50.6 & [119] & 35.4 & 45.4 &  2.5 & W10a & 75$\pm$8\tablenotemark{{\it l}} & \nodata & 195$\pm$14$^{nr}$ & W10a & \nodata\\
166 & 100/- & 27.9 & 22 & 10.6 & 17.0 &  9.8 & 123$\pm$5 & \nodata & \nodata & [132$\pm$66]$^{nr}$ & 8.76$\pm$0.02 & 2310--2650\\
167 & 159/140* & 54.7 & 256 & 18.5 & 26.1 &  1.9 & 544$\pm$6 & 181$\pm$12\tablenotemark{{\it l}} & \nodata & 333$\pm$31$^{nr}$ & 9.97$\pm$0.03 & 8400--9100\\
168 & 80/75 & 66.1 & 180 & 11.8 & 29.5 &  1.6 & W10a & 100$\pm$10\tablenotemark{{\it l}} & 157 & 199$\pm$21$^{ir}$ & W10a & \nodata\\
170 & 12/- & 42.5 & 92 & 17.4 & 34.0 &  3.0 & W10a & \nodata & \nodata & 187$\pm$27$^{nr}$ & W10a & \nodata\\
171 & 140/140 & 32.0(42.5) & [71] & 13.2 & 15.9 & 17.9 & W10a & \nodata & \nodata & [161$\pm$40]$^{nr}$ & W10a & \nodata\\
172 & 94/100 & 26.4 & 44 &  6.0 & 18.1 & 11.2 & 194$\pm$5 & 98$\pm$9 & \nodata & 139$\pm$13$^{sd}$ & 9.52$\pm$0.02 & 4249--4524\\
173 & -/100 & 45.6 & \nodata &  6.9 & \nodata & \nodata & \nodata & 205$\pm$14 & \nodata & 290$\pm$20$^{sd}$ & $<$9.07 & 11987--12387\\
174 & 66/66 & 46.3 & 173 & 20.2 & 45.4 &  3.5 & W10a & 133$\pm$10\tablenotemark{{\it l}} & \nodata & 270$\pm$31$^{nr}$ & W10a & \nodata\\
175 & 177/178 & 78.0 & 101 & 10.9 & 15.9 &  2.4 & \nodata & 34$\pm$13\tablenotemark{{\it l}} & \nodata & 112$\pm$12$^{ir}$ & $<$8.40\tablenotemark{k} & 2115--2354\\
176 & -/140 & 33.0 & \nodata & 16.7 & \nodata & \nodata & [113$\pm$15] & 314$\pm$18 & \nodata & 444$\pm$26$^{sd}$ & 8.77$\pm$0.08$^{\dagger}$ & 5880--6088\\
177 & 140/140 & 45.6 & \nodata & 17.1 & \nodata & \nodata & W10a & 189$\pm$11 & \nodata & 267$\pm$16$^{sd}$ & W10a & \nodata\\
178 & -/2 & 81.2 & \nodata & 10.8 & \nodata & \nodata & W10a & 134$\pm$9 & 182 & 203$\pm$19$^{sr}$ & W10a & \nodata\\
179 & 140*/- &  0.0 & 99 & 20.1 & 30.6 &  3.5 & W10a & \nodata & \nodata & \nodata & W10a & \nodata\\
180 & 1/2 & 49.5 & [238] & 19.0 &  7.9 &  3.8 & \nodata & 236$\pm$14\tablenotemark{{\it l}} & 163 & [333$\pm$20]$^{sd}$ & $<$9.47 & 10039--10631\\
181 & 44*/45* & 40.0(0.0) & 32 &  7.0 & 10.2 &  5.5 & [W10a] & 38$\pm$14\tablenotemark{{\it l}} & \nodata & [71$\pm$11]$^{ir}$ & W10a$^{\ddagger}$ & \nodata\\
182 & -/20 & 47.4 & \nodata & 18.8 & \nodata & \nodata & [W10a] & 268$\pm$15 & \nodata & 379$\pm$22$^{sd}$ & W10a$^{\ddagger}$ & \nodata\\
183 & 145/145.1 & 49.5 & 79 &  9.9 & 15.9 &  2.8 & W10a & 28$\pm$7\tablenotemark{{\it l}} & 68 & 117$\pm$36$^{ir}$ & W10a & \nodata\\
184 & 88/90 & 74.2 & 151 & 24.8 & 53.3 &  2.9 & W10a & 82$\pm$10\tablenotemark{{\it l}} & 146 & 171$\pm$18$^{sr}$ & W10a & \nodata\\
185 & 153/153 & 67.2 & 167 & 12.6 & 21.5 &  4.9 & W10a & 124$\pm$9 & \nodata & 184$\pm$21$^{ir}$ & W10a & \nodata\\
186 & 10/- & 61.0 & 110 & 50.6 & 73.7 &  9.1 & W10a & \nodata & \nodata & 134$\pm$11$^{ir}$ & W10a & \nodata\\
187 & 90/90 & 57.0(42.5) & 101 &  5.5 &  7.9 &  2.8 & [196$\pm$7] & 78$\pm$13\tablenotemark{{\it l}} & 104 & 145$\pm$19$^{sr}$ & 9.65$\pm$0.03$^{\dagger}$ & 5605--5846\\
188 & -/90* &  0.0 & \nodata & 28.1 & \nodata & \nodata & \nodata & 275$\pm$16 & \nodata & 389$\pm$22$^{sd}$ & $<$9.42\tablenotemark{k} & 7233--7611\\
189 & 45/45 & 25.2 & 91 & 17.9 & 30.6 &  2.2 & W10a & 45$\pm$12\tablenotemark{{\it l}} & \nodata & [236$\pm$109]$^{ir}$ & W10a & \nodata\\
190 & 79/90* & 82.4 & 146 & 24.4 & 62.4 &  3.1 & W10a & 66$\pm$9\tablenotemark{{\it l}} & \nodata & 155$\pm$5$^{nr}$ & W10a & \nodata\\
191 & 90*/90* & 37.8 & 74 & 15.0 & 22.7 & 13.8 & W10a & 182$\pm$12 & \nodata & 258$\pm$16$^{sd}$ & W10a & \nodata\\
192 & 158/158.1 & 78.0 & 153 & 22.4 & 38.6 &  1.6 & W10a & 70$\pm$7\tablenotemark{{\it l}} & 154 & 177$\pm$18$^{sr}$ & W10a & \nodata\\
193 & 158*/158* & 23.1 & 88 & 10.5 & 21.5 &  3.4 & W10a & 85$\pm$9\tablenotemark{{\it l}} & \nodata & [303$\pm$129]$^{nr}$ & W10a & \nodata\\
194 & -/90* & 31.7 & \nodata & 18.4 & \nodata & \nodata & W10a & 297$\pm$17 & \nodata & 420$\pm$25$^{sd}$ & W10a & \nodata\\
195 & 94/94 & 79.9 & 216 & 28.3 & 49.9 &  5.7 & W10a & 127$\pm$13\tablenotemark{{\it l}} & \nodata & 219$\pm$11$^{ir}$ & W10a & \nodata\\
196 & 113/110 & 66.1 & 115 &  9.7 & 19.3 &  9.1 & [W10a] & 77$\pm$30\tablenotemark{{\it l}} & \nodata & 134$\pm$16$^{ir}$ & W10a$^{\ddagger}$ & \nodata\\
197 & -/180 & 44.5 & \nodata & 12.9 & \nodata & \nodata & \nodata & 231$\pm$14 & \nodata & 326$\pm$20$^{sd}$ & $<$8.69 & 6741--7183\\
198 & 12/20 & 73.2 & 73 & 32.4 & 61.2 & 11.2 & W10a & \nodata & \nodata & 88$\pm$10$^{ir}$ & W10a & \nodata\\
\enddata
\tablecomments{$^{\rm{a}}$ Slit PAs of observations used to derive ionized gas rotation curve and stellar absorption line kinematics, respectively. $^{\rm{b}}$ Parentheses indicate inclination estimates from KFF, usually inferior except for UGC9562 and NGC3499, for which the new estimate applies only to the gas, not the stars. $^{\rm{c}}$ Bracketed numbers denote unreliable velocities (\S\ref{sec:optkin}).  Uncertainties on $V_{\rm pmm}$ are set to 11\,km~s$^{-1}$ following KFF (\S\ref{sec:optkin}). $^{\rm{d}}$ Half-light radius in the $B$ band from Jansen et al. (2000b), converted from the authors' geometric mean aperture radius convention to a major axis radius convention. This $B$-band radius is denoted $r_e$ as distinct from the $r$-band half-light radius denoted $r_{50}^r$ in Table~1.$^{\rm{e}}$ Bracketed numbers denote unreliable linewidths (\S\ref{sec:optkin}). $^{\rm{f}}$ Uncertainties in V$_*$ are set to 17 km\,s$^{-1}$ (\S\ref{sec:optkin}). $^{\rm{g}}$ Bracketed numbers denote unreliable linewidths. $^{\rm{h}}$ Mass includes factor of 1.4 to account for He. $^{\rm{j}}$ Range of heliocentric velocities used in the HI flux measurement or upper limit determination. $^{\rm{k}}$ HI data from ALFALFA survey, with upper limits estimated from the survey sensitivity limit (Haynes et al. 2011).$^l$ Stellar velocity dispersion may be an unreliable metric of characteristic velocity $V$, based on either late type morphology or low dispersion (\S\ref{sec:optkin}). * PA uncertain or slit PA misaligned by more than 10$^{\circ}$ with the galaxy major axis. $^{\dagger}$ Confirmed likely confused; HI flux corrected as described in \S\ref{sec:hi} and HI linewidth designated as unreliable. $^{\ddagger}$ Literature HI data possibly subject to confusion based on identification of a close companion or ongoing interaction/merger by KFF. $^{\dagger \dagger}$ Linewidth inconsistent with much larger ALFALFA value; both come from low S/N ($\sim$6) spectra and are inconsistent with the Tully-Fisher relation.}
\end{deluxetable*}

\end{document}